\documentclass[aps,nobibnotes,nofootinbib,reprint,superscriptaddress,amsmath,amssymb]{revtex4-1}

\usepackage{subfigure}

\usepackage{dsfonts}
\usepackage{color}
\usepackage{graphicx}
\usepackage{dcolumn}
\usepackage{bm}
\usepackage{multirow}

\usepackage{hyperref}
\usepackage{natbib}

\newcommand{\fsl}[1]{#1\kern-0.55em/}
\DeclareMathOperator{\tr}{tr}

\DeclareMathOperator{\STr}{STr}

\DeclareMathAlphabet{\EuRoman}{U}{cmsy}{m}{n}
\SetMathAlphabet{\EuRoman}{bold}{U}{cmsy}{b}{n}
\newcommand{\matheu}{\EuRoman}

\begin{document}

\title{Effect of short-range interactions on the quantum critical behavior of spinless fermions on the honeycomb lattice}

\author{D. Mesterh\'azy}
\email{d.mesterhazy@thphys.uni-heidelberg.de}
\author{J. Berges}
\email{j.berges@thphys.uni-heidelberg.de}
\affiliation{
  Institut f\"ur Theoretische Physik, Universit\"at Heidelberg\\ 
  Philosophenweg 16, 69120 Heidelberg, Germany
}
\author{L. von Smekal}
\email{lorenz.smekal@physik.tu-darmstadt.de}
\affiliation{
  Institut f\"ur Kernphysik, Technische Universit\"at Darmstadt\\ 
  Schlossgartenstra{\ss}e 2, 64289 Darmstadt, Germany
}

\date{\today}

\begin{abstract}
We present a functional renormalization group investigation of an Euclidean three-dimensional matrix Yukawa model with $U(N)$ symmetry, which describes $N = 2$ Weyl fermions that effectively interact via a short-range repulsive interaction. This system relates to an effective low-energy theory of spinless electrons on the honeycomb lattice and can be seen as a simple model for suspended graphene. We find a continuous phase transition characterized by large anomalous dimensions for the fermions and composite degrees of freedom. The critical exponents define a new universality class distinct from Gross-Neveu type models, typically considered in this context.
\end{abstract}

\maketitle

\section{Introduction}
\label{Sec:Introduction}

Since the experimental realization of graphene \cite{Novoselov:2005kj, *Geim:2007} there has been a tremendous activity leading to new intriguing phenomena in condensed matter physics \cite{Geim:2009}. The characteristic feature of graphene is the presence of the so-called Dirac points at the corners of the first Brillouin zone. At these special points, a linear dispersion for the low-energy excitations occurs \cite{CastroNeto:2009zz}, closely resembling that of massless relativistic fermions. The massless relativistic dispersion leads to remarkable electronic properties. A prominent example is the observation of the anomalous quantum Hall effect corresponding to a pseudospin--$\tfrac{1}{2}$ Berry phase \cite{Novoselov:2005kj, Zhang:2005}. Moreover, graphene may serve as a simple toy model for studying long sought-after quantum relativistic effects \cite{Geim:2009} as, e.\,g.\ Klein tunneling \cite{Katsnelson:2006b, Beenakker:2008} and Zitterbewegung \cite{Katsnelson:2006a}. These phenomena can be understood in the framework of non-interacting relativistic Dirac fermions which are realized in monolayer graphene on a substrate. However, for suspended graphene \cite{Meyer:2007, *Bolotin:2008, *Du:2008} the situation is different and the system is strongly influenced by the large unscreened Coulomb interactions \cite{Kotov:2010}. In what way the dynamics modifies the low-energy behavior of the excitations in graphene is an important open question. This parallels the situation in strongly interacting quantum field theories, as e.\,g.\ quantum chromodynamics (QCD) where the interaction at low energies leads to the spontaneous breaking of chiral symmetry \cite{Braun:2011}. In that sense, graphene can be seen as a laboratory for strongly interacting fermions. For suspended graphene the essential question is whether the Coulomb interactions are strong enough to drive the system close to an interacting fixed point. In the vicinity of a fixed point the system is governed only by the low-energy modes. The details of the underlying lattice theory are no longer relevant, and the theory drastically simplifies. There it often occurs that one has additional symmetries that are not present in the lattice theory \cite{Sachdev:2001}. Striking examples being the effective relativistic dispersion and the effective chiral symmetry for the low-energy theory.

In this work, we consider the situation where the low-energy theory is defined in the vicinity of an interacting fixed point and we inquire specifically about its critical properties. In fact, suspended graphene may be expected to be close to a non-trivial quantum critical point if the coupling is sufficiently strong \cite{Son:2007ja}. While in the perturbative regime short-range interactions are irrelevant for the dynamics, at strong coupling this is not necessarily so. Local four fermion interactions can be generated dynamically and may play an important role even for the long-range correlated system \cite{Drut:2007zx, Juricic:2009px, *Herbut:2009vu, Semenoff:2012}. In the past, the role of the short-range repulsive interactions has been studied in the framework of the tight-binding model on the honeycomb lattice where, depending on the strength of the interactions, a competition between the staggered density and non-trivial topological phases was found \cite{Raghu:2008}. Although both types of order are conceivable, for the case of suspended graphene, one expects a semimetal-Mott insulator phase transition \cite{Gorbar:2002iw, *Gamayun:2009em, Drut:2008rg, *Drut:2009zi, Herbut:2006cs, *Herbut:2009qb, Juricic:2009px, *Herbut:2009vu, Semenoff:2012}, where the chiral symmetry is broken spontaneously by a non-zero vacuum expectation value of the chiral condensate. This corresponds to a type of staggered density phase \cite{Semenoff:1984dq, Herbut:2006cs, *Herbut:2009qb} that alternates on the two sublattices of the bipartite honeycomb lattice.

Here, we specifically address the critical properties for this chiral phase transition using the non-perturbative functional renormalization group \cite{Wetterich:1992yh, *Berges:2000ew}. In particular, we neglect the influence of the long-range Coulomb interactions and characterize the properties of the short-range repulsive quantum critical point. Our approach circumvents the problems of a purely perturbative approach close to criticality, and for the first time, allows us to follow the flow of this model into the broken phase. Introducing composite degrees of freedom for the order parameters, we show that our model has a continuous phase transition in the universality class of a three-dimensional matrix Yukawa model with $U(2)$ symmetry.

The structure of this paper is as follows: In Sec.\ \ref{Sec:LowEnergyTheory} and \ref{Sec:SymmetryProperties} we introduce the low-energy theory for the relativistic quasiparticles in graphene and discuss its symmetry properties. We consider the behavior of fermion bilinears under the discrete parity $\matheu{P}$, charge conjugation $\matheu{C}$, and time-reversal $\matheu{T}$ transformations and identify their role in the underlying lattice theory. In Sec.\ \ref{Sec:SimpleModel} we introduce the low-energy theory of spinless fermions on the honeycomb lattice which serves as a model for the dynamics of fermions interacting via a short-range repulsive interaction. In Sec.\ \ref{Sec:FunctionalRenormalizationGroup}, within the framework of the functional renormalization group, we derive the flow equations for the partially bosonized model. Finally, in Sec.\ \ref{Sec:Results}, we characterize the properties of the continuous chiral phase transition and determine the critical scaling exponents.

\section{Low-energy theory}
\label{Sec:LowEnergyTheory}

On a substrate the low-energy theory of graphene is described by the free Lagrangian 
\begin{equation}
\mathcal{L} = i \bar{\psi}^{a} \gamma_{\mu} \partial_{\mu} \psi^{a} ~,
\label{Eq:FreeLagrangian}
\end{equation}
with the linear dispersion of Dirac quasiparticles. The flavor index takes the values $a = 1 , \ldots , N_{f}$ and characterizes the physical spin of the quasiparticles. For a single layer of graphene, the number of Dirac fermions is $N_{f} = 2$. In this work, we take $N_{f} = 1$, which corresponds to a system of spinless fermions on the honeycomb lattice whose band structure can also be modeled by photonic crystals \cite{Haldane:2008a, *Haldane:2008b, Sepkhanov:2007, *Sepkhanov:2008, Bittner:2010, *Bittner:2012}. In the following, we will leave the value $N_{f}$ unspecified as long as not stated otherwise. The low-energy excitations on the honeycomb lattice in two space dimensions are described in terms a Lagrangian in \mbox{$d = 3$} Euclidean space-time dimensions, with the index $\mu = 0, 1, 2$. That is, throughout this work we assume full Euclidean rotational invariance. For the 2+1-dimensional relativistic theory, this translates to the statement that the dynamical critical exponent is assumed to be $z = 1$ and that the Fermi velocity $v_{F}$ is non-critical. In fact, it has been argued that close to the semimetal-insulator critical point Lorentz-symmetry breaking perturbations are irrelevant and that a description in terms of a Euclidean-invariant low-energy theory is viable \cite{Herbut:2009qb,*Herbut:2009vu,Roy:2011pg}. In Euclidean space-time\footnote{For our Euclidean conventions see e.\,g.\ \cite{Wetterich:2010ni}.} the Lagrangian \eqref{Eq:FreeLagrangian} satisfies Osterwalder-Schrader reflection positivity \cite{Osterwalder:1973dx}, and the spinors $\psi^{\dagger} \equiv i \bar{\psi} \gamma_{0}$ are not conjugate to $\psi$, but instead define independent degrees of freedom. Furthermore, we use a reducible chiral representation for the fermions where the gamma matrices satisfy the Dirac algebra
\begin{equation}
\{ \gamma_{\mu} , \gamma_{\nu} \} = 2 \delta_{\mu \nu} ~, \quad \mu, \nu = 0, 1 ,2 ~,  
\end{equation}
and are given explicitly by
\begin{equation}
\gamma_{0} = \begin{pmatrix} 0 & - i \sigma_{3} \\ i \sigma_{3} & 0 \end{pmatrix} ~, \quad \gamma_{k} = \begin{pmatrix} 0 & - i \sigma_{k} \\ i \sigma_{k} & 0 \end{pmatrix} ~, \quad k = 1,2 ~,
\label{Eq:DiracRepresentation}
\end{equation}
where $\sigma_{k}$, $k = 1,2,3$ denote the $2\times 2 $ Pauli matrices. Apart from these matrices the Dirac algebra consists of the two matrices
\begin{equation}
\gamma_{3} = \begin{pmatrix} 0 ~ & ~ \mathds{1}  \\ \mathds{1} ~ & ~ 0 \end{pmatrix} ~, \qquad \gamma_{5} = \begin{pmatrix} \mathds{1}  & 0 \\ 0 & -\mathds{1} \end{pmatrix} ~,
\end{equation}
that anticommute with all $\gamma_{\mu}$, $\mu = 0,1,2$, and with each other, as well as their combination $\gamma_{35} = \tfrac{i}{2} [\gamma_{3} , \gamma_{5}]$. Note, that these matrices do not appear in the Lagrangian \eqref{Eq:FreeLagrangian} which gives rise to a certain freedom to define the discrete symmetries \cite{Gomes:1990ed} (see Sec. \ref{Sec:SymmetryProperties}).

\begin{figure}
\centering
\includegraphics[width=0.5\textwidth]{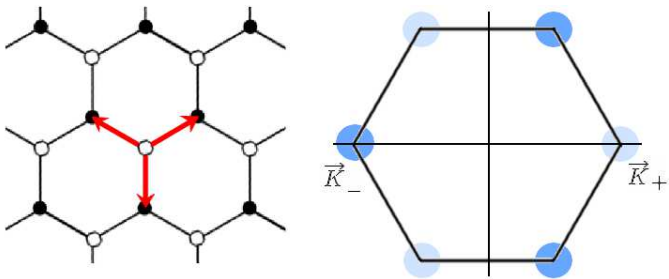}
\caption{\label{Fig:Graphene} {(Left) The bipartite hexagonal lattice with the two sublattices $A$ and $B$ indicated by full and open dots. The red arrows denote the nearest neighbor hopping in the tight-binding Hamiltonian \eqref{Eq:TightBindingHamiltonian}. (Right) The two inequivalent Dirac points $\vec{K}_{+}$ and $\vec{K}_{-}$ at opposite corners of the first Brillouin zone.}}
\end{figure}

In the chiral representation the states with definite chirality
\begin{equation}
\gamma_{5} \psi_{\pm} = \pm \psi_{\pm} ~,
\end{equation}
are taken to define the excitations around the two distinct Dirac points $\vec{K}_{+}$ and $\vec{K}_{-} = -\vec{K}_{+}$ at opposite corners of the first Brillouin zone. It is exactly at these two points where the one-particle spectrum becomes linear and can be modeled by a theory of relativistic Dirac fermions \eqref{Eq:FreeLagrangian}. The remaining components of the chiral left- and right-handed fermions essentially characterize the excitations on the two triangular sublattices $A$ and $B$ of the bipartite honeycomb lattice. To make this mapping explicit we give the connection to the one-particle fermion operators that describe the hopping of electrons on the honeycomb lattice. The free tight-binding Hamiltonian
\begin{equation}
H_{0} = - t \sum_{\langle i , j \rangle}  \left( u^{\dagger a}(\vec{r}_{i}) v^{a} (\vec{r}_{j}) + \textrm{H.\,c.} \right) ~,
\label{Eq:TightBindingHamiltonian}
\end{equation}
defines the dynamics, where a summation over the spin (flavor) indices $a = 1, \ldots, N_{f}$ is implied (recall that the flavor index relates to the physical spin of the particles on the honeycomb lattice). Here, $t$ is the hopping parameter and the sum is taken over all nearest neighbor sites on the honeycomb lattice. The operators $u^{a}(\vec{r}_{i})$ and $v^{a} (\vec{r}_{j})$ anticommute and define the fermionic excitations on the two sublattices $A$ and $B$. In the low-energy limit, where we consider only the linear excitations around the two Dirac points, this model reproduces the free Dirac Lagrangian \eqref{Eq:FreeLagrangian} (in units where $v_{F} = t a \sqrt{3} / 2 = 1$, with $a$ being the lattice spacing). It is this limit that provides the connection between the microscopic degrees of freedom that enter the dynamics \eqref{Eq:TightBindingHamiltonian} and the low-energy excitations of the continuum theory \eqref{Eq:FreeLagrangian}. Following this correspondence, the Dirac spinor $\psi$ has a direct representation in terms of the single-particle fermionic operators $u_{\pm}^{a}$ and $v_{\pm}^{a}$, defined at the two Dirac points $\vec{K}_{\pm}$, respectively. It is given by
\begin{eqnarray}
\psi &=& \begin{pmatrix} u_{+} \\ i v_{+} \\ i v_{-} \\ u_{-} \end{pmatrix} ~, \quad \bar{\psi} = - \left( i v_{-}^{\dagger} , \, u_{-}^{\dagger} ,\, u_{+}^{\dagger} ,\, i v_{+}^{\dagger} \right) ~,
\label{Eq:DiracComponents}
\end{eqnarray}
up to a global phase factor, where we have dropped the flavor indices for simplicity, and we will write them explicitly when needed. The relative phases follow from the chosen representation of the Dirac algebra \eqref{Eq:DiracRepresentation}. It immediately follows that the two-component chiral left- and right-handed fermions can be identified as
\begin{equation}
\psi_{+} = \begin{pmatrix} u_{+} \\ i v_{+} \end{pmatrix} ~, \quad \psi_{-} = \begin{pmatrix} i v_{-} \\ u_{-} \end{pmatrix} ~.
\end{equation}
Note that this identification of the spinor degrees of freedom in the chiral representation is equivalent to the one given, e.\,g.\ in \cite{Gusynin:2007}. Both the decomposition of the honeycomb lattice into the two sublattices, and the two inequivalent Dirac points in the first Brillouin zone are illustrated in Fig.\ \ref{Fig:Graphene}.

The low-energy theory of Dirac fermions \eqref{Eq:FreeLagrangian} has a continuous $U(2)$ chiral symmetry which is not apparent on the level of the microscopic tight-binding model \eqref{Eq:TightBindingHamiltonian}. It is generated by the matrices $\mathds{1}$, $\gamma_{3}$, $\gamma_{5}$, and $\gamma_{3 5}$ with following transformation properties
\begin{eqnarray}
U_{\mathds{1}}(1) : && \qquad \psi \rightarrow e^{i \theta} \psi ~, \!\qquad\quad \bar{\psi} \rightarrow \bar{\psi} e^{- i \theta} ~, \\
U_{\gamma_{3}}(1) : && \qquad \psi \rightarrow e^{i \gamma_{3} \theta} \psi ~, \quad\quad \bar{\psi} \rightarrow \bar{\psi} e^{i \gamma_{3} \theta} ~, \\
U_{\gamma_{5}}(1) : && \qquad \psi \rightarrow e^{i \gamma_{5} \theta} \psi ~, \quad\quad \bar{\psi} \rightarrow \bar{\psi} e^{i \gamma_{5} \theta} ~, \\
U_{\gamma_{3 5}}(1) : && \qquad \psi \rightarrow e^{i \gamma_{3 5} \theta} \psi ~, \!\!\quad\quad \bar{\psi} \rightarrow \bar{\psi} e^{- i \gamma_{3 5} \theta} ~,
\end{eqnarray}
where $\theta$ is a real parameter. In fact, this $U(2)$ chiral symmetry leads to a global $U(2 N_{f})$ symmetry which plays an important role for the dynamics, as it constrains the possible interactions for the low-energy theory close to the critical point. Specifically, for the local four fermion interactions it is possible to define a complete set of operators that respect the $U(2 N_{f})$ flavor symmetry \cite{Herbut:2006cs, *Herbut:2009qb, Gies:2010st}. The corresponding operators are quasi-local in the microscopic lattice description (i.\,e.\ next neighbor and next-to-nearest neighbor). We have four such interactions, two of which are of flavor-singlet type, with the vector Thirring-like interaction 
\begin{equation}
(\bar{\psi}^{a} \gamma_{\mu} \psi^{a})^{2} ~,
\end{equation} 
and the scalar Gross-Neveu-like interaction
\begin{equation}
(\bar{\psi}^{a} \gamma_{3 5} \psi^{a})^{2} ~.
\end{equation} 
Furthermore, there is a flavor-nondiagonal generalized Nambu-Jona-Lasinio interaction
\begin{equation}
(\bar{\psi}^{a} \psi^{b})^{2} - ( \bar{\psi}^{a} \gamma_{3} \psi^{b})^{2} - (\bar{\psi}^{a} \gamma_{5} \psi^{b})^{2} + (\bar{\psi}^{a} \gamma_{3 5} \psi^{b})^{2} ~,
\label{Eq:NJLInteraction}
\end{equation}
and another flavor-nondiagonal interaction of vector-type
\begin{eqnarray}
&& \hspace{-1cm} (\bar{\psi}^{a} \gamma_{\mu} \psi^{b})^{2} + \left( \bar{\psi}^{a} \frac{\sigma_{\mu \nu}}{\sqrt{2}}  \psi^{b} \right)^{2} \nonumber\\
&& -\: (\bar{\psi}^{a} i \gamma_{\mu} \gamma_{3} \psi^{b})^{2} - (\bar{\psi}^{a} i \gamma_{\mu} \gamma_{5} \psi^{b})^{2} ~,
\end{eqnarray}
where $\sigma_{\mu \nu} = \frac{i}{2} [\gamma_{\mu}, \gamma_{\nu}]$ and $(\bar{\psi}^{a} \Gamma^{(j)} \psi^{b})^{2} \equiv \bar{\psi}^{a} \Gamma^{(j)} \psi^{b} \bar{\psi}^{b} \Gamma^{(j)} \psi^{a}$, with $\Gamma^{(j)}$ some element of the Dirac algebra. This defines all interactions invariant under the global $U(2 N_{f})$ symmetry \cite{Herbut:2006cs, *Herbut:2009qb, Gies:2010st}. However, this set of operators is over-complete. By Fierz transformations it is possible to show that only two of the above operators are linearly independent. Thus, we may choose to write any $U(2 N_{f})$-complete theory in terms of only the scalar and vector flavor-singlet interactions
\begin{eqnarray}
\mathcal{L} &=& i \bar{\psi}^{a} \gamma_{\mu} \partial_{\mu} \psi^{a} + \frac{\bar{g}_{V}}{2 N_{f}} \left( \bar{\psi}^{a} \gamma_{\mu} \psi^{a} \right)^{2} \nonumber\\
&& +\: \frac{\bar{g}_{S}}{2 N_{f}} \left( \bar{\psi}^{a} \gamma_{3 5} \psi^{a} \right)^{2}  ~,
\label{Eq:MicroscopicModelFourFermion}
\end{eqnarray}
which fully parametrize the short-range interactions. This theory is invariant also under the discrete parity, charge, and time reversal operations.

\section{Symmetry properties}
\label{Sec:SymmetryProperties}

A possible instability triggered by the local four fermion interactions can lead to very different patterns of spontaneous symmetry breaking. Such an instability essentially leads to a non-vanishing vacuum expectation value of some fermion bilinear $\langle \bar{\psi} \Gamma^{(j)} \psi \rangle$, with $\Gamma^{(j)}$ an element of the Dirac algebra. To identify the physical role of the order parameters in the low-energy theory we review the properties of generic fermion bilinears under discrete symmetry transformations. This enables us to map the order parameters in the low-energy theory onto the corresponding quantities in the microscopic lattice model. Such an identification is necessary, as different representations of the Dirac algebra may lead to very different interpretations for the order parameters and for the fermion masses that are generated dynamically by the interactions. It is this mapping that allows us to understand the properties of the possible phases and their relevance for graphene.

\subsection{$\matheu{P}$, $\matheu{C}$, and $\matheu{T}$}
\label{Sec:PCT}

Here, we take the parity transformation to reverse both spatial coordinates. This choice reflects the direct mapping to the tight-binding model where the parity operation is defined with respect to the center of the first Brillouin zone (see e.\,g.\ the discussion in \cite{Gusynin:2007}). It is important to emphasize that reversing both spatial coordinates does not necessarily correspond to a rotation in the two-dimensional plane, since the generators for both transformations in spinor space do not have to coincide.

For the fermions parity acts in the following way
\begin{equation}
\matheu{P} \psi(x) \matheu{P}^{-1} = P \psi (\tilde{x})~, \qquad \tilde{x} = ( x_{0} , - x_{1} , - x_{2} ) ~,
\label{Eq:ParityTransformation}
\end{equation}
with the unitary matrix $P$ acting on the spinor components. Parity transformations should leave the kinetic term invariant, and we see that any operator of the form
\begin{equation}
P \in \{ \gamma_{0} , i \gamma_{1} \gamma_{2} , i \gamma_{0} \gamma_{3} , i \gamma_{0} \gamma_{5} \} ~,
\label{Eq:ParityTransformationOperator}
\end{equation}
will do the trick. However, each of these possibilities can lead to very different transformation properties for the fermion bilinears $\bar{\psi} \Gamma^{(j)} \psi$. E.\,g.\ , in principle we could obtain a mass term $i \bar{\psi} \psi$ that is parity-odd. This is in contrast to the usual situation in three-dimensional relativistic field theories where one doubles the degrees of freedom to define a parity-even mass term for the fermions \cite{Jackiw:1980kv,Gomes:1990ed}. Here, we want to find those order parameters that correspond to the physical excitations on the honeycomb lattice. Therefore, we define the discrete symmetry operations in such a way that they are consistent with the identification of the spinor components with the one-particle fermion operators on the honeycomb lattice. It is clear that an inversion about the center of the first Brillouin zone should exchange both, the Dirac points and the sublattices. Since, in the chiral representation, the states with definite chirality correspond to the excitations around the two inequivalent Dirac points $\vec{K}_{+}$ and $\vec{K}_{-} = -\vec{K}_{+}$, this leaves only two possibilities for the operator $P$, namely those that exchange states with opposite chirality: $\gamma_{0}$ and $i \gamma_{0} \gamma_{5}$. In principle, we could choose any one of the two. We define $P = \gamma_{0}$ which yields the same transformation properties for the fermion bilinears $\bar{\psi} \Gamma^{(j)} \psi$ as in \cite{Gusynin:2007} so that the mass term $i \bar{\psi} \psi$ and $i \bar{\psi} \gamma_{3 5} \psi$ are parity-even, whereas $\bar{\psi} \gamma_{3} \psi$ and $\bar{\psi} \gamma_{5} \psi$ are parity-odd.\footnote{Note, that the generator for parity transformations $P = \gamma_{0}$ does not correspond to the generator of rotations $\frac{1}{4} [\gamma_{1} , \gamma_{2}]$ in the two-dimensional plane, even though in both cases the sign of both spatial coordinates is flipped.} The components of the Dirac spinor \eqref{Eq:DiracComponents} transform under parity according to
\begin{equation}
\begin{pmatrix} u_{+} \\ i v_{+} \\ i v_{-} \\ u_{-} \end{pmatrix} \underset{\matheu{P}}{\longrightarrow} \begin{pmatrix} v_{-} \\ i u_{-} \\ i u_{+} \\ v_{+} \end{pmatrix} ~,
\end{equation}
where it is understood that the transformed spinor has reversed spatial coordinates.

Charge conjugation is defined as
\begin{equation}
\matheu{C} \psi \,\matheu{C}^{-1} = ( \bar{\psi}  C )^{T} ~.
\label{Eq:ChargeConjugation}
\end{equation}
with the unitary operator $C$ being any one of the following possibilities
\begin{equation}
C \in \{ \gamma_{2} , i \gamma_{0} \gamma_{1} , i \gamma_{2} \gamma_{3} , i \gamma_{2} \gamma_{5} \} ~.
\label{Eq:ChargeConjugationOperator}
\end{equation}
This follows from the requirement that under charge conjugation $i \bar{\psi} \gamma_{\mu} \psi \rightarrow - i \bar{\psi} \gamma_{\mu} \psi$ should hold. Again, the question is how to constrain this set of operators. It is clear that charge conjugation should leave the two Dirac points invariant. However, it exchanges the sublattices $A$ and $B$ as it transforms particles into antiparticles. We are left with two possibilities for the operator $C$: $\gamma_{2}$ and $i \gamma_{2} \gamma_{5}$. Here, we define $C = i \gamma_{2} \gamma_{5}$ where the fermion bilinears $i \bar{\psi} \psi$, $\bar{\psi} \gamma_{3} \psi$, and $i \bar{\psi} \gamma_{3 5} \psi$ are even under charge conjugation, and $\bar{\psi} \gamma_{5} \psi$ is odd. For the components of the Dirac spinor charge conjugation acts as
\begin{equation}
\begin{pmatrix} u_{+} \\ i v_{+} \\ i v_{-} \\ u_{-} \end{pmatrix} \underset{\matheu{C}}{\longrightarrow} \begin{pmatrix} - v_{+}^{\dagger} \\ - i u_{+}^{\dagger} \\ i u_{-}^{\dagger}  \\ v_{-}^{\dagger} \end{pmatrix} ~.
\end{equation}
Notice, that the chiral left- and right-handed components transform with a relative phase factor.

We are left with the antiunitary time reversal\footnote{Note, that in Euclidean space time reversal simply complex conjugates $c$-numbers without changing the sign of spatial momentum components or Euclidean time.} 
\begin{equation}
\matheu{T} \psi \matheu{T}^{-1} = T \psi ~, 
\label{Eq:TimeConjugation}
\end{equation}
where the unitary matrix $T$ is given by
\begin{equation}
T \in \{ i \gamma_{2} , \gamma_{0} \gamma_{1} , \gamma_{2} \gamma_{3} , \gamma_{2} \gamma_{5} \} ~.
\label{Eq:TimeConjugationOperator}
\end{equation}
Time reversal changes both the momentum and spin of the quasiparticles where we neglect the part of the operator that acts on the physical spin, given by some non-diagonal matrix in flavor space). As it reverses the momentum it should exchange the two inequivalent Dirac points at opposing corners of the first Brillouin zone $\vec{K}_{+}$ and $\vec{K}_{-} = -\vec{K}_{+}$. Thus, it appears that we again have two possibilities: $i \gamma_{2}$ and $\gamma_{2} \gamma_{5}$. We take $T = i \gamma_{2}$ for which the bilinears $i \bar{\psi} \psi$, $\bar{\psi} \gamma_{3} \psi$, and $\bar{\psi} \gamma_{5} \psi$ are even, and $i \bar{\psi} \gamma_{3 5} \psi$ is odd under time reversal. The action of the transformation \eqref{Eq:TimeConjugation} on the components of the Dirac spinor is then given by
\begin{equation}
\begin{pmatrix} u_{+} \\ i v_{+} \\ i v_{-} \\ u_{-} \end{pmatrix} \underset{\matheu{T}}{\longrightarrow} -i \begin{pmatrix} u_{-} \\ i v_{-} \\ i v_{+} \\ u_{+} \end{pmatrix} ~.
\end{equation}
However, it should be kept in mind that for simplicity we neglect the transformation that acts on the true spin (i.e. flavor) indices.

From these considerations it follows that in the chiral representation the mass term $i \bar{\psi} \psi$ is invariant separately under $\matheu{P}$, $\matheu{C}$, and $\matheu{T}$. All other mass terms break at least one of the discrete symmetries. The properties of the various fermion bilinears are summarized in \mbox{Tab.\ \ref{Tab:PCT}}.

{\renewcommand{\arraystretch}{1.3}
\renewcommand{\tabcolsep}{0.5cm}
\begin{table}[!t]
\centering
\begin{tabular}{c | ccc}
  & $\matheu{P}$ & $\matheu{C}$ & $\matheu{T}$ \\\hline\hline
  $i \bar{\psi} \psi$                 &  $+$  &  $+$  &  $+$\\
  $\bar{\psi} \gamma_{3} \psi$         &  $-$  &  $+$  &  $+$\\
  $\bar{\psi} \gamma_{5} \psi$         &  $-$  &  $-$  &  $+$\\
  $i \bar{\psi} \gamma_{3 5} \psi$     &  $+$  &  $+$  &  $-$\\\hline
  $i \bar{\psi} \gamma_{\mu} \psi$     &  $i \bar{\psi} \tilde{\gamma}_{\mu} \psi$  &  $- i \bar{\psi} \gamma_{\mu} \psi$  &  $i \bar{\psi} \gamma_{\mu} \psi$  \\\hline
\end{tabular}
\caption{\label{Tab:PCT}Transformation properties of fermion bilinears under $\matheu{P}$, $\matheu{C}$, and $\matheu{T}$ where $\tilde{\gamma}_{\mu} = (\gamma_{0}, - \gamma_{1} , - \gamma_{2})$.}
\end{table}}

\subsection{Antiunitary symmetries}
\label{Sec:AntiunitarySymmetries}

Apart from the antiunitary time reversal symmetry
\begin{equation}
[ \mathcal{D} , i \gamma_{2} K ] = 0 ~,
\label{Eq:AntiUnitary1}
\end{equation}
that was defined in the previous section, the Euclidean Dirac operator with a possible mass term $\mathcal{D} = \gamma_{\mu} \partial_{\mu} + m$ has another antiunitary symmetry $\matheu{S} = -i \gamma_{0} \gamma_{1} K$ which is written as
\begin{equation}
[ \mathcal{D} , -i \gamma_{0} \gamma_{1} K ] = 0 ~.
\label{Eq:AntiUnitary2}
\end{equation}
Here, the operator $K$ denotes complex conjugation. In terms of the Dirac spinor components the symmetry \eqref{Eq:AntiUnitary2} exchanges the excitations on the two sublattices $A$ and $B$, and also the physical spin of the quasiparticles (as for the time reversal symmetry, we will neglect the part acting on the spin components in the following). In contrast to the time reversal \eqref{Eq:TimeConjugation} however, it does not exchange the two Dirac points. That is, it reverses the momentum of the chiral left- and right-handed excitations independently, and in that sense eq.\ \eqref{Eq:AntiUnitary2} can be seen as a time reversal acting separately at the two inequivalent Dirac points \cite{Beenakker:2008}. While the time reversal symmetry satisfies $(i \gamma_{2} K)^{2} = 1$, and therefore defines an orthogonal symmetry here (an additional minus sign arises when including the true spin components), we have $(-i \gamma_{0} \gamma_{1} K)^{2} = -1$ for the antiunitary operator \eqref{Eq:AntiUnitary2}, which corresponds to a symplectic symmetry (in the spinless case). These two different antiunitary symmetries are essentially due to the fact that one has an even number of two-component Weyl fermions in the low-energy theory.

The commutator of both antiunitary operators $\matheu{T}$ and $\matheu{S}$ vanishes, and therefore their product
\begin{equation}
\matheu{T} \matheu{S} = i \gamma_{3 5} ~,
\end{equation}
gives a well-defined unitary operator, exchanging both the two Dirac points and sublattices $A$ and $B$. Clearly, if both $\matheu{T}$ and $\matheu{S}$ are symmetries of the theory then the discrete chiral transformation $\matheu{T}\matheu{S}$ also defines a symmetry operation. In Tab.\ \ref{Tab:Antiunitary} we have collected the transformation properties of the fermion bilinears under the antiunitary and discrete chiral transformations.

Let us comment on the importance of the antiunitary symmetries. Typically, in QCD-like theories the antiunitary symmetry of the Dirac operator is related to the (pseudo)reality of the fermion color representation. Though three-color QCD with quarks in the fundamental representation does not fall into this class, examples are two-color QCD, adjoint QCD, or the $G_{2}$ gauge theory \cite{Kogut:1999iv, *Kogut:2000ek, vonSmekal:2012vx}. In these theories the antiunitary symmetry is responsible for an enlargement of the $SU(N_{f}) \times SU(N_{f}) \times U(1)$ chiral and baryon number symmetries to a global $SU(2 N_{f})$ extended flavor symmetry when the fermions are massless. Furthermore, it determines the dynamics of the low-energy excitations giving rise to different patterns of spontaneous symmetry breaking \cite{Kogut:1999iv, *Kogut:2000ek, vonSmekal:2012vx}. Considering the low-energy theory of free massless fermions \eqref{Eq:FreeLagrangian} with the antiunitary symmetries \eqref{Eq:AntiUnitary1} and \eqref{Eq:AntiUnitary2} and the global $SU(2 N_{f})$ flavor symmetry, one is very much reminded of the situation in QCD-like theories with real or pseudoreal fermion color representations. Here, however, the extended flavor symmetry is a simple consequence of the reducible four-dimensional representation for the fermions in three space-time dimensions. As far as the the antiunitary symmetries of the Dirac operator are concerned, one has to ask whether they are relevant for the low-energy dynamics in presence of interactions or disorder \cite{Heinzner:2004xj, *Altland:2006, *McCann:2006, *Ostrovsky:2007, *DasSarma:2010}. In fact, when the fermions are charged and couple to an abelian $U(1)$ gauge field, the Dirac operator in the gauge-field background does not have the antiunitary symmetries \eqref{Eq:AntiUnitary1} and \eqref{Eq:AntiUnitary2}. Such a 2+1 dimensional QED dynamics can be modeled in the context of Random Matrix Theory by a chiral Gaussian Unitary ensemble (chGUE), that belongs to the class AIII after Cartan's classification of symmetric spaces \cite{Verbaarschot:1993pm, *Verbaarschot:1994qf, *Verbaarschot:2000dy, Zirnbauer:1996zz, *Zirnbauer:2010gg}. The spontaneous breaking of the antiunitary symmetries is then ruled out. Of course, in the instantaneous approximation, the Coulomb field alone would not break the time-reversal invariance. However, close to the charge neutral point this approximation breaks down when the Fermi velocity increases due to the strong electron-electron interaction \cite{Gonzalez:1993uz, Elias:2011}. Therefore, in the following we are interested especially in the chiral symmetry breaking mass term which leaves the antiunitary symmetries unchanged.

{\renewcommand{\arraystretch}{1.3}
\renewcommand{\tabcolsep}{0.7cm}
\begin{table}[!t]
\centering
\begin{tabular}{c | ccc}
  & $\matheu{T}$  & $\matheu{S} $ & $\matheu{T} \matheu{S} $ \\\hline\hline
  $i \bar{\psi} \psi$                  &  $+$  &  $+$  &  $+$\\
  $\bar{\psi} \gamma_{3} \psi$          &  $+$  &  $-$  &  $-$\\
  $\bar{\psi} \gamma_{5} \psi$          &  $+$  &  $-$  &  $-$\\
  $i \bar{\psi} \gamma_{3 5} \psi$      &  $-$  &  $-$  &  $+$\\\hline
  $i \bar{\psi} \gamma_{\mu} \psi$      &  $+$  &  $+$  &  $+$\\\hline
\end{tabular}
\caption{\label{Tab:Antiunitary}Transformation properties of fermion bilinears under the antiunitary and discrete chiral transformations.}
\end{table}}

\subsection{Order parameters}
\label{Sec:OrderParameters}

From the above discussion it follows that the expectation value $i \langle \bar{\psi} \psi \rangle$ is invariant under $\matheu{P}$, $\matheu{C}$, and $\matheu{T}$ whereas both $\langle \bar{\psi} \gamma_{3} \psi\rangle$ and $\langle \bar{\psi} \gamma_{5} \psi \rangle$ are parity-odd (compare Tab.\ \ref{Tab:PCT}). All three order parameters break the extended $U(2 N_{f})$ flavor symmetry, generating a dynamical mass for the fermions. The symmetry breaking pattern is given by \cite{Pisarski:1984dj, *Polychronakos:1987rf, *Dunne:2003ji}
\begin{equation}
U(2 N_{f}) \rightarrow U(N_{f}) \times U(N_{f}) ~.
\end{equation}
It is clear how to identify these order parameters with the excitations in the underlying lattice model. They can be mapped onto the staggered density phase where one has an alternating density on the two different sublattices $A$ and $B$ \cite{Semenoff:1984dq, Araki:2012gs, Herbut:2006cs, *Herbut:2009qb}, and a bond ordered phase that corresponds to a hopping texture on the nearest-neighbor links in the language of the tight-binding model \cite{Hou:2006qc}. Indeed, a non-vanishing staggered density on the two sublattices breaks parity and we may associate the order parameter $\langle \bar{\psi} \gamma_{3} \psi \rangle$ with such a phase. The corresponding bilinear measures the imbalance in the local densities of the two sublattices and therefore does not mix the chiral modes. This is immediately apparent when we write the bilinear directly in terms of the one-particle fermion operators on the honeycomb lattice:
\begin{equation}
\langle \bar{\psi} \gamma_{3} \psi \rangle \rightarrow \langle v_{+}^{\dagger} v_{+} \rangle + \langle v_{-}^{\dagger} v_{-} \rangle - \langle u_{+}^{\dagger} u_{+} \rangle - \langle u_{-}^{\dagger} u_{-} \rangle ~.
\end{equation}
The order parameter for the bond ordered phase however, should couple excitations both at the two inequivalent Dirac points $\vec{K}_{+}$ and $\vec{K}_{-}$ and on the two sublattices $A$ and $B$ \cite{Hou:2006qc,Gusynin:2007}, which is accomplished by the bilinear $i \langle \bar{\psi} (\cos \alpha + \gamma_{5} \sin \alpha) \psi \rangle$. The parameter $\alpha$ controls the hopping texture, and depending on its value, one obtains either a parity-conserving, or a parity-breaking type of order (see \mbox{Tab.\ \ref{Tab:PCT}}). Again, switching to the language of the single-particle excitations on the honeycomb lattice this corresponds to the bilinear
\begin{eqnarray}
&& \hspace{-0.5cm}i \langle \bar{\psi} (\cos \alpha + \gamma_{5} \sin \alpha) \psi \rangle \nonumber\\
&& \hspace{0cm} \rightarrow \: \left( \cos \alpha + \sin \alpha \right) \left( \langle v_{-}^{\dagger} u_{+} \rangle + \langle u_{-}^{\dagger} v_{+} \rangle \right) \nonumber\\ 
&& \hspace{1cm} +\: \left( \cos \alpha - \sin \alpha \right) 
\left( \langle u_{+}^{\dagger} v_{-} \rangle + \langle v_{+}^{\dagger} u_{-} \rangle \right) .
\end{eqnarray}
The condensate $i \langle \bar{\psi} \gamma_{3 5} \psi \rangle$ is in a sense special, as it leaves the chiral symmetry intact. However, it does break the time reversal symmetry. This corresponds to a topologically non-trivial phase that relates to counter-propagating currents on the two different types of sublattices \cite{Haldane:1988zza, *Bernevig:2006zz}. In terms of the one-particle fermion operators it is written as
\begin{equation}
i \langle \bar{\psi} \gamma_{3 5} \psi \rangle \rightarrow - \langle v_{+}^{\dagger} v_{+} \rangle + \langle v_{-}^{\dagger} v_{-} \rangle + \langle u_{+}^{\dagger} u_{+} \rangle - \langle u_{-}^{\dagger} u_{-} \rangle ~.
\end{equation}
The identification of the order parameters given here is equivalent to the one proposed in \cite{Gusynin:2007} where a chiral representation for the Dirac algebra was used also. A complete classification of all possible bilinears is given in \cite{Chamon:2007hx,Ryu:2009}.

\section{A simple model: Spinless fermions on the honeycomb lattice}
\label{Sec:SimpleModel}

Here, we focus on the case of a single flavor Dirac Lagrangian with a local four fermion interaction. This essentially corresponds to a low-energy theory of spinless fermions on the honeycomb lattice with a nearest neighbor and next-to-nearest neighbor coupling \cite{Raghu:2008}. In the purely fermionic description, one may expect that above some critical value for the coupling, the short-range repulsive interactions give rise to an instability. Here, we go beyond a mean field approach by introducing a set of composite fields that allows us to follow the system into the ordered phase. Even for this simple model there is a complex phase diagram with different types of order: Depending on the strength of the interactions there is a competition between a staggered density phase and a topologically non-trivial phase \cite{Raghu:2008}. The global symmetry group for this model is $U(2)$ and the symmetry breaking pattern for the chiral transition is given by $SU(2) \rightarrow U(1)$ while the topological phase transition leaves the $U(2)$ symmetry intact but breaks time reversal invariance.

For this simple model the relevant dynamics of the $SU(2) \rightarrow U(1)$ chiral phase transition is adequately described by taking into account only the fluctuations in a generalized Nambu-Jona-Lasinio-type channel \eqref{Eq:NJLInteraction}. To illustrate this point we make a Fierz-transformation to write the flavor-singlet four fermion interactions in terms of a vector- and a NJL-type interaction:
\begin{eqnarray}
&&\frac{\bar{g}_{V}}{2 N_{f}} ( \bar{\psi}^{a} \gamma_{\mu} \psi^{a} )^{2} + \frac{\bar{g}_{S}}{2 N_{f}} ( \bar{\psi}^{a} \gamma_{3 5} \psi^{a })^{2} \nonumber\\ 
&& \hspace{0.2cm} =\: \frac{\bar{g}_{V} - \bar{g}_{S}}{2 N_{f}} ( \bar{\psi}^{a} \gamma_{\mu} \gamma^{a})^{2} - \frac{\bar{g}_{S}}{2 N_{f}} \left\{ (\bar{\psi}^{a} \psi^{b})^{2} - ( \bar{\psi}^{a} \gamma_{3} \psi^{b})^{2} \right. \nonumber\\
&& \hspace{0.5cm} \left. -\: (\bar{\psi}^{a} \gamma_{5} \psi^{b})^{2} + (\bar{\psi}^{a} \gamma_{3 5} \psi^{b})^{2}\right\} ~.
\end{eqnarray}
One may recognize that line defined by $\bar{g}_{V}-\bar{g}_{S} = 0$ in the $( \bar{g}_{V}$, $\bar{g}_{S} )$-coupling plane defines a theory where the vector-like interaction $(\bar{\psi}^{a} \gamma_{\mu} \psi^{a})^{2}$ becomes irrelevant and the flavor-nondiagonal NJL-type interaction dominates. In fact, this scenario is realized for small numbers of fermion flavors $N_f$ close to a non-Gaussian fixed point as has been demonstrated in a functional renormalization group investigation of the generalized Thirring model \eqref{Eq:MicroscopicModelFourFermion} in three dimensions \cite{Gies:2010st}. Thus, we expect that the critical properties of spinless fermions on the honeycomb lattice in the vicinity of the chiral critical point are well-described by a generalized NJL-model. Of course, strictly speaking, this model may not be in the same universality class as the full $U(2)$-symmetric single flavor model \eqref{Eq:MicroscopicModelFourFermion} which is characterized by three different interacting fixed points that describe very different types of critical behavior. The continuous chiral phase transition, however, corresponds to a Thirring-like fixed point which is very close to being the pure NJL-type interaction with $\bar g_V=\bar g_S$ for $N_f=1$, but which moves towards the Thirring axis $\bar g_S=0$ for $N_f\to\infty$ \cite{Gies:2010st}.

To see how the repulsive interactions drive the system into the broken phase we integrate out the flavor non-diagonal NJL-type interaction via a Hubbard-Stratonovich transformation. In this way we obtain a matrix Yukawa model with a $U(N)$ symmetry for $N = 2$ species of massless, two-component Weyl fermions $\psi^{a}$ and $\bar{\psi}^{a}$, $a = 1, \ldots, N$. For the spinless fermions on the honeycomb lattice $N = 2$, which corresponds to a single Dirac fermion $N_{f} = 1$ in the reducible representation, as e.\,g.\ modeled in microwave photonic crystals \cite{Bittner:2010, *Bittner:2012}. The Weyl fermions couple to a Hermitian matrix field $\Phi_{a b}$ and the action of this model is given by\footnote{We use the notation $\int_{x} \equiv \int d^{d}x$ for the space and $\int_{p} \equiv \frac{d^{d}p}{(2 \pi)^{d}}$ for the momentum integrals.}
\begin{equation}
S[\Phi,\bar{\psi},\psi] = \int_{x} \left\{ \bar{\psi}^{a} i \fsl{\partial} \psi^{a} + \bar{h} \,\bar{\psi}^{a} i \Phi_{a b} \psi^{b} + \frac{1}{2} \bar{m}^{2} \tr{\Phi^{2}} \right\} ~,
\label{Eq:MicroscopicActionAnsatz}
\end{equation}
where the trace $\tr ( \cdots )$ acts on the indices of the matrix field. Here, and in the following, we define $\fsl{\partial} \equiv \sigma_{\mu} \partial_{\mu}$ which belongs to the irreducible representation $\gamma_{0} = \sigma_{3}$, and $\gamma_{k} = \sigma_{k}$, $k = 1, 2$. Eq.\ \eqref{Eq:MicroscopicActionAnsatz} constitutes the starting point for our investigation in the framework of the functional renormalization group.

\section{Functional renormalization group}
\label{Sec:FunctionalRenormalizationGroup}

The full information about the quantum dynamics of a theory is given by the quantum effective action $\Gamma$, which is the generating functional for one-particle irreducible correlation functions. The functional renormalization group is a non-perturbative approach to determine the quantum effective action, taking into account all quantum fluctuations. Implementing Wilson`s renormalization group idea \cite{Wilson:1971bg, *Wilson:1973jj}, the fluctuations are included successively by integrating out the higher modes. Thereby, one obtains the effective average action $\Gamma_{k}$ with all the fluctuations included above the characteristic momentum scale $k$. The scale-dependence is implemented by an infrared regulator $\mathcal{R}_{k}$ to suppress the fluctuations of the low-momentum modes in the theory. In the limit $k \rightarrow 0$, when all quantum fluctuations are included, the functional renormalization group yields the full effective action $\Gamma$.

The flow equation for the effective average action is given by \cite{Wetterich:1992yh, *Bagnuls:2000ae, *Berges:2000ew, *Aoki:2000wm, *Polonyi:2001se, *Pawlowski:2005xe, *Gies:2006wv, *Delamotte:2007pf, *Rosten:2010vm, *Kopietz:2010zz, *Metzner:2011cw}
\begin{equation}
\partial_{t} \Gamma_{k} [\chi] = \frac{1}{2} \STr \left\{ \partial_{t} \mathcal{R}_{k} \left( \Gamma_{k}^{(1,1)} [\chi] + \mathcal{R}_{k} \right)^{-1} \right\} ~,
\label{Eq:FlowEquation}
\end{equation}
where $t = \ln k / \Lambda$ defines the scale parameter, and $\Lambda$ is some appropriate ultraviolet scale where we impose the microscopic dynamics. The supertrace $\STr$ in \eqref{Eq:FlowEquation} denotes a summation over fields and possible internal indices, as well as an integration over momentum, while it provides a minus sign in the fermionic sector. The second functional derivatives of the effective average action $\Gamma_{k}^{(1,1)}$ define the fluctuation matrix.
In the momentum representation, we have
\begin{equation}
\left( \Gamma_{k} ^{(1,1)} [\chi] \right)_{I J}(p,q) \equiv \frac{\overrightarrow{\delta}}{\delta \chi_{I}^{T} (-p)} \, \Gamma_{k} [\chi] \, \frac{\overleftarrow{\delta}}{\delta \chi_{J}(q)} ~,
\end{equation}
where the indices $I,J$ label the different components of the auxiliary field $\chi$ that contains the complete field content of our model: the matrix field $\Phi_{a b}$ and $N$ species of Weyl fermions $\psi^{a}$ and $\bar{\psi}^{a}$, i.\,e.\ \footnote{Since the matrix-valued field $\Phi$ is Hermitian, $\Phi$ and $\Phi^{\ast}$ are not independent degrees of freedom.}
\begin{equation}
\chi^{T} (-p) \equiv \left( \Phi^{T}(-p) \, , \, \psi^{T}(-p) \, , \, \bar{\psi}(p) \right) ~,
\end{equation}
and
\begin{equation}
\chi(q) \equiv \begin{pmatrix} \Phi(q) \\
\psi(q) \\
\bar{\psi}^{T}(-q) \end{pmatrix}~,
\end{equation}
where we have suppressed the flavor indices. Together with the infrared regulator $\mathcal{R}_{k}$ it represents the full regularized inverse propagator $\left(\Gamma_{k}^{(1,1)} + \mathcal{R}_{k}\right)$ at the scale $k$.

The regulator function $\mathcal{R}_{k}$ takes the following form
\begin{equation}
\mathcal{R}_{k}(p) = \begin{pmatrix} R_{B,k}(p) & 0 & 0 \\ 0 & 0 & R_{F,k}(p) \\ 0 & R^{T}_{F, k}(p) & 0 \end{pmatrix} ~,
\end{equation}
where in the bosonic and fermionic sector, we have
\begin{eqnarray}
R_{B,k}(p) &=& Z_{B,k} p^{2} r_{B,k} (p^{2}) ~, 
\label{Eq:BosonRegulatorFunction} \\
R_{F,k} (p) &=& Z_{F,k} \fsl{p} \, r_{F,k} (p^{2}) ~.
\label{Eq:FermionRegulatorFunction}
\end{eqnarray}
Both are fully described by the regulator shape functions $r_{B,k}$ and $r_{F,k}$ that characterize the scheme-dependence of the renormalization procedure. Since they depend only on the dimensionless ratio $y = p^{2} / k^{2}$ we will drop the index $k$ in the following. The fermion regulator shape function $r_{F}$ is taken to satisfy the constraint $p^{2} (1 + r_{B}) = p^{2} (1 + r_{F})^{2}$ and thus, is completely determined by the choice of $r_{B}$.

In practice, to solve \eqref{Eq:FlowEquation} one is bound to rely on approximations for the effective average action $\Gamma_{k}$ where one truncates the set of possible operators following an expansion scheme, e.\,g.\ in powers of derivatives \cite{Canet:2002gs, *Morris:1994ie, *Morris:1997xj}. However, such an approximation also induces a spurious dependence on the regulator for the full quantum effective action when the scale $k$ is sent to zero \cite{Litim:2000ci, *Litim:2001up, Latorre:2000qc}. In this work we therefore employ two different types of infrared regulators to test the regulator scheme-dependence of our results in the physical limit. We consider the optimized regulator \cite{Litim:2000ci, *Litim:2001up}
\begin{equation}
r_{B} (y) = \left( \frac{1}{y} - 1 \right) \theta(1 - y)~,
\label{Eq:OptimizedRegulator}
\end{equation}
and also an exponential-type regulator
\begin{equation}
r_{B} (y) = \left( \exp(y) - 1 \right)^{-1} ~.
\label{Eq:ExponentialRegulator}
\end{equation}

\subsection{Effective average action}
\label{Sec:EffectiveAverageAction}

Our ansatz for the effective average action is given by
\begin{eqnarray}                                                                                                                   
\Gamma_{k}[\Phi,\bar{\psi},\psi] &=& \int_{x} \Big\{ Z_{F,k} \,\bar{\psi}^{a} i \fsl{\partial} \psi^{a} + \frac{1}{2} Z_{B,k} \tr \left( \partial_{\mu} \Phi \right)^{2} \nonumber \\ 
&& +\: \bar{h}_{k} \,\bar{\psi}^{a} i \Phi_{a b} \psi^{b} + U_{k}(\Phi) \Big\} ~. 
\label{Eq:EffectiveAverageActionAnsatz}
\end{eqnarray}
In contrast to the microscopic model, we have a kinetic term for the composite field and a scale-dependent wavefunction renormalization both for the fermions and the bosons. Thus, we include the bosonic fluctuations that give a non-trivial momentum structure for the fermion interactions \cite{Braun:2010tt, Braun:2011}. Clearly, our ansatz \eqref{Eq:EffectiveAverageActionAnsatz} goes far beyond a simple mean-field approximation where one neglects the fluctuations from the composite degrees of freedom.

For the $U(N)$ matrix-model the effective average potential $U_{k}(\Phi)$ is a function of the invariants of the $U(N)$ symmetry group. For system of $N = 2$ Weyl fermions there are exactly two invariants, one that is linear $\bar{\sigma}_{k} = \tr \Phi$, and another one that is quadratic $\bar{\rho}_{k} = \tfrac{1}{2} \tr \Phi^{2}$ in the fields. A non-vanishing vacuum expectation value for the composite field $\Phi_{a b}$ signals the dynamical generation of a mass for the fermions. Depending on the flavor structure of the matrix field we have different types of order: The chirally broken phase corresponds to a vacuum configuration which is either non-diagonal, or diagonal non-uniform (so that the trace vanishes, i.\,e.\ $\tr \Phi = 0$). On the other hand, $\Phi_{a b} \sim \delta_{a b}$ in the non-trivial topological phase which we will not consider here. Close to the phase transition only those fluctuations of the $\Phi$-field will play a significant role that give a contribution to the quadratic invariant $\bar{\rho}_{k}$ -- the $\bar{\sigma}_{k}$-field is irrelevant there. Thus, to investigate the nature of the chiral phase transition we may neglect the fluctuations of the $\bar{\sigma}_{k}$-field. However, we want to emphasize that this approximation is no way essential for the following calculations.

We expand the effective average potential in powers of $\bar{\rho}_{k}$ around the minimum $\bar{\rho}_{0,k}$, given by the scale-dependent vacuum expectation value:
\begin{equation}
U_{k}(\bar{\rho}_{k}) = \bar{m}_{k}^{2} ( \bar{\rho}_{k} - \bar{\rho}_{0,k} ) + \sum_{n=2}^{n_{max}} \frac{\bar{\lambda}_{n,k}}{n!} ( \bar{\rho}_{k} - \bar{\rho}_{0,k} )^{n} ~. 
\label{Eq:EffectiveAveragePotentialAnsatz}
\end{equation}
This approximation captures all the relevant fluctuation at the chiral phase transition.
In the symmetric regime the vacuum expectation value $\bar{\rho}_{0,k}$ is zero whereas in the chirally broken phase $\bar{\rho}_{0,k} \neq 0$ and the mass $\bar{m}_{k}^{2}$ becomes zero. In our ansatz we include the first set of irrelevant operators according to a naive power counting with respect to the canonical mass dimension.

\subsection{Flow equation for the effective average potential}
\label{Sec:FlowEquationEffectiveAveragePotential}

To extract the flow equations for the parameters and couplings in the effective average action \eqref{Eq:EffectiveAverageActionAnsatz} one has to project the functional flow given by the r.\,h.\,s.\ of \eqref{Eq:FlowEquation} onto the corresponding operators. For the couplings that appear in the effective average potential this is done by evaluating the second functional derivative $\Gamma_{k}^{(1,1)}$ in a constant background configuration of the matrix field $\Phi_{a b}$.

The flow equation for the effective average potential receives contributions both from the boson and fermion degrees of freedom
\begin{equation}
\partial_{t} U_{k}(\Phi) = \partial_{t} U_{B,k} (\Phi) + \partial_{t} U_{F,k} (\Phi) ~.
\label{Eq:FlowEquationEffectivePotential}
\end{equation}
To keep the notation clear, we will drop the $k$-index in the following. Where necessary, we will revert to our original notation.

From \eqref{Eq:FlowEquation} we obtain the boson contribution to the effective average potential
\begin{equation}
\partial_{t} U_{B} = \frac{1}{2} \int_{q} \, \partial_{t} R_{B} \sum_{i} P_{B} ( \bar{M}_{B i} ) ~,
\label{Eq:FlowEquationEffectivePotentialBoson}
\end{equation}
where the full regularized boson propagator $P_{B}$ is given in the Appendix \ref{App:Propagators}. We sum over all mass eigenvalues $\bar{M}^{2}_{B i}$ of the mass matrix as given by the second derivatives of the potential
\begin{equation}
\bar{M}_{B}^{2} (\Phi)_{\, a b , c d} (p,q) \equiv \frac{\overrightarrow{\delta}}{\delta \Phi^{T \, a b} (-p)} \int_{x} \, U_{k}(\Phi) \, \frac{\overleftarrow{\delta}}{\delta \Phi^{ c d } (q)} ~.
\label{Eq:MassMatrix}
\end{equation}
In general this expression will be momentum-dependent. However, for the calculation of the effective average potential with a constant background configuration $\Phi_{a b}$ all momentum-dependence in \eqref{Eq:MassMatrix} drops out. The mass matrices are given in Appendix \ref{App:MassMatrices}.

For illustrational purposes we perform the derivation of the flow equation in the fermion sector explicitly by integrating out the fermions in the action. To evaluate the fermion contribution to the effective average potential it is useful to write the action in terms of the four-component spinors 
\begin{equation}
\Psi(q) = \begin{pmatrix} \psi(q) \\ \bar{\psi}^{T}(-q) \end{pmatrix} ~,
\label{Eq:SpinorBasis}
\end{equation}
which is constructed from the two independent degrees of freedom $\psi$ and $\bar{\psi}$. In that case, the fermion bilinear part of the action takes the form
\begin{equation}
\Gamma_{k, \Psi\Psi} = \int_{q, (q_{0} > 0)} \Psi^{T \, a} (-q) \mathcal{D}_{a b}(q) \Psi^{b} (q) ~,
\end{equation}
in momentum space. Note, that the domain of integration is restricted to positive frequencies $q_{0} > 0$ to counteract the doubling of degrees of freedom that comes from switching to the four-component spinors \eqref{Eq:SpinorBasis}. In this basis, the inverse regularized fermion propagator is given by
\begin{eqnarray}
\mathcal{D}_{a b} (q) &=& Z_{F} (1 + r_{F}) \begin{pmatrix} & ~\fsl{q}^{T}~ \\ ~\fsl{q}~ &  \end{pmatrix} \delta_{a b} \nonumber\\
&&\hspace{0.2cm} +\: i \bar{h}(q) \begin{pmatrix} & - \Phi^{T}_{a b} \\ \Phi_{a b} & \end{pmatrix} ~,
\end{eqnarray}
where $\bar{h}(q) \equiv \bar{h}(-q,q)$ denotes the momentum-dependent Yukawa coupling.\footnote{For the momentum dependent Yukawa coupling $\bar{h}(-p , q)$ the momenta $p$ and $q$ denote the incoming fermion momenta at the Yukawa vertex.} As for the boson contribution, we evaluate the inverse propagator $\mathcal{D}_{a b}(q)$ in a constant background field $\Phi_{a b}$.
Performing the integration over the Grassmann fields, the fermion contribution to the potential takes the form
\begin{equation}
U_{F} = - \int_{q, (q_{0} > 0)} \, \ln \det \mathcal{D}(q)~,
\label{Eq:EffectivePotentialFermion}
\end{equation}
where the determinant acts on the flavor and spinor indices. To evaluate this expression we put $\mathcal{D}_{a b} (q)$ in standard diagonal form. That is, by a unitary transformation we diagonalize $\Phi_{a b} = \Phi_{a} \delta_{a b}$, so that
$\mathcal{D}_{a b}(q) = \mathcal{D}_{a} (q) \delta_{a b}$, and the determinant in \eqref{Eq:EffectivePotentialFermion} can be written as
\begin{equation}
\det \mathcal{D}_{a} = \left( Z_{F}^{2} ( 1 + r_{F} )^{2} q^{2} + \bar{h} (q)^{2} \Phi_{a}^{2} \right)^{2}~.
\end{equation}
With this result, the fermion contribution to the flow equation \eqref{Eq:FlowEquationEffectivePotential} becomes
\begin{equation}
\partial_{t} U_{F} = - 2 \int_{q} \, q^{2} \, Z_{F} (1+r_{F}) \, \partial_{t} \big( Z_{F} r_{F} \big) \, \sum_{a}  \widetilde{P}_{F} ( \bar{M}_{F a} ) ~,
\label{Eq:FlowEquationEffectivePotentialFermion}
\end{equation}
where $\widetilde{P}_{F}( \bar{M}_{F a} ) = \left( \det \mathcal{D}_{a} \right)^{-\frac{1}{2}}$. One may easily verify that this is just the result that is obtained when the supertrace in \eqref{Eq:FlowEquation} is computed directly, using the definition of the full regularized propagators (see Appendix \ref{App:Propagators}), and the regulator $R_{F}$. Here, the masses $\bar{M}_{F a}$ denote the $N$ eigenvalues of the $N \times N$ matrix $\bar{h}(q) \Phi$.

Eq.\ \eqref{Eq:FlowEquationEffectivePotentialBoson} and \eqref{Eq:FlowEquationEffectivePotentialFermion} together give the full contribution to the effective average potential. To investigate the critical properties at the phase transition, however, it is convenient to bring the flow equations to a form where one may easily identify possible fixed point solutions. For that purpose, we switch to dimensionless renormalized quantities $\rho = k^{2-d} Z_{B} \bar{\rho}$ 
and $u (\rho) = k^{-d} U_{k} (\rho)$. Then, the flow equation for the potential is given by
\begin{equation}
\partial_{t} u =
- d u + \left. k^{-d} \partial_{t} U_{k} \right|_{\bar{\rho}} + \left. ( d - 2 + \eta_{B} ) \rho k^{-d} \frac{\partial U_{k}}{\partial \rho} \right|_{t} ~,
\end{equation}
where we have introduced the scalar anomalous dimension $\eta_{B} \equiv - \partial_{t} \ln Z_{B}$. Substituting our previous result this finally gives
\begin{eqnarray}
\partial_{t} u &=& - d u + ( d - 2 + \eta_{B} ) \rho u' \nonumber\\
&& +\: 2 v_{d} \, \bigg\{ ( N^{2} - 2 ) l_{0}^{(B)} (u' ; \eta_{B}) + l_{0}^{(B)} (u' + 2 \rho u'' ; \eta_{B}) \nonumber\\
&& -\: 2 N l_{0}^{(F)} \left( \frac{2}{N} \rho h^{2} ; \eta_{F} \right) \bigg\} ~, 
\label{Eq:FlowEquationDimensionalessRegularizedPotential}
\end{eqnarray}
where the prime $u' \equiv \left. \frac{\partial u}{\partial \rho}\right|_{t}$ denotes differentiation with respect to the dimensionless renormalized field $\rho$, and $v_{d} = \left( 2^{d+1} \pi^{d/2} \Gamma\left( d / 2 \right) \right)^{-1}$. Furthermore, $\eta_{F} \equiv - \partial_{t} \ln Z_{F}$ and $h^{2} = k^{d-4} Z_{F}^{-2} Z_{B}^{-1} \bar{h}^{2}$ is the dimensionless renormalized Yukawa coupling. Here, we have introduced the threshold functions $l_{0}^{(B)}$ and $l_{0}^{(F)}$ that parametrize the boson and fermion one-loop integrals contributing to the effective average potential. They are defined in Appendix \ref{App:ThresholdFunctions} where their form is given explicitly for the optimized regulator \eqref{Eq:OptimizedRegulator}. The corresponding expressions for the exponential regulator \eqref{Eq:ExponentialRegulator} can be found in, e.\,g.\ \cite{Berges:2000ew, Braun:2011}. The threshold functions carry the full scheme-dependence of the renormalization group equations. In that sense, the flow equations are universal -- only the dimensionality and symmetries determine the flow and the regulator-dependence resides solely in the threshold functions.

In the symmetric regime, we may derive the flow equations for the dimensionless renormalized couplings $\epsilon = k^{-2} Z_{B}^{-1} \bar{m}^{2}$ and $\lambda_{n} = k^{(n-1) d - 2 n} Z_{B}^{-n} \bar{\lambda}_{n}$, $n = 2, \ldots, n_{max} $, from \eqref{Eq:FlowEquationDimensionalessRegularizedPotential} simply by differentiating with respect to field $\rho$, that is, we have $\epsilon = u'$ for the mass parameter and $\lambda_{n} = u^{(n)}$ for the couplings. The derivatives of the threshold functions are evaluated as
\begin{equation}
\frac{\partial}{\partial w} l^{(B)}_{n} (w ; \eta_{B}) = - \left( n + \delta_{n,0} \right) l^{(B)}_{n+1}(w ; \eta_{B}) ~,
\end{equation}
and equivalently for $l^{(F)}_{n}(w; \eta_{F})$. Here, we give the flow equations for the mass parameter $\epsilon$, and the couplings $\lambda_{2}$, and $\lambda_{3}$ in the symmetric phase:
\begin{widetext}
\begin{eqnarray}
\partial_{t} \epsilon &=& ( - 2 + \eta_{B}) \epsilon - 2 v_{d} \left\{ \left( N^{2} + 1 \right) \lambda_{2} \, l_{1}^{(B)} \left( \epsilon ; \eta_{B} \right) - 4 h^{2} \, l_{1}^{(F)} \left( 0 ; \eta_{F} \right) \right\} ~, \\
\partial_{t} \lambda_{2} &=& (d - 4 + 2 \eta_{B}) \lambda_{2} + 2 v_{d} \left\{ \left( N^{2} + 7 \right) \lambda_{2}^{2} \, l_{2}^{(B)} \left( \epsilon ; \eta_{B} \right) - \left( N^{2} + 3 \right) \lambda_{3} \, l_{1}^{(B)} \left( \epsilon ; \eta_{B} \right) - \frac{8}{N} h^{4} \, l_{2}^{(F)} \left( 0 ; \eta_{F} \right)  \right\} ~, \\
\partial_{t} \lambda_{3} &=& (2 d - 6 + 3 \eta_{B} ) \lambda_{3} - 2 v_{d} \left\{ 2 \left( N^{2} + 25 \right) \lambda_{2}^{3} \, l_{3}^{(B)} \left( \epsilon ; \eta_{B} \right) - 3 \left( N^{2} + 13 \right) \lambda_{2} \lambda_{3} \, l_{2}^{(B)} \left( \epsilon ; \eta_{B} \right) \right. \nonumber\\
&& +\: \left. \left( N^{2} + 5 \right) \lambda_{4} \, l_{1}^{(B)} \left( \epsilon ; \eta_{B} \right) - \frac{32}{N^{2}} h^{6} \, l_{3}^{(F)} \left( 0 ; \eta_{F} \right) \right\} ~.
\end{eqnarray}
\end{widetext}
The flow equations for the higher order couplings can be obtained by a simple differentiation with respect to the field, and are not given here explicitly.

As the system goes over into the broken phase, the scale-dependent mass parameter $\epsilon$ goes to zero, and the field assumes a non-vanishing expectation value $\rho_{0} \neq 0$, defined by $u'(\rho_{0}) = 0$. Due to the scale-dependence of $\rho_{0}$, which is given by
\begin{equation}
\partial_{t} \rho_{0} = - \frac{1}{\lambda_{2}} \partial_{t} u'(\rho_{0})  ~,
\end{equation}
we get an additional contribution to the flow \eqref{Eq:FlowEquationDimensionalessRegularizedPotential} in the broken phase:
\begin{equation}
\partial_{t} \lambda_{n} = \left. \partial_{t} \lambda_{n} \right|_{\rho_{0}} + \lambda_{n+1} \partial_{t} \rho_{0} ~.
\label{Eq:FlowEquationCouplingsBrokenPhase}
\end{equation}
There, the flow equations for $\rho_{0}$, and the couplings $\lambda_{2}$, and $\lambda_{3}$ are given by
\begin{widetext}
\begin{eqnarray}
\partial_{t} \rho_{0} &=& (2 - d - \eta_{B}) \rho_{0} + 2 v_{d} \left\{ \left( N^{2} - 2 \right) l_{1}^{(B)} \left( 0 ; \eta_{B} \right) + \left( 3 + \frac{2 \rho_{0} \lambda_{3}}{\lambda_{2}} \right) l_{1}^{(B)} \left( 2 \rho_{0} \lambda_{2} ; \eta_{B} \right) - \frac{4}{\lambda_{2}} h^{2} \, l_{1}^{(F)} \left( \frac{2}{N} \rho_{0} h^{2} ; \eta_{F} \right) \right\} \! , \\
\partial_{t} \lambda_{2} &=& (d - 4 + 2 \eta_{B}) \lambda_{2} + 2 v_{d} \bigg\{ \left( N^{2} - 2 \right) \lambda_{2}^{2} \, l_{2}^{(B)} \left( 0 ; \eta_{B} \right) 
+\left( 3 \lambda_{2} + 2 \rho_{0} \lambda_{3} \right)^{2} l_{2}^{(B)} \left( 2 \rho_{0} \lambda_{2} ; \eta_{B} \right) \nonumber\\
&& -\: \left( 2 \lambda_{3} + 2 \rho_{0} \lambda_{4}  - \frac{ 2 \rho_{0} \lambda_{3}^{2}}{\lambda_{2}} \right) l_{1}^{(B)} \left( 2 \rho_{0} \lambda_{2} ; \eta_{B} \right)
- \frac{8}{N} h^{4} \, l_{2}^{(F)} \left( \frac{2}{N} \rho_{0} h^{2} ; \eta_{F} \right) - 4 \frac{\lambda_{3}}{\lambda_{2}} h^{2} \, l_{1}^{(F)} \left( \frac{2}{N} \rho_{0} h^{2} ; \eta_{F} \right) \bigg\} ~,
\\ 
\partial_{t} \lambda_{3} &=& (2 d - 6 + 3 \eta_{B} ) \lambda_{3} - 2 v_{d} \bigg\{ \left( N^{2} - 2 \right) \left( 2 \lambda_{2}^{3} \, l_{3}^{(B)} \left( 0 ; \eta_{B} \right) - 3 \lambda_{2} \lambda_{3} \, l_{2}^{(B)} \left( 0 ; \eta_{B} \right) \right) + 2 \left( 3 \lambda_{2} + 2 \rho_{0} \lambda_{3} \right)^{3} l_{3}^{(B)} \left( 2 \rho_{0} \lambda_{2} ; \eta_{B} \right) \nonumber\\
&& -\: 3 \left( 3 \lambda_{2} + 2 \rho_{0} \lambda_{3} \right) \left( 5 \lambda_{3} + 2 \rho_{0} \lambda_{4} \right) l_{2}^{(B)} \left( 2 \rho_{0} \lambda_{2} ; \eta_{B} \right) + \left( 4 \lambda_{4} + 2 \rho_{0} \lambda_{5} - \frac{ 2 \rho_{0} \lambda_{3} \lambda_{4}}{\lambda_{2}} \right) l_{1}^{(B)} \left( 2 \rho_{0} \lambda_{2}  ; \eta_{B} \right) \nonumber\\
&& -\: \frac{32}{N^{2}} h^{6} \, l_{3}^{(F)} \left( \frac{2}{N} \rho_{0} h^{2} ; \eta_{F} \right) + 4 \frac{\lambda_{4}}{\lambda_{2}} h^{2} \, l_{1}^{(F)} \left( \frac{2}{N} \rho_{0} h^{2} ; \eta_{F} \right) \bigg\} \, .
\end{eqnarray}
\end{widetext}
The flow equations for the higher order couplings can easily be obtained via \eqref{Eq:FlowEquationCouplingsBrokenPhase}.

Recall that the relevant symmetry breaking pattern for the simple model (Sec.\ \ref{Sec:SimpleModel}) is given by $SU(2) \rightarrow U(1)$. This is in direct correspondence to the $O(3) \rightarrow O(2)$ transition in the three-component vector model \cite{Wetterich:1991be, Tetradis:1993ts}. Building on our previous remark concerning the universality of the renormalization group flow, we observe that in the bosonic sector (neglecting the fermion contributions) the flow equations for the effective potential are identical to the flow equations for the three-dimensional $O(3)$ vector model \cite{Tetradis:1993ts}. We want to emphasize that this is a simplification that occurs only for the special case where $N = 2$. In the general case, the flow equations for the effective potential correspond to the matrix Yukawa model with $U(N)$ symmetry.

\subsection{Boson anomalous dimension}
\label{Sec:BosonAnomalousDimension}

For the computation of the boson anomalous dimension $\eta_{B} = - \partial_{t} \ln Z_{B}$ we first evaluate the flow equation \eqref{Eq:FlowEquation} in a spatially varying field configuration $\Phi(x)$. This is necessary for the projection onto the kinetic term and the wavefunction renormalization $Z_{B}$. In particular, we consider a distortion around the non-diagonal vacuum configuration \mbox{$\Phi_{a b} = \widehat{\Phi}_{0} \Sigma_{a b}$} characterized by a non-vanishing momentum $Q$, i.\,e.
\begin{equation}
\Phi_{a b} (x) = \widehat{\Phi}_{0} \Sigma_{a b} + \underbrace{\left(\delta \Phi e^{-i Q x} + \textrm{c.c.} \right)}_{\equiv \,\Delta(x)} \Lambda_{a b} ~,
\label{Eq:NonDiagonalVacuumConfigurationPerturbed}
\end{equation}
where the Hermitian matrices $\Sigma$ and $\Lambda$ satisfy the properties $\Sigma^{T} = - \Sigma$ and $\Lambda^{T} = \Lambda$. Clearly, the fluctuations in the $\Lambda$-direction are orthogonal to the ground state orientation. Though we take only one of the possible orthogonal directions for the fluctuations into account this still yields a complete description of those contributions coming from the Goldstone modes. Of course, in the broken phase, we also have fluctuations $\sim \Delta'(x) \Sigma_{a b}$ from the radial mode that give additional contributions to the boson anomalous dimension.

In momentum space the configuration \eqref{Eq:NonDiagonalVacuumConfigurationPerturbed} reads
\begin{equation}
\Phi_{a b}(p) = \widehat{\Phi}_{0} \delta(p,0) \Sigma_{a b} + \Delta(p,Q) \Lambda_{a b} ~,
\end{equation}
where we define $\delta(p,q) \equiv (2 \pi)^{d} \delta^{(d)} ( p - q )$, and the amplitude is given by $\Delta(p,Q) = \left( \delta \Phi \delta(p,Q) + \delta \Phi^{\ast} \delta(-p,Q)\right)$. Taking the ansatz \eqref{Eq:NonDiagonalVacuumConfigurationPerturbed} one may easily verify that
\begin{equation}
\partial_{t} Z_{B} \equiv \frac{1}{N} \lim_{Q\rightarrow 0} \frac{\partial}{\partial Q^{2}} \left[ \lim_{\delta \Phi \rightarrow 0} \frac{\partial}{\partial (\delta \Phi \delta \Phi^{\ast})} \partial_{t} \Gamma_{k} \right] ~,
\label{Eq:ProjectionBosonWavefunctionRenormalization}
\end{equation}
gives us the flow equation for the momentum-independent part of the wavefunction renormalization $Z_{B}$. Here, we neglect all momentum-dependence of the wavefunction renormalization. Eq.\ \eqref{Eq:ProjectionBosonWavefunctionRenormalization} defines a projection of the flow equation \eqref{Eq:FlowEquation} onto the flow of the wavefunction renormalization, i.\,e.\ $\left. \partial_{t} \Gamma_{k} \right|_{Z_{B}} \equiv \partial_{t} Z_{B}$. In the following, we will use this notation frequently.

To evaluate \eqref{Eq:ProjectionBosonWavefunctionRenormalization} we make use of a series expansion of the flow equation. Using the decomposition of the full inverse regularized propagator
\begin{equation}
\Gamma_{k}^{(1,1)} + \mathcal{R}_{k} = \mathcal{P}_{k}^{-1} + \mathcal{F}_{k}~,
\end{equation}
which is written in terms of $\mathcal{F}_{k}$ containing all field-dependent fluctuations around the background field configuration, and the inverse background field propagator $\mathcal{P}_{k}^{-1}$ that carries the explicit regulator dependence, we may write the flow equation \eqref{Eq:FlowEquation} as a series expansion in powers of the fields:
\begin{eqnarray}
\partial_{t} \Gamma_{k} &=& \frac{1}{2} \STr \, \partial_{t} \mathcal{R}_{k} \,\mathcal{P}_{k} + \frac{1}{2} \STr \hat{\partial}_{t} \big( \mathcal{P}_{k} \mathcal{F}_{k} \big) \nonumber \\
&& -\: \frac{1}{4} \STr \hat{\partial}_{t} \left( \mathcal{P}_{k} \mathcal{F}_{k} \right)^{2} + \mathcal{O}(\mathcal{F}_{k}^{3}) ~. 
\label{Eq:SeriesExpansion}
\end{eqnarray}
Here we have defined the formal derivative operator
\begin{equation}
\hat{\partial}_{t} \equiv \partial_{t} \mathcal{R}_{k} \frac{\partial}{\partial \left( \mathcal{P}_{k}^{-1} \right)} ~,
\label{Eq:HatDerivative}
\end{equation}
that acts on the inverse regularized matrix propagator $\mathcal{P}_{k}^{-1}$ (see Appendix \ref{App:Propagators}). In terms of \eqref{Eq:NonDiagonalVacuumConfigurationPerturbed} the leading contribution to the fluctuation is $\mathcal{F}_{k}^{n} \sim \mathcal{O} (\Delta^{n})$ and for the calculation of the anomalous dimension only the second order term $\delta^{(2)} \Gamma_{k}$ in the fluctuation $\mathcal{F}_{k}$ is important. Clearly, the lowest order term will not contribute, as it is independent of $\delta \Phi$ and $\delta \Phi^{\ast}$ and thus yields a vanishing contribution to \eqref{Eq:ProjectionBosonWavefunctionRenormalization}. The $\delta^{(1)} \Gamma_{k}$ term does include combinations of the type $\propto \delta \Phi \delta \Phi^{\ast}$, however, they are independent of momentum $Q^{2}$. Thus, the lowest order term relevant in \eqref{Eq:ProjectionBosonWavefunctionRenormalization} is the second order term $\delta^{(2)} \Gamma_{k}$ in the series expansion, and specifically, we will need those terms in $\mathcal{F}_{k}$ that are of linear order in the amplitude $\Delta$. Apart from the bosonic background $\Phi_{a b}$ we need to specify the fermionic background configuration, where we take 
\begin{equation}
\psi = \bar{\psi} = 0 ~.
\end{equation}
Then, the matrix of second functional derivatives $\Gamma_{k}^{(1,1)}$ becomes block-diagonal in the boson and fermion subspaces and we may treat the boson and fermion contributions to \eqref{Eq:ProjectionBosonWavefunctionRenormalization} separately.

We start with the bosonic sector. For matrix-valued fields, it is convenient to work in the non-diagonal basis for the propagators and the fluctuations \cite{Jungnickel:1995fp, *Berges:1996ja}. We follow the outline given above, and evaluate the second order term $\delta^{(2)} \Gamma_{B} = -\frac{1}{4} \STr \hat{\partial}_{t} \left( P_{B} F_{B} \right)^{2}$ in the series expansion where the index $B$ denotes the corresponding quantities in the bosonic sector where we have dropped the $k$-index for clarity, i.\,e.\ $\Gamma_{B, k} \equiv \Gamma_{B}$, $P_{B,k} \equiv P_{B}$, etc. For that we need the boson propagator in the constant background configuration $\Phi_{a b} = \widehat{\Phi}_{0} \Sigma_{a b}$ which takes the following form
\begin{equation}
\left( P_{B} \right)_{a b, c d} (p) = \frac{1}{A(p) }\left( \delta_{a c} \delta_{b d} - \frac{B}{A(p) + N B} \Sigma_{a b} \Sigma^{T}_{c d} \right) ~,
\label{Eq:BosonPropagatorBackground}
\end{equation}
where we have introduced the quantities 
\begin{equation}
A(p) = Z_{B} ( 1 + r_{B} ) p^{2} + \frac{\partial U_{k}}{\partial \bar{\rho}} ~, \quad B = \widehat{\Phi}_{0}^{2} \, \frac{\partial^{2} U_{k}}{\partial \bar{\rho}^{2}} ~.
\label{Eq:ApAndB}
\end{equation}
Furthermore, we need the contribution from the fluctuations to linear order in $\Delta$, which is given by
\begin{eqnarray}
( F_{B} )_{a b , c d} (p,q) &=& \Delta(p-q,Q) \, \widehat{\Phi}_{0}\frac{\partial^{2} U_{k}}{\partial \bar{\rho}^{2}} \nonumber\\
&& \hspace{-1cm} \times\: \left( \Sigma_{a b} \Lambda^{T}_{c d} + \Lambda_{a b} \Sigma^{T}_{c d} \right) + \mathcal{O}(\Delta^{2}) ~.
\end{eqnarray}
With these results we can immediately compute the trace in $\delta^{(2)} \Gamma_{B}$ and evaluate the projection onto the kinetic term:
\begin{eqnarray}
\left. \delta^{(2)} \Gamma_{B} \right|_{Z_{B}} &=& - N \widehat{\Phi}_{0}^{2} \left( \frac{\partial^{2} U_{k}}{\partial \bar{\rho}^{2}} \right)^{2} \nonumber\\
&& \hspace{-1cm}\times\:\lim_{Q \rightarrow 0} \frac{\partial}{\partial Q^{2}} \int_{p} \hat{\partial}_{t} \Big\{ P_{B 1} ( p ) P_{B 2} ( p+Q ) \Big\} \, . \nonumber\\
&& \label{Eq:BosonAnomalousDimensionProjectionBoson}
\end{eqnarray}
The indices on the propagators $P_{B}$ refer to the corresponding eigenvalues of the mass matrix $\bar{M}_{B}^{2}$, i.\,e.\ $P_{B 1} \equiv P_{B} (\bar{M}_{B 1})$ etc.\ that are given by (see Appendix \ref{App:MassMatrices}):
\begin{equation}
\bar{M}_{B 1}^{2} = \frac{\partial U_{k}}{\partial \bar{\rho}} ~, \quad \bar{M}_{B 2}^{2} = \frac{\partial U_{k}}{\partial \bar{\rho}} + N \widehat{\Phi}_{0}^{2} \, \frac{\partial^{2} U_{k}}{\partial \bar{\rho}^{2}} ~.
\end{equation}
These propagators $P_{B 1}$ and $P_{B 2}$ belong to the Goldstone modes and radial mode, respectively. We want to emphasize that the derivative operator $\hat{\partial}_{t}$ appearing in \eqref{Eq:BosonAnomalousDimensionProjectionBoson} is different from the one defined in \eqref{Eq:HatDerivative}. That is, we slightly abuse the notation and take
\begin{eqnarray}
\hat{\partial}_{t} ~ &\equiv& ~ \partial_{t} R_{B} \frac{\partial}{\partial \left( P_{B}^{-1} \right)} \nonumber\\
&& \hspace{0.2cm} +\: \frac{2}{Z_{F}} \frac{\widetilde{P}_{F}^{-1}(0)}{1 + r_{F}} \partial_{t} \left( Z_{F} r_{F} \right) \frac{\partial}{\partial \left( \widetilde{P}_{F}^{-1} \right)} ~,
\end{eqnarray}
from now on, where it is understood that $\widetilde{P}_{F}^{-1}(0)$ is simply the kinetic part of $\widetilde{P}_{F}^{-1}$ (evaluated at zero mass). Since \eqref{Eq:HatDerivative} is a matrix operator there is no risk of confusion. Notice, that \eqref{Eq:BosonAnomalousDimensionProjectionBoson} is proportional to the vacuum amplitude $\widehat{\Phi}_{0}$ and thus, the boson contribution to the wavefunction renormalization vanishes in the symmetric regime.

We evaluate the contribution to the flow equivalently in the fermion subspace. For the non-diagonal background configuration \eqref{Eq:NonDiagonalVacuumConfigurationPerturbed} the fermion propagator is given by
\begin{equation}
\left( P_{F} \right)_{a b} = \begin{pmatrix} 0 & \left( P_{F}^{(+)} \right)_{a b} \\ \left( P_{F}^{(-) \, T} \right)_{a b} & 0 \end{pmatrix} ~,
\end{equation}
where the components 
\begin{eqnarray}
&& \hspace{-0.7cm}\big( P_{F}^{(\pm)} \big)_{a b} (p) \nonumber\\
&& \hspace{-0cm} =\: \widetilde{P}_{F} (p) \left( Z_{F} (1 + r_{F}) \fsl{p} \, \delta_{a b} \mp i \bar{h}(p) \widehat{\Phi}_{0} \Sigma_{a b} \right) ~, 
\label{Eq:FermionPropagatorBackground}
\end{eqnarray}
and
\begin{equation}
\widetilde{P}_{F} (p) = \left(Z_{F}^{2} (1 + r_{F})^{2} p^{2} + \bar{h}(p)^{2} \widehat{\Phi}_{0}^{2}\right)^{-1} ~.
\end{equation}
In the fermion subspace the fluctuations take the form
\begin{eqnarray}
&& \hspace{-0.5cm} \left( F_{F} \right)_{a b} (p,q) = i \bar{h}(-p,q) \nonumber\\
&& \times\:\Delta(p-q,Q) \begin{pmatrix} 0 & - \Lambda^{T}_{a b} \\ \Lambda_{a b} & 0 \end{pmatrix} + \mathcal{O}(\Delta^{2}) ~.
\end{eqnarray}
Going through the same steps as above, that is, computing the trace in $\delta^{(2)}\Gamma_{F}$, and evaluating the projection \eqref{Eq:ProjectionBosonWavefunctionRenormalization} we obtain
\begin{eqnarray}
\left. \delta^{(2)} \Gamma_{F} \right|_{Z_{B}} &=& - 2 \, \lim_{Q\rightarrow 0} \frac{\partial}{\partial Q^{2}} \int_{p} \left[ \bar{h}(-p,p+Q) \right]^{2} \nonumber\\
&& \hspace{-1.5cm} \times\: \hat{\partial}_{t} \, \Big\{ Z_{F} ( p ) ( 1 + r_{F} (p) ) \widetilde{P}_{F}(p) \nonumber\\
&& \hspace{-0.5cm} \times\: Z_{F} ( p + Q ) \left( 1 + r_{F} ( p + Q ) \right) \widetilde{P}_{F} (p + Q) \nonumber\\
&& \hspace{-0.5cm} +\: \widehat{\Phi}_{0}^{2} \, \bar{h}(p) \bar{h}(p+Q) \widetilde{P}_{F} (p) \widetilde{P}_{F} (p+Q) \Big\} ~, 
\label{Eq:BosonAnomalousDimensionProjectionFermion}
\end{eqnarray}
where $\left[ \bar{h}(-p,p+Q) \right]^{2} \equiv \bar{h}(-p,p+Q) \bar{h}(-p-Q, p)$. Putting both results \eqref{Eq:BosonAnomalousDimensionProjectionBoson} and \eqref{Eq:BosonAnomalousDimensionProjectionFermion} together \mbox{$\partial_{t} Z_{B} = \left. \delta^{(2)} \Gamma_{B} \right|_{Z_{B}} + \left. \delta^{(2)} \Gamma_{F} \right|_{Z_{B}}$} and using the definition $\eta_{B} = - \partial_{t} \ln Z_{B}$ we obtain the evolution equation for the boson anomalous dimension:
\begin{eqnarray}
\eta_{B} &=& 16 \frac{v_{d}}{d} \, \bigg\{ \rho_{0} \, \lambda_{2}^{2} \, m_{2,2}^{(B)} (\epsilon,\epsilon + 2 \rho_{0} \lambda_{2} ; \eta_{B} ) \nonumber\\
&& +\:  h^{2} \, m_{4}^{(F)} \left( \frac{2}{N} \rho_{0} h^{2} ; \eta_{F} \right) \nonumber\\
&& +\: \frac{2}{N} \rho_{0} h^{4} \, m_{2}^{(F)} \left( \frac{2}{N} \rho_{0} h^{2} ; \eta_{F} \right) \bigg\} ~.
\label{Eq:FlowEquationAnomalousDimensionBoson}
\end{eqnarray}
Here we have introduced the threshold functions $m_{2,2}^{(B)}$, $m_{2}^{(F)} $ and $m_{4}^{(F)}$ that define the one-loop contribution appearing in the calculation of the wavefunction renormalization. They are given explicitly in Appendix \ref{App:ThresholdFunctions}.

In the broken phase the wavefunction renormalization will get additional contributions from the radial mode \cite{Wetterich:1991be, Tetradis:1993ts}. In principle, these terms should be taken into account, however, since most of the running takes place in the symmetric regime, we do not expect them to play any major role.

\subsection{Fermion anomalous dimension and Yukawa coupling}
\label{Sec:FermionAnomlousDimension}

The derivation of the flow equation for the fermion anomalous dimension $\eta_{F} \equiv - \partial_{t} \ln Z_{F}$ and the Yukawa coupling $h^{2}$ proceeds in the same way as explained in \mbox{Sec.\ \ref{Sec:BosonAnomalousDimension}}. Here, the only difference is, that we need to choose a non-homogeneous configuration for the fermion fields, where in the momentum-representation we have
\begin{equation}
\psi(q) = \psi \delta(q,Q) ~, \qquad \bar{\psi}(q) = \bar{\psi} \delta(q,Q) ~.
\label{Eq:FermionConfigurationNonHomogeneous}
\end{equation}
The matrix field is evaluated in the constant background configuration $\Phi_{a b} = \widehat{\Phi}_{0} \Sigma_{a b}$, where the boson and fermion propagators are given by \eqref{Eq:BosonPropagatorBackground} and \eqref{Eq:FermionPropagatorBackground}, respectively.

Starting from the ansatz for the fermions \eqref{Eq:FermionConfigurationNonHomogeneous} we evaluate the fluctuation matrix
\begin{widetext}
\begin{eqnarray}
\hspace{-1cm} \mathcal{F}_{k}(p,q) = \begin{pmatrix} 0 & i \bar{h}(p - q,q) \, \bar{\psi}_{b}(q-p) \delta_{a c'} & - i \bar{h}(q,-q+p) \, \psi_{a}^{T} (-q+p) \delta_{b c'} \\
-i \bar{h}(-q + p, -p) \, \bar{\psi}_{a'}^{T} (q - p) \delta_{b' c}  & 0 & 0 \\
i \bar{h}(-p, p -q) \, \psi_{b'} (- q + p) \delta_{a' c} & 0 & 0
\end{pmatrix} ~, \nonumber\\
&&
\end{eqnarray}
\end{widetext}
where we indicate the flavor indices on the right-hand side explicitly. Recall, that the fluctuation is defined as the field-dependent part of the second functional derivative of the effective average action. The functional derivative from the left is taken with respect to the fields $\left(\Phi_{a b}^{T}, \psi_{c}^{T}, \bar{\psi}_{c} \right)$ which defines the row indices. The primed column indices are defined equivalently via the right-hand functional derivative. Together with the background field propagator $\mathcal{P}_{k}$ (see Appendix \ref{App:Propagators}) we evaluate the second order contribution \mbox{$\delta \Gamma_{k}^{(2)} = - \tfrac{1}{4} \STr \hat{\partial}_{t}\left( \mathcal{P}_{k} \mathcal{F}_{k} \right)^{2}$} in the series expansion. A short calculation yields
\begin{eqnarray}
\delta \Gamma_{k}^{(2)} &=& \frac{1}{2} \int_{p,q} \hat{\partial}_{t} \,\bigg\{ \left[ \bar{h} ( p - q , q ) \right]^{2} \left( P_{B} \right)_{a b, c d} (p) \nonumber\\
&& \times\: \bar{\psi}_{d}(q - p) \big( P_{F}^{(+)} \big)_{c a} (q) \psi_{b}(q - p) \nonumber\\ 
&& +\: \left[ \bar{h} ( q , p - q ) \right]^{2} \left( P_{B} \right)_{a b, d c} (p) \nonumber\\
&& \times\: \psi^{T}_{d}(-q + p) \big( P_{F}^{(-)} \big)^{T}_{c b} (q) \bar{\psi}^{T}_{a}(p - q) \bigg\} ~.
\end{eqnarray}
Inserting the expressions the boson and fermion propagators \eqref{Eq:BosonPropagatorBackground} and \eqref{Eq:FermionPropagatorBackground} this can be written in the form:
\begin{eqnarray}
\delta \Gamma_{k}^{(2)} &=& \frac{1}{N} \int_{p} \hat{\partial}_{t} \,\bigg\{ \left[ \bar{h} ( Q , p ) \right]^{2} \widetilde{P}_{F} (p) \Big[ Z_{F} \left( 1 + r_{F} \right) \nonumber\\ 
&& \hspace{-1.3cm} \times\: \bar{\psi}^{a} \,\fsl{p} \,\big\{\left(N^{2} - 2 \right) P_{B 1} (p-Q) + P_{B 2} (p-Q) \big\} \psi^{a} \nonumber\\
&& \hspace{-1.3cm} +\: i \bar{h}(Q) \, \bar{\psi}^{a} \,\widehat{\Phi}_{0} \Sigma_{a b} \left. \big\{ 2 P_{B 1} (p-Q) - P_{B 2} (p-Q) \right\} \psi^{b} \Big] \bigg\} ~. \nonumber\\
\end{eqnarray}
Projecting this equation onto the corresponding operators in the ansatz for the effective average action we obtain the evolution equation for the fermion anomalous dimension
\begin{eqnarray}
\eta_{F} &=& \frac{8}{N} \frac{v_{d}}{d} h^{2} \, \bigg\{ (N^{2} - 2) m_{1,2}^{(FB)} \left( \frac{2}{N} \rho_{0} h^{2} , \epsilon ; \eta_{F} , \eta_{B} \right) \nonumber\\
&& +\: m_{1,2}^{(FB)} \left( \frac{2}{N} \rho_{0} h^{2} ,  \epsilon + 2 \rho_{0} \lambda_{2} ; \eta_{F} , \eta_{B} \right) \bigg\} ~,
\label{Eq:FlowEquationAnomalousDimensionFermion}
\end{eqnarray}
and the flow equation for the momentum-independent part of the Yukawa coupling
\begin{widetext}
\begin{eqnarray}
\partial_{t} h^{2} &=& (d - 4 + 2 \eta_{F} + \eta_{B} ) h^{2} - \frac{8}{N} h^{4} v_{d} \left\{ 2 \, l^{(FB)}_{1,1} \left( \frac{2}{N} \rho_{0} h^{2} , \epsilon ; \eta_{F} , \eta_{B} \right) - l^{(FB)}_{1,1} \left( \frac{2}{N} \rho_{0} h^{2} , \epsilon + 2 \rho_{0} \lambda_{2} ; \eta_{F} , \eta_{B} \right) \right\} \nonumber\\
&& +\: \frac{16}{N} \rho_{0} h^{4} v_{d} \left\{ 2 \lambda_{2} \, l^{(FB)}_{1,2} \left( \frac{2}{N} \rho_{0} h^{2} , \epsilon ; \eta_{F} , \eta_{B} \right) - \left(2 \lambda_{2} + 2 \rho_{0} \lambda_{3} \right) l^{(FB)}_{1,2} \left( \frac{2}{N} \rho_{0} h^{2} , \epsilon + 2 \rho_{0} \lambda_{2} ; \eta_{F} , \eta_{B} \right) \right\} \nonumber\\
&& +\: \frac{32}{N^{2}} \rho_{0} h^{6} v_{d} \left\{ 2 \, l^{(FB)}_{2,1} \left( \frac{2}{N} \rho_{0} h^{2} , \epsilon ; \eta_{F} , \eta_{B} \right) - l^{(FB)}_{2,1} \left( \frac{2}{N} \rho_{0} h^{2} , \epsilon + 2 \rho_{0} \lambda_{2} ; \eta_{F} , \eta_{B} \right) \right\} ~.
\label{Eq:FlowEquationYukawa}
\end{eqnarray}
\end{widetext}
Here, we have defined the threshold functions $m_{1,2}^{(FB)}$, $l^{(FB)}_{1,1}$, $l^{(FB)}_{1,2}$, and $l^{(FB)}_{2,1}$ that are given explicitly in Appendix \ref{App:ThresholdFunctions}. Together with the flow equations for the parameters and couplings of the effective average potential and the anomalous dimension for the bosons they constitute a closed set of differential equations that can be solved numerically.

\section{Results for the quantum phase transition}
\label{Sec:Results}

\begin{figure*}[t]
\centering
\hspace{-0.5cm}
\subfigure{
\centering
\includegraphics[width=0.46\textwidth]{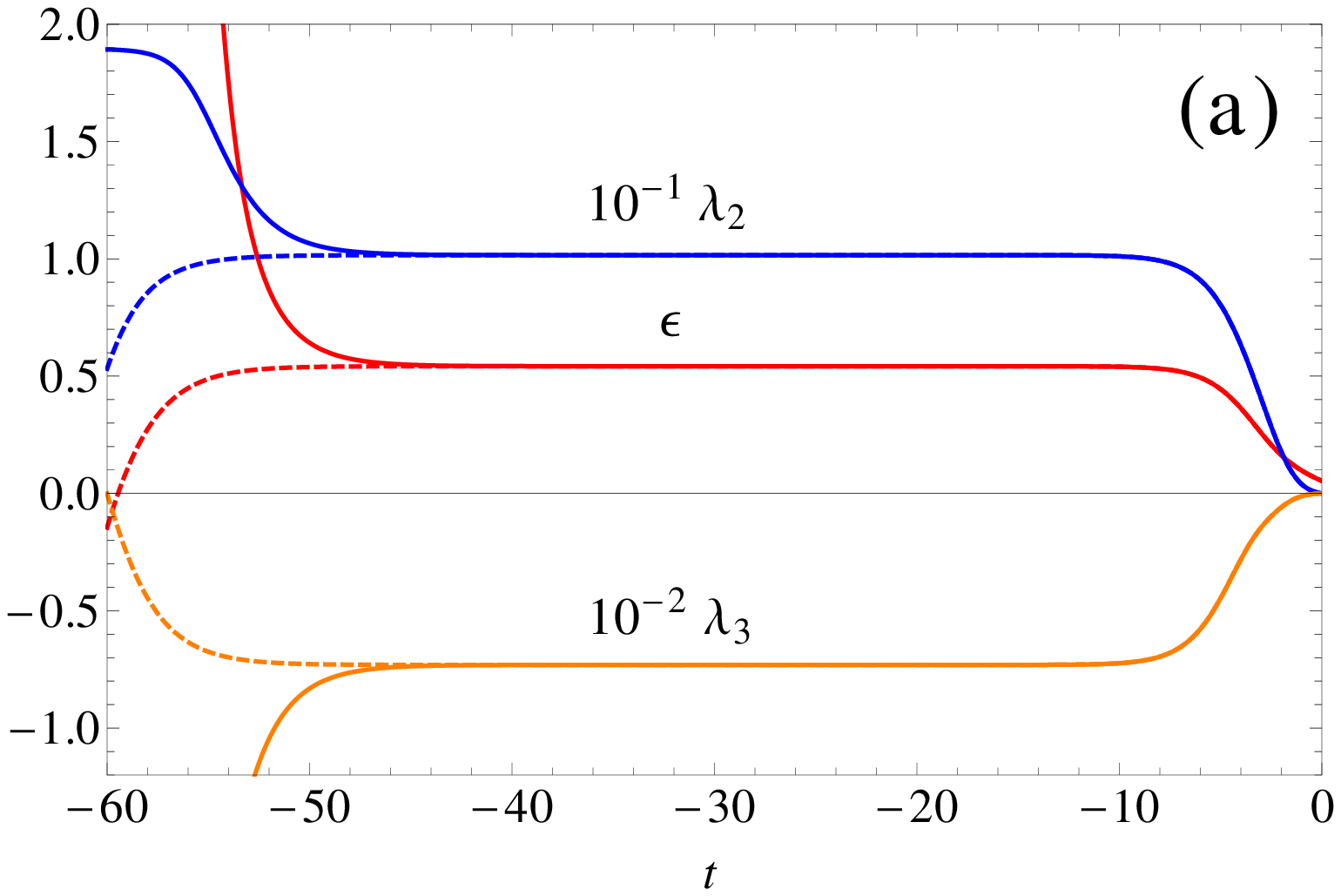}
}
\quad
\subfigure{
\centering
\includegraphics[width=0.45\textwidth]{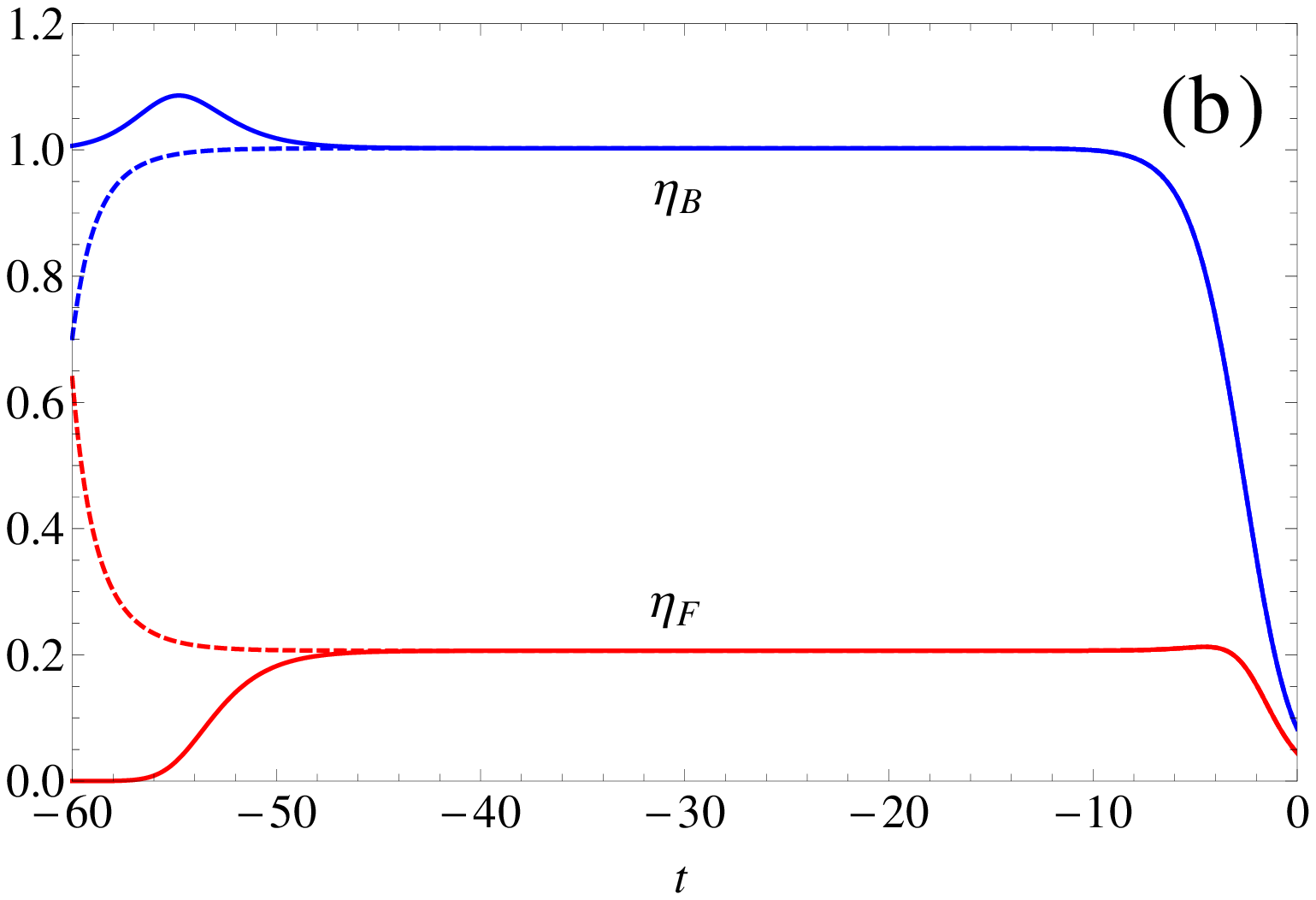}
}
\caption{\label{Fig:ScalingSolution} Renormalization group flow (a) for the dimensionless renormalized parameter $\epsilon$, the (rescaled) dimensionless renormalized couplings $10^{-1} \lambda_{2}$, and $10^{-2} \lambda_{3}$, and (b) for the boson and fermion anomalous dimensions $\eta_{B}$ and $\eta_{F}$ as a function of the scale parameter $t = \ln k/ \Lambda$ close to the critical trajectory. The full and dashed curves refer to initial conditions just above and below the critical parameter $\epsilon_{\Lambda,(cr)}$, respectively.}
\end{figure*}

\renewcommand{\arraystretch}{1.5}
\renewcommand{\tabcolsep}{0.5cm}
\begin{table*}[t]
\centering
\begin{tabular}{c | c c c c}
                                        &   4th order   &   6th order   &   8th order  &  10th order  \\ \hline\hline
$\eta_{B}$		&	$0.989$    &   $0.999$   &   $1.003$   &  $1.000$  \\ \hline
$\eta_{F} $		&	$0.223$    &   $0.211$   &   $0.207$   &  $0.210$  \\ \hline
$\nu $			&	$1.922$    &   $1.936$   &   $1.791$   &  $1.874$  \\ \hline
$\gamma $		&	$1.942$    &   $1.939$   &   $1.786$   &  $1.875$  \\ \hline
$\nu (2 - \eta_{B}) $	&	$1.942$    &   $1.939$   &   $1.786$   &  $1.875$  \\ \hline
$\beta $		&	$1.911$    &   $1.935$   &    $1.793$   &   $1.874$  \\	\hline
$\frac{1}{2} \nu (d - 2 + \eta_{B})$	&	$1.911$    &   $1.935$   &    $1.793$   &   $1.874$  \\ \hline
\end{tabular}
\caption{\label{Tab:CriticalExponents} $N = 2$ critical exponents for different orders in the series expansion.}
\end{table*}

We evolve the flow equations starting from an appropriate ultraviolet scale $\Lambda$ to the physical limit $k \rightarrow 0$. The following results have been obtained for the set of initial conditions: $Z_{B, \Lambda} = 10^{-10}$, $Z_{F, \Lambda} = 1$, $\bar{h}^{2}_{\Lambda} = \Lambda$, where the mass at the ultraviolet scale $\bar{m}^{2}_{\Lambda}$ is taken as a free parameter, and all higher order couplings are set to zero, i.\,e.\ $\bar{\lambda}_{n, \Lambda} = 0$, $n = 2 , \ldots , n_{max}$. By varying the dimensionless mass $\epsilon_{\Lambda}$ at the ultraviolet scale we may tune the system across a quantum phase transition. Here, $\delta \epsilon = | \epsilon_{\Lambda} - \epsilon_{\Lambda,(cr)}|$ measures the deviation of the parameter $\epsilon_{\Lambda}$ from its critical value $\epsilon_{\Lambda, (cr)}$. Close to $\epsilon_{\Lambda, (cr)}$ we find a fixed point solution for the dimensionless renormalized parameters and couplings. That is, the parameters and couplings stay nearly constant over a wide range of scales as illustrated in Fig.\ \ref{Fig:ScalingSolution}. This is a clear indication for the presence of a continuous phase transition where the system displays a universal scaling behavior. For $\delta \epsilon > 0$, where the mass parameter $\epsilon_{\Lambda}$ is above the critical value $\epsilon_{\Lambda, (cr)}$, the solution stays in the symmetric regime. However, starting just below $\epsilon_{\Lambda,(cr)}$ the scale-dependent mass eventually becomes negative which signals the transition into the broken phase. This result does not depend on the special choice of initial conditions, that is, we have checked the stability of our results for different initial values of $Z_{B, \Lambda}$, $\bar{h}^{2}_{\Lambda}$, and $\bar{\lambda}_{n, \Lambda}$. Furthermore, the fixed point solution exists for all considered truncations of the effective average potential (see Tab.\ \ref{Tab:CriticalCouplings}), where we have taken the Taylor series expansion of the effective potential up to the tenth order in the field $\Phi$. The independence of the scaling solution both on the initial conditions and higher order operators in the effective potential is a manifestation of universality near a continuous phase transition.

\renewcommand{\arraystretch}{1.5}
\renewcommand{\tabcolsep}{0.5cm}
\begin{table*}[t]
\centering
\begin{tabular}{c | c c c c}
     &   4th order   &   6th order   &   8th order   &  10th order  \\ \hline\hline
$\epsilon_{\ast}$   &    0.4842   &   0.5242   &    0.5424   &   0.5288  \\
$\lambda_{2, \ast}$  &   10.7678   &  10.3744    &   10.1573   &  10.3210  \\
$\lambda_{3, \ast}$  &   &  -48.5405   &   -73.0962   &  -54.6552  \\
$\lambda_{4, \ast}$  &    &    &   -1956.82   &  -485.084  \\ 
$\lambda_{5, \ast}$  &    &    &    &   219713   \\ \hline
$h^{2}_{\ast}$   &   12.8622   &   12.9203   &   12.9438   &  12.9264  \\ \hline
\end{tabular}
\caption{\label{Tab:CriticalCouplings} $N = 2$ fixed point values for different orders in the series expansion.}
\end{table*}

In the symmetric phase both the wavefunction renormalization $Z_{B}$ and the renormalized mass $m_{R}^{2} = Z_{B}^{-1} \bar{m}^{2}$ at the scale $k$ receive large contributions from the massless fermions. That is $\eta_{B} \rightarrow 1$ for $k\rightarrow 0$ even far from the phase transition which can be clearly seen in Fig.\ \ref{Fig:ScalingSolution}(b) where the boson anomalous dimension assumes a value close to one in the symmetric phase. To compute the critical scaling we introduce the renormalized mass $\tilde{m}_{R}^{2}$ at a fixed scale $k_{c}$:
\begin{equation}
\tilde{m}^{2}_{R} (k_{c} , \delta \epsilon) = k_{c}^{2} \left( u_{k_{c}}'(0) - u_{k_{c}, (cr)}'(0) \right) ~.
\label{Eq:TildeRenormalizedMass}
\end{equation}
It is given in terms of the first derivatives of the effective average potential in the symmetric phase, where the scale \mbox{$k_{c} = r_{c} \tilde{m}_{R}$} is defined via the parameter $r_{c}$ in a standard way \cite{Rosa:2000ju, *Hofling:2002hj}.

The critical exponent $\nu$ characterizes the divergence of the correlation length at the critical point. Here, the correlation length is identified with the inverse renormalized mass \cite{Tetradis:1993ts, Berges:2000ew} as given in \eqref{Eq:TildeRenormalizedMass} and the critical exponent $\nu$ is defined as \cite{Rosa:2000ju, *Hofling:2002hj}
\begin{eqnarray}
\nu &=& \frac{1}{2} \lim_{\delta \epsilon \rightarrow 0} \frac{\partial \ln \tilde{m}^{2}_{R} (k_{c}, \delta \epsilon) }{\partial \ln \delta \epsilon} \nonumber\\
&=& \lim_{\delta \epsilon \rightarrow 0} \left( \hat{\nu}(k_{c},\delta \epsilon) + \tilde{\nu}(k_{c},\delta \epsilon) \nu \right) ,
\end{eqnarray}
where
\begin{eqnarray}
\hat{\nu}(k, \delta \epsilon) &=& \frac{1}{2} \left. 
\frac{\partial \ln \tilde{m}^{2}_{R} (k , \delta \epsilon) }{\partial \ln \delta \epsilon} \right|_{t} 
~, \\
\tilde{\nu}(k, \delta \epsilon) &=& \frac{1}{2} \left. \frac{\partial \ln \tilde{m}^{2}_{R} (k , \delta \epsilon)}{\partial t} \right|_{\delta \epsilon} ~.
\end{eqnarray}
The value for the critical exponent $\nu$ is independent of the parameter $r_{c}$, as long as $r_{c} \lesssim 1$. This essentially corresponds to the requirement that the scale $k_{c}$ is sufficiently close to the limiting value $k \rightarrow 0$. In our calculations we have taken $r_{c} \simeq 0.01$.

The critical exponent $\gamma$ determines the divergence of the susceptibility which is encoded in the non-renormalized mass $\bar{m}^{2} = Z_{B} m_{R}^{2}$ \cite{Tetradis:1993ts,Berges:2000ew}. Although it is evaluated in the symmetric phase, it is not affected by the fluctuations of the fermions. We have
\begin{equation}
\gamma = \lim_{\delta \epsilon \rightarrow 0} \frac{\partial \ln \bar{m}^{2} (\delta \epsilon)}{\partial \ln \delta \epsilon} ~.
\end{equation}

Finally, the critical exponent $\beta$ measures the fluctuations of the renormalized order parameter $\rho_{0,k}$ and is defined in the broken phase:
\begin{equation}
\beta = \frac{1}{2} \lim_{\delta \epsilon \rightarrow 0} \frac{\partial \ln \rho_{0}^{2}}{\partial \ln \delta \epsilon} ~.
\end{equation}

We extract the anomalous dimensions $\eta_{B}$ and $\eta_{F}$ the same way as the critical couplings. Close to the critical parameter $\epsilon_{\Lambda,(cr)}$ the renormalization group flow approaches the fixed point solution where the system is scale-invariant. That is, the solutions to the flow equation stay constant over a wide range of scales where we may extract the corresponding quantities. The values of the anomalous dimensions $\eta_{B}$ and $\eta_{F}$ are defined at the critical point in the window where we have a plateau (see Fig.\ \ref{Fig:ScalingSolution}). Our results are summarized in Tab.\ \ref{Tab:CriticalExponents} where we show the values of the critical exponents. They are given for different orders of the series expansion for the effective average potential and were obtained using the optimized regulator \eqref{Eq:OptimizedRegulator}. We have also calculated the critical exponents for the exponential regulator. For that calculation, however, we neglect the dependence on the anomalous dimensions in the threshold functions. Since the anomalous dimensions are of order one, this gives a very rough estimate of the systematic error for our results. We find an agreement of the critical exponents on the $10\%$ level. The scaling relations $\gamma = \nu \left(2 - \eta_{B}\right)$ and $\beta = \tfrac{1}{2} \nu \left( d - 2 + \eta_{B} \right)$ for the critical exponents provide a consistency check of our calculations and are also given in Tab.\ \ref{Tab:CriticalExponents}. We see that our results show a reasonable convergence in the series expansion.

In Tab.\ \ref{Tab:CriticalCouplings} we give the values of the critical parameters and couplings, where the asterisk denotes the fixed point values, i.\,e.\ $\epsilon_{\ast} \equiv u'_{\ast}$, $\lambda_{2,\ast} \equiv u^{(2)}_{\ast}$ etc.\ These quantities are not universal and depend on the particular renormalization group scheme. It is important to comment on their behavior in the series expansion of the effective average potential. Taking into account only the relevant operators, that is, expanding the potential to fourth order, yields a reasonably good result for the scaling exponents. This can also be seen directly in Tab.\ \ref{Tab:CriticalCouplings} where the inclusion of higher order irrelevant operators does not significantly alter the values for the relevant critical couplings, in contrast to the higher order couplings, that vary strongly for different orders of the expansion. The relevant couplings are completely stable and show that the important physical information is captured already in the lowest truncation with all relevant operators included.

\section{Conclusions}
\label{Sec:Conclusions}

We have calculated the critical exponents at the quantum critical point for the three-dimensional matrix Yukawa type model with $U(2)$ symmetry, which describes $N=2$ species of Weyl fermions. This theory captures the relevant fluctuations close to the chiral phase transition for a low-energy effective model of spinless fermions on the honeycomb lattice. We have shown that the calculated critical exponents at the continuous quantum critical point define a new universality class distinct from Gross-Neveu or Neveu-Yukawa type models. In particular this system is special in the sense that it is characterized by large values of the anomalous dimensions. Similar results have been obtained in a single Dirac cone model where the semimetal-superfluid transition was investigated using functional renormalization group techniques \cite{Strack:2009ia, *Obert:2011}. There, also a second order phase transition was found with large values for the anomalous dimensions, both for the anomalous dimensions of composite and fermion fields. In the context of compact three-dimensional QED one also observes a large value for the anomalous dimension of the gauge field $\eta_{A} = 1$, where the result holds exactly due to gauge invariance \cite{Herbut:1996ut, Hove:2000zz}. Whether these non-trivial properties can be found in suspended graphene is still an important open question. To see if these results are indeed relevant for graphene requires us to include the long-range Coulomb interactions. In that case one has to ask whether the instantaneous interaction is relevant for the critical dynamics, or if one has an effective restoration of Euclidean rotational symmetry. Although, there are indications for such a behavior in the critical region of a Gross-Neveu-Yukawa fixed point for the semimetal-insulator transition \cite{Herbut:2009qb,*Herbut:2009vu} until now this is an open issue.

\begin{acknowledgments}

We thank J.\ Braun, J.\ Drut, L.\ Janssen, J.\ Pawlowski, and D.\ Scherer for discussions. This work was supported by the Deutsche Forschungsgemeinschaft within the \mbox{SFB 634}.

\end{acknowledgments}

\appendix

\section{Definition of propagators}
\label{App:Propagators}

In our calculations we frequently need the full regularized propagator $\mathcal{P}_{k}$ evaluated in a constant background field. For the inverse regularized propagator we have
\begin{widetext}
\begin{equation}
\mathcal{P}^{-1}_{k}(p) = \begin{pmatrix} Z_{B} (1 + r_{B}) p^{2} + \bar{M}_{B}^{2} & 0 & 0 \\ 0 & 0 & Z_{F,k} ( 1 + r_{F} ) \fsl{p}^{T} - i \bar{M}_{F} \\ 0 & Z_{F} ( 1 + r_{F} ) \fsl{p} + i \bar{M}_{F} & 0
\end{pmatrix} ~,
\label{Eq:InversePropagator}
\end{equation}
\end{widetext}
where $\bar{M}_{B}^{2}$ and $\bar{M}_{F}$ define the scale-dependent mass matrices that depend on the particular background field configuration (see Appendix \ref{App:MassMatrices}).

The background field propagator takes the form
\begin{equation}
\mathcal{P}_{k} (p) \equiv \begin{pmatrix} P_{B} ( p ) & 0 & 0 \\ 0 & 0 & P_{F}^{(+)} ( p ) \\ 0 & P_{F}^{(-) \,T} ( p ) & 0
\end{pmatrix} ~,
\label{Eq:Propagator}
\end{equation}
where the boson propagator is given by
\begin{equation}
P_{B} ( p ) = \left( Z_{B} (1 + r_{B}) p^{2} + \bar{M}_{B}^{2} \right)^{-1} ~,
\end{equation}
and the fermion propagator 
\begin{equation} 
P_{F}^{(\pm)} ( p  ) = \widetilde{P}_{F} ( \bar{M}_{F} ) \left( Z_{F} (1 + r_{F}) \fsl{p} \mp  i \bar{M}_{F} \right) ~,
\end{equation}
with
\begin{equation}
\widetilde{P}_{F} ( p ) = \left( Z_{F}^{2} (1 + r_{F})^{2} p^{2} + \bar{M}_{F}^{2} \right)^{-1} ~.
\label{mathPF}
\end{equation}
Since the propagators are functions of the mass matrices $\bar{M}_{B}$ and $\bar{M}_{F}$, they do not necessarily have to be diagonal in flavor space. While it is easy to evaluate the flow equation for the effective potential in the diagonal basis, it is useful to keep the propagators in their nondiagonal form for the computation of flow equations for the anomalous dimensions and Yukawa coupling.

\section{Threshold functions}
\label{App:ThresholdFunctions}

For generic regulators the threshold functions are defined by
\begin{eqnarray}
&& \hspace{-0.6cm}l_{n}^{(B)} (w ; \eta_{B}) = \frac{\delta_{n,0} + n}{2} \int_{0}^{\infty} \!\! dy \, y^{\frac{d}{2} - 1} \nonumber\\
&& \hspace{0.3cm} \times\: \frac{1}{Z_{B}} \frac{\partial R_{B}}{\partial t} \left[ Z_{B} P_{B} \left( Z_{B} w \right) \right]^{n+1} \\
&& \hspace{-0.6cm} l_{n}^{(F)} (w ; \eta_{F}) = \left( \delta_{n,0} + n \right) \int_{0}^{\infty} \!\! dy \, y^{\frac{d}{2}} \nonumber\\
&& \hspace{0.3cm} \times\:  Z_{F} \left( 1+r_{F} \right) \frac{\partial}{\partial t} \left( Z_{F} r_{F} \right) \left[ Z_{F}^{2} \widetilde{P}_{F} \left( Z_{F}^{2} w \right) \right]^{n+1} \\
&& \hspace{-0.6cm} l_{n_{1},n_{2}}^{(FB)} (w_{1},w_{2} ; \eta_{F} , \eta_{B})  = - \frac{1}{2} \int_{0}^{\infty} \!\! dy \, y^{\frac{d}{2}-1} \, \nonumber\\
&& \hspace{0.3cm} \times\: \frac{\hat{\partial}}{\partial t} \Big\{ \left[ Z_{F}^{2} \widetilde{P}_{F} \left( Z_{F}^{2} w_{1} \right) \right]^{n_{1}} \left[ Z_{B} P_{B} \left( Z_{B} w_{2} \right) \right]^{n_{2}} \Big\} \nonumber\\
&& \\
&& \hspace{-0.6cm} m_{2}^{(F)} (w ; \eta_{F}) = - \frac{1}{2} \int_{0}^{\infty} \!\! dy \, y^{\frac{d}{2}-1} \nonumber\\
&& \hspace{0.3cm} \times\: \frac{\hat{\partial}}{\partial t} \left\{ \left[ Z_{F}^{2} \widetilde{P}_{F} \left( Z_{F}^{2} w \right) \right]^{2}  \frac{\partial}{\partial y} \left[ Z_{F}^{2} \widetilde{P}_{F} \left( Z_{F}^{2} w \right) \right] \right\}^{2}  \nonumber\\
&& \\
&& \hspace{-0.6cm} m_{4}^{(F)} (w ; \eta_{F}) =- \frac{1}{2}  \int_{0}^{\infty} \!\! dy \, y^{\frac{d}{2}+1} \nonumber\\
&& \hspace{0.3cm} \times\:\frac{\hat{\partial}}{\partial t} \left\{ \frac{\partial}{\partial y} \Big[  \left( 1+r_{F} \right) Z_{F}^{2} \widetilde{P}_{F} \left( Z_{F}^{2} w\right) \Big] \right\}^{2} \\
&& \hspace{-0.6cm} m_{n_{1},n_{2}}^{(FB)} (w_{1},w_{2} ; \eta_{F} , \eta_{B}) = - \frac{1}{2} \int_{0}^{\infty} \!\! dy\, y^{\frac{d}{2}} \nonumber\\
&& \hspace{0.3cm} \times\: \frac{\hat{\partial}}{\partial t} \bigg\{ \left( 1+r_{F} \right) \left[ Z_{F}^{2} \widetilde{P}_{F}\left( Z_{F}^{2} w_{1} \right) \right]^{n_{1}} \nonumber\\
&& \hspace{1.1cm} \times\: \left[ Z_{B} P_{B} \left( Z_{B} w_{2} \right) \right]^{n_{2}} \frac{\partial}{\partial y} \left[ Z_{B} P_{B} \left( Z_{B} w_{2} \right) \right] \bigg\} \nonumber\\
&&
\end{eqnarray}
where we have defined the dimensionless quantity $y = q^{2} / k^{2}$. Here, it is understood that the regulators and propagators are taken as functions of $y$, i.\,e.\ $R_{B}(y) \equiv R_{B} (q^{2}) / k^{2}$, $P_{B}(y) \equiv P_{B} (q^{2}) k^{2}$, etc. and the parameters $w$, $w_{1}$, and $w_{2}$  denote dimensionless renormalized quantities. Furthermore we use the formal scale derivative 
\begin{eqnarray}
\frac{\hat{\partial}}{\partial t} ~ &\equiv& ~ \frac{\partial_{t} R_{B}}{\partial t} \frac{\partial}{\partial \left( P_{B}^{-1} \right)} \nonumber\\
&& \hspace{0.2cm} +\: \frac{2}{Z_{F}} \frac{\widetilde{P}_{F}^{-1}(0)}{1 + r_{F}} \frac{\partial}{\partial t} \left( Z_{F} r_{F} \right) \frac{\partial}{\partial \left( \widetilde{P}_{F}^{-1} \right)} ~, 
\end{eqnarray}
that includes the scale-dependence of the regulator functions.

For the three-dimensional optimized regulator the shape functions are given by
\begin{eqnarray}
r_{B} (y) &=& \left( \frac{1}{y} - 1 \right) \theta(1-y) ~, \\
r_{F} (y) &=& \left( \frac{1}{\sqrt{y}} -1 \right) \theta(1-y) ~,
\end{eqnarray}
and the threshold functions can be calculated analytically. They take the following form:
\begin{eqnarray}
&& \hspace{-0.6cm} l_{n}^{(B)} (w ; \eta_{B}) \nonumber\\ 
&& \hspace{-0.3cm} =\: \frac{2 (\delta_{n,0} + n)}{d} \left( 1 - \frac{\eta_{B}}{d+2} \right) \frac{1}{(1  + w)^{n+1}} \\
&& \hspace{-0.6cm} l_{n}^{(F)} (w ; \eta_{F}) \nonumber\\ 
&& \hspace{-0.3cm} =\: \frac{2 (\delta_{n,0} + n)}{d} \left( 1 - \frac{\eta_{F}}{d+1} \right) \frac{1}{(1  + w)^{n+1}} \\
&& \hspace{-0.6cm} l^{(FB)}_{n_{1},n_{2}} (w_{1} , w_{2} ; \eta_{F} , \eta_{B}) \nonumber\\
&& \hspace{-0.4cm} = \frac{2}{d} \frac{1}{(1+w_{1})^{n_{1}} (1+w_{2})^{n_{2}}} \left\{ \frac{n_{1}}{1+w_{1}} \left( 1 - \frac{\eta_{F}}{d+1} \right) \right. \nonumber\\
&& \hspace{0.3cm} \left. +\: \frac{n_{2}}{1+w_{2}} \left( 1 - \frac{\eta_{B}}{d+2} \right) \right\} \\
&& \hspace{-0.6cm} m_{2}^{(F)}(w;\eta_{F}) = \frac{1}{(1+w)^{4}} \\
&& \hspace{-0.6cm} m_{4}^{(F)}(w; \eta_{F}) = \frac{1}{(1+w)^{4}} + \frac{1 - \eta_{F}}{d-2} \frac{1}{(1+w)^{3}} \nonumber\\
&& \hspace{0.3cm} -\: \left(\frac{1-\eta_{F}}{2 d - 4} + \frac{1}{4}\right) \frac{1}{(1+w)^{2}} \\
&& \hspace{-0.6cm} m^{(FB)}_{n_{1},n_{2}} (w_{1} , w_{2} ; \eta_{F} , \eta_{B}) \nonumber\\
&& \hspace{-0.3cm} =\: \left( 1- \frac{\eta_{B}}{d + 1} \right) \frac{1}{(1+w_{1})^{n_{1}} (1+w_{2})^{n_{2}}}
\end{eqnarray}

\section{Mass matrices}
\label{App:MassMatrices}

The mass matrix $\bar{M}_{B}$ is defined via the second functional derivatives of the effective average potential:
\begin{equation}
\bar{M}_{B}^{2}(\Phi)_{\, a b, c d} (p,q) \equiv \frac{\overrightarrow{\delta}}{\delta \Phi^{T \, a b}(-p)} \int_{x} U_{k}(\Phi) \,\frac{\overleftarrow{\delta}}{\delta \Phi^{c d}(q)} ~.
\label{Eq:MassMatrixDefinition}
\end{equation}
For the effective average potential $U_{k}$ we consider only the dependence on the quadratic invariant $\bar{\rho} = \tfrac{1}{2} \tr{ \Phi^{2} }$. To evaluate \eqref{Eq:MassMatrixDefinition}, we need the following functional derivatives of $\bar{\rho}$ which are given given by
\begin{eqnarray}
\frac{\delta \bar{\rho} (p')}{\delta \Phi^{T\, a b} (-p)} &=& \Phi_{a b}(p+p') ~, \\
\frac{\delta \bar{\rho} (p')}{\delta \Phi^{a b} (q)} &=& \Phi_{a b}^{T}(p'-q) ~, \\
\frac{\delta^{2} \bar{\rho} (p')}{\delta \Phi^{T\, a b} (-p) \delta \Phi^{c d} (q)} &=& \delta_{a c} \delta_{b d}  \delta(p',-p + q) ~.
\end{eqnarray}
We obtain
\begin{eqnarray}
&& \hspace{-1cm} \bar{M}_{B}^{2}(\Phi)_{\, a b, c d} (p,q) \nonumber\\
&=& \int_{x,p',q'} \frac{\delta^{2} U_{k}}{\delta \bar{\rho}(p') \delta \bar{\rho}(q')} \frac{\delta \bar{\rho} (p')}{\delta \Phi^{T\, a b} (-p)} \frac{\delta \bar{\rho} (q')}{\delta \Phi^{c d} (q)} \nonumber\\
&& +\: \int_{x,p'} \frac{\delta U_{k}}{\delta \bar{\rho}(p')} \frac{\delta^{2} \bar{\rho}(p')}{\delta \Phi^{T \, a b}(-p) \delta \Phi^{c d}(q)} ~.
\end{eqnarray}
In the non-diagonal constant background configuration $\Phi_{a b} = \widehat{\Phi}_{0} \Sigma_{a b}$ the above expression simplifies:
\begin{eqnarray}
&& \hspace{-1cm} \bar{M}_{B}^{2}(\Phi)_{\, a b, c d} (p,q) \nonumber\\
&=& \left[ \frac{\partial U_{k}}{\partial \bar{\rho}} \delta_{a c} \delta_{b d} + \widehat{\Phi}_{0}^{2} \frac{\partial^{2} U_{k}}{\partial \bar{\rho}^{2}} \Sigma_{a b} \Sigma_{c d}^{T}\right] \delta(p,q) ~.
\end{eqnarray}
The eigenvalues of this matrix are given by
\begin{equation}
\bar{M}_{B 1}^{2} = \frac{\partial U_{k}}{\partial \bar{\rho}} ~, \quad \bar{M}_{B 2}^{2} = \frac{\partial U_{k}}{\partial \bar{\rho}} + N \widehat{\Phi}_{0}^{2} \frac{\partial^{2} U_{k}}{\partial \bar{\rho}^{2}} ~,
\end{equation}
and correspond to the masses of the Goldstone and radial modes.

In the background configuration $\Phi_{a b} = \widehat{\Phi}_{0} \Sigma_{a b}$ the fermion mass matrix is given by
\begin{equation}
\bar{M}_{F a b} (p, q) = \bar{h}(q) \widehat{\Phi}_{0} \, \Sigma_{a b} \delta (p,q) ~.
\end{equation}

\bibliography{references}

\begin{thebibliography}{10}%
\makeatletter
\providecommand \@ifxundefined [1]{%
 \ifx #1\undefined \expandafter \@firstoftwo
 \else \expandafter \@secondoftwo
\fi
}%
\providecommand \@ifnum [1]{%
 \ifnum #1\expandafter \@firstoftwo
 \else \expandafter \@secondoftwo
\fi
}%
\providecommand \enquote [1]{``#1''}%
\providecommand \bibnamefont  [1]{#1}%
\providecommand \bibfnamefont [1]{#1}%
\providecommand \citenamefont [1]{#1}%
\providecommand\href[0]{\@sanitize\@href}%
\providecommand\@href[1]{\endgroup\@@startlink{#1}\endgroup\@@href}%
\providecommand\@@href[1]{#1\@@endlink}%
\providecommand \@sanitize [0]{\begingroup\catcode`\&12\catcode`\#12\relax}%
\@ifxundefined \pdfoutput {\@firstoftwo}{%
 \@ifnum{\z@=\pdfoutput}{\@firstoftwo}{\@secondoftwo}%
}{%
 \providecommand\@@startlink[1]{\leavevmode\special{html:<a href="#1">}}%
 \providecommand\@@endlink[0]{\special{html:</a>}}%
}{%
 \providecommand\@@startlink[1]{%
  \leavevmode
  \pdfstartlink
   attr{/Border[0 0 1 ]/H/I/C[0 1 1]}%
   user{/Subtype/Link/A<</Type/Action/S/URI/URI(#1)>>}%
  \relax
 }%
 \providecommand\@@endlink[0]{\pdfendlink}%
}%
\providecommand \url  [0]{\begingroup\@sanitize \@url }%
\providecommand \@url [1]{\endgroup\@href {#1}{\urlprefix}}%
\providecommand \urlprefix [0]{URL }%
\providecommand \Eprint[0]{\href }%
\@ifxundefined \urlstyle {%
  \providecommand \doi [1]{doi:\discretionary{}{}{}#1}%
}{%
  \providecommand \doi [0]{doi:\discretionary{}{}{}\begingroup
  \urlstyle{rm}\Url }%
}%
\providecommand \doibase [0]{http://dx.doi.org/}%
\providecommand \Doi[1]{\href{\doibase#1}}%
\providecommand \bibAnnote [3]{%
  \BibitemShut{#1}%
  \begin{quotation}\noindent
    \textsc{Key:}\ #2\\\textsc{Annotation:}\ #3%
  \end{quotation}%
}%
\providecommand \bibAnnoteFile [2]{%
  \IfFileExists{#2}{\bibAnnote {#1} {#2} {\input{#2}}}{}%
}%
\providecommand \typeout [0]{\immediate \write \m@ne }%
\providecommand \selectlanguage [0]{\@gobble}%
\providecommand \bibinfo [0]{\@secondoftwo}%
\providecommand \bibfield [0]{\@secondoftwo}%
\providecommand \translation [1]{[#1]}%
\providecommand \BibitemOpen[0]{}%
\providecommand \bibitemStop [0]{}%
\providecommand \bibitemNoStop [0]{.\EOS\space}%
\providecommand \EOS [0]{\spacefactor3000\relax}%
\providecommand \BibitemShut [1]{\csname bibitem#1\endcsname}%
\bibitem{Novoselov:2005kj}%
  \BibitemOpen
  \bibfield{author}{%
  \bibinfo {author} {\bibfnamefont{K.}~\bibnamefont{Novoselov}}, \bibinfo
  {author} {\bibfnamefont{A.}~\bibnamefont{Geim}}, \bibinfo {author}
  {\bibfnamefont{S.}~\bibnamefont{Morozov}}, \bibinfo {author}
  {\bibfnamefont{D.}~\bibnamefont{Jiang}}, \bibinfo {author}
  {\bibfnamefont{M.}~\bibnamefont{Katsnelson}}, \emph{et~al.},\ }%
  \bibfield{journal}{%
  \Doi{10.1038/nature04233}{\bibinfo {journal} {Nature}}\ }%
  \textbf{\bibinfo {volume} {438}},\ \bibinfo {pages} {197} (\bibinfo {year}
  {2005}),\
  \Eprint{http://arxiv.org/abs/cond-mat/0509330}{arXiv:cond-mat/0509330
  [cond-mat.mes-hall]}%
  \bibAnnoteFile{NoStop}{Novoselov:2005kj}%
\bibitem{Geim:2007}%
  \BibitemOpen
  \bibfield{author}{%
  \bibinfo {author} {\bibfnamefont{A.~K.}\ \bibnamefont{{Geim}}}\ and\ \bibinfo
  {author} {\bibfnamefont{K.~S.}\ \bibnamefont{{Novoselov}}},\ }%
  \bibfield{journal}{%
  \Doi{10.1038/nmat1849}{\bibinfo {journal} {Nature Materials}}\ }%
  \textbf{\bibinfo {volume} {6}},\ \bibinfo {pages} {183} (\bibinfo {month}
  {Mar.}\ \bibinfo {year} {2007})%
  \bibAnnoteFile{NoStop}{Geim:2007}%
\bibitem{Geim:2009}%
  \BibitemOpen
  \bibfield{author}{%
  \bibinfo {author} {\bibfnamefont{A.~K.}\ \bibnamefont{{Geim}}},\ }%
  \bibfield{journal}{%
  \Doi{10.1126/science.1158877}{\bibinfo {journal} {Science}}\ }%
  \textbf{\bibinfo {volume} {324}},\ \bibinfo {pages} {1530} (\bibinfo {month}
  {Jun.}\ \bibinfo {year} {2009}),\
  \Eprint{http://arxiv.org/abs/0906.3799}{arXiv:0906.3799 [cond-mat.mes-hall]}%
  \bibAnnoteFile{NoStop}{Geim:2009}%
\bibitem{CastroNeto:2009zz}%
  \BibitemOpen
  \bibfield{author}{%
  \bibinfo {author} {\bibfnamefont{A.}~\bibnamefont{Castro~Neto}}, \bibinfo
  {author} {\bibfnamefont{F.}~\bibnamefont{Guinea}}, \bibinfo {author}
  {\bibfnamefont{N.}~\bibnamefont{Peres}}, \bibinfo {author}
  {\bibfnamefont{K.}~\bibnamefont{Novoselov}},\ and\ \bibinfo {author}
  {\bibfnamefont{A.}~\bibnamefont{Geim}},\ }%
  \bibfield{journal}{%
  \bibinfo {journal} {Rev.Mod.Phys.}\ }%
  \textbf{\bibinfo {volume} {81}},\ \bibinfo {pages} {109} (\bibinfo {year}
  {2009})%
  \bibAnnoteFile{NoStop}{CastroNeto:2009zz}%
\bibitem{Zhang:2005}%
  \BibitemOpen
  \bibfield{author}{%
  \bibinfo {author} {\bibfnamefont{Y.}~\bibnamefont{{Zhang}}}, \bibinfo
  {author} {\bibfnamefont{Y.-W.}\ \bibnamefont{{Tan}}}, \bibinfo {author}
  {\bibfnamefont{H.~L.}\ \bibnamefont{{Stormer}}},\ and\ \bibinfo {author}
  {\bibfnamefont{P.}~\bibnamefont{{Kim}}},\ }%
  \bibfield{journal}{%
  \Doi{10.1038/nature04235}{\bibinfo {journal} {\nat}}\ }%
  \textbf{\bibinfo {volume} {438}},\ \bibinfo {pages} {201} (\bibinfo {month}
  {Nov.}\ \bibinfo {year} {2005}),\
  \Eprint{http://arxiv.org/abs/arXiv:cond-mat/0509355}{arXiv:cond-mat/0509355}%
  \bibAnnoteFile{NoStop}{Zhang:2005}%
\bibitem{Katsnelson:2006b}%
  \BibitemOpen
  \bibfield{author}{%
  \bibinfo {author} {\bibfnamefont{M.~I.}\ \bibnamefont{{Katsnelson}}},
  \bibinfo {author} {\bibfnamefont{K.~S.}\ \bibnamefont{{Novoselov}}},\ and\
  \bibinfo {author} {\bibfnamefont{A.~K.}\ \bibnamefont{{Geim}}},\ }%
  \bibfield{journal}{%
  \Doi{10.1038/nphys384}{\bibinfo {journal} {Nature Physics}}\ }%
  \textbf{\bibinfo {volume} {2}},\ \bibinfo {pages} {620} (\bibinfo {month}
  {Sep.}\ \bibinfo {year} {2006}),\
  \Eprint{http://arxiv.org/abs/arXiv:cond-mat/0604323}{arXiv:cond-mat/0604323}%
  \bibAnnoteFile{NoStop}{Katsnelson:2006b}%
\bibitem{Beenakker:2008}%
  \BibitemOpen
  \bibfield{author}{%
  \bibinfo {author} {\bibfnamefont{C.~W.~J.}\ \bibnamefont{Beenakker}},\ }%
  \bibfield{journal}{%
  \Doi{10.1103/RevModPhys.80.1337}{\bibinfo {journal} {Rev. Mod. Phys.}}\ }%
  \textbf{\bibinfo {volume} {80}},\ \bibinfo {pages} {1337} (\bibinfo {month}
  {Oct}\ \bibinfo {year} {2008})%
  \bibAnnoteFile{NoStop}{Beenakker:2008}%
\bibitem{Katsnelson:2006a}%
  \BibitemOpen
  \bibfield{author}{%
  \bibinfo {author} {\bibfnamefont{M.~I.}\ \bibnamefont{{Katsnelson}}},\ }%
  \bibfield{journal}{%
  \Doi{10.1140/epjb/e2006-00203-1}{\bibinfo {journal} {European Physical
  Journal B}}\ }%
  \textbf{\bibinfo {volume} {51}},\ \bibinfo {pages} {157} (\bibinfo {month}
  {May}\ \bibinfo {year} {2006}),\
  \Eprint{http://arxiv.org/abs/arXiv:cond-mat/0512337}{arXiv:cond-mat/0512337}%
  \bibAnnoteFile{NoStop}{Katsnelson:2006a}%
\bibitem{Meyer:2007}%
  \BibitemOpen
  \bibfield{author}{%
  \bibinfo {author} {\bibfnamefont{J.~C.}\ \bibnamefont{{Meyer}}}, \bibinfo
  {author} {\bibfnamefont{A.~K.}\ \bibnamefont{{Geim}}}, \bibinfo {author}
  {\bibfnamefont{M.~I.}\ \bibnamefont{{Katsnelson}}}, \bibinfo {author}
  {\bibfnamefont{K.~S.}\ \bibnamefont{{Novoselov}}}, \bibinfo {author}
  {\bibfnamefont{T.~J.}\ \bibnamefont{{Booth}}},\ and\ \bibinfo {author}
  {\bibfnamefont{S.}~\bibnamefont{{Roth}}},\ }%
  \bibfield{journal}{%
  \Doi{10.1038/nature05545}{\bibinfo {journal} {\nat}}\ }%
  \textbf{\bibinfo {volume} {446}},\ \bibinfo {pages} {60} (\bibinfo {month}
  {Mar.}\ \bibinfo {year} {2007}),\
  \Eprint{http://arxiv.org/abs/arXiv:cond-mat/0701379}{arXiv:cond-mat/0701379}%
  \bibAnnoteFile{NoStop}{Meyer:2007}%
\bibitem{Bolotin:2008}%
  \BibitemOpen
  \bibfield{author}{%
  \bibinfo {author} {\bibfnamefont{K.~I.}\ \bibnamefont{{Bolotin}}}, \bibinfo
  {author} {\bibfnamefont{K.~J.}\ \bibnamefont{{Sikes}}}, \bibinfo {author}
  {\bibfnamefont{Z.}~\bibnamefont{{Jiang}}}, \bibinfo {author}
  {\bibfnamefont{M.}~\bibnamefont{{Klima}}}, \bibinfo {author}
  {\bibfnamefont{G.}~\bibnamefont{{Fudenberg}}}, \bibinfo {author}
  {\bibfnamefont{J.}~\bibnamefont{{Hone}}}, \bibinfo {author}
  {\bibfnamefont{P.}~\bibnamefont{{Kim}}},\ and\ \bibinfo {author}
  {\bibfnamefont{H.~L.}\ \bibnamefont{{Stormer}}},\ }%
  \bibfield{journal}{%
  \Doi{10.1016/j.ssc.2008.02.024}{\bibinfo {journal} {Solid State
  Communications}}\ }%
  \textbf{\bibinfo {volume} {146}},\ \bibinfo {pages} {351} (\bibinfo {month}
  {Jun.}\ \bibinfo {year} {2008}),\
  \Eprint{http://arxiv.org/abs/0802.2389}{arXiv:0802.2389 [cond-mat.mes-hall]}%
  \bibAnnoteFile{NoStop}{Bolotin:2008}%
\bibitem{Du:2008}%
  \BibitemOpen
  \bibfield{author}{%
  \bibinfo {author} {\bibfnamefont{X.}~\bibnamefont{{Du}}}, \bibinfo {author}
  {\bibfnamefont{I.}~\bibnamefont{{Skachko}}}, \bibinfo {author}
  {\bibfnamefont{A.}~\bibnamefont{{Barker}}},\ and\ \bibinfo {author}
  {\bibfnamefont{E.~Y.}\ \bibnamefont{{Andrei}}},\ }%
  \bibfield{journal}{%
  \Doi{10.1038/nnano.2008.199}{\bibinfo {journal} {Nature Nanotechnology}}\ }%
  \textbf{\bibinfo {volume} {3}},\ \bibinfo {pages} {491} (\bibinfo {month}
  {Aug.}\ \bibinfo {year} {2008}),\
  \Eprint{http://arxiv.org/abs/0802.2933}{arXiv:0802.2933}%
  \bibAnnoteFile{NoStop}{Du:2008}%
\bibitem{Kotov:2010}%
  \BibitemOpen
  \bibfield{author}{%
  \bibinfo {author} {\bibfnamefont{V.~N.}\ \bibnamefont{{Kotov}}}, \bibinfo
  {author} {\bibfnamefont{B.}~\bibnamefont{{Uchoa}}}, \bibinfo {author}
  {\bibfnamefont{V.~M.}\ \bibnamefont{{Pereira}}}, \bibinfo {author}
  {\bibfnamefont{F.}~\bibnamefont{{Guinea}}},\ and\ \bibinfo {author}
  {\bibfnamefont{A.~H.}\ \bibnamefont{{Castro Neto}}},\ }%
  \bibfield{journal}{%
  \bibinfo {journal} {ArXiv e-prints}}%
   (\bibinfo {month} {Dec.}\ \bibinfo {year} {2010}),\
  \Eprint{http://arxiv.org/abs/1012.3484}{arXiv:1012.3484 [cond-mat.str-el]}%
  \bibAnnoteFile{NoStop}{Kotov:2010}%
\bibitem{Braun:2011}%
  \BibitemOpen
  \bibfield{author}{%
  \bibinfo {author} {\bibfnamefont{J.}~\bibnamefont{Braun}},\ }%
  \bibfield{journal}{%
  \Doi{10.1088/0954-3899/39/3/033001}{\bibinfo {journal} {J.Phys.G}}\ }%
  \textbf{\bibinfo {volume} {G39}},\ \bibinfo {pages} {033001} (\bibinfo {year}
  {2012}),\ \Eprint{http://arxiv.org/abs/1108.4449}{arXiv:1108.4449 [hep-ph]}%
  \bibAnnoteFile{NoStop}{Braun:2011}%
\bibitem{Sachdev:2001}%
  \BibitemOpen
  \bibfield{author}{%
  \bibinfo {author} {\bibfnamefont{S.}~\bibnamefont{{Sachdev}}},\ }%
  \emph{\bibinfo {title} {Quantum Phase Transitions, by Subir Sachdev,
  pp.~368.~ISBN 0521004543.~Cambridge, UK: Cambridge University Press, April
  2001.}}\ (\bibinfo {year} {2001})%
  \bibAnnoteFile{NoStop}{Sachdev:2001}%
\bibitem{Son:2007ja}%
  \BibitemOpen
  \bibfield{author}{%
  \bibinfo {author} {\bibfnamefont{D.}~\bibnamefont{Son}}}%
   (\bibinfo {year} {2007}),\
  \Eprint{http://arxiv.org/abs/cond-mat/0701501}{arXiv:cond-mat/0701501
  [cond-mat.str-el]}%
  \bibAnnoteFile{NoStop}{Son:2007ja}%
\bibitem{Drut:2007zx}%
  \BibitemOpen
  \bibfield{author}{%
  \bibinfo {author} {\bibfnamefont{J.~E.}\ \bibnamefont{Drut}}\ and\ \bibinfo
  {author} {\bibfnamefont{D.~T.}\ \bibnamefont{Son}},\ }%
  \bibfield{journal}{%
  \Doi{10.1103/PhysRevB.77.075115}{\bibinfo {journal} {Phys.Rev.}}\ }%
  \textbf{\bibinfo {volume} {B77}},\ \bibinfo {pages} {075115} (\bibinfo {year}
  {2008}),\ \Eprint{http://arxiv.org/abs/0710.1315}{arXiv:0710.1315
  [cond-mat.str-el]}%
  \bibAnnoteFile{NoStop}{Drut:2007zx}%
\bibitem{Juricic:2009px}%
  \BibitemOpen
  \bibfield{author}{%
  \bibinfo {author} {\bibfnamefont{V.}~\bibnamefont{Juricic}}, \bibinfo
  {author} {\bibfnamefont{I.~F.}\ \bibnamefont{Herbut}},\ and\ \bibinfo
  {author} {\bibfnamefont{G.~W.}\ \bibnamefont{Semenoff}},\ }%
  \bibfield{journal}{%
  \bibinfo {journal} {Phys.Rev.}\ }%
  \textbf{\bibinfo {volume} {B80}},\ \bibinfo {pages} {081405} (\bibinfo {year}
  {2009}),\ \Eprint{http://arxiv.org/abs/0906.3513}{arXiv:0906.3513
  [cond-mat.str-el]}%
  \bibAnnoteFile{NoStop}{Juricic:2009px}%
\bibitem{Herbut:2009vu}%
  \BibitemOpen
  \bibfield{author}{%
  \bibinfo {author} {\bibfnamefont{I.~F.}\ \bibnamefont{Herbut}}, \bibinfo
  {author} {\bibfnamefont{V.}~\bibnamefont{Juricic}},\ and\ \bibinfo {author}
  {\bibfnamefont{O.}~\bibnamefont{Vafek}},\ }%
  \bibfield{journal}{%
  \Doi{10.1103/PhysRevB.80.075432}{\bibinfo {journal} {Phys.Rev.}}\ }%
  \textbf{\bibinfo {volume} {B80}},\ \bibinfo {pages} {075432} (\bibinfo {year}
  {2009}),\ \Eprint{http://arxiv.org/abs/0904.1019}{arXiv:0904.1019
  [cond-mat.str-el]}%
  \bibAnnoteFile{NoStop}{Herbut:2009vu}%
\bibitem{Semenoff:2012}%
  \BibitemOpen
  \bibfield{author}{%
  \bibinfo {author} {\bibfnamefont{G.~W.}\ \bibnamefont{{Semenoff}}},\ }%
  \bibfield{journal}{%
  \Doi{10.1088/0031-8949/2012/T146/014016}{\bibinfo {journal} {Physica Scripta
  Volume T}}\ }%
  \textbf{\bibinfo {volume} {146}},\ \bibinfo {pages} {014016} (\bibinfo
  {month} {Jan.}\ \bibinfo {year} {2012}),\
  \Eprint{http://arxiv.org/abs/1108.2945}{arXiv:1108.2945 [hep-th]}%
  \bibAnnoteFile{NoStop}{Semenoff:2012}%
\bibitem{Raghu:2008}%
  \BibitemOpen
  \bibfield{author}{%
  \bibinfo {author} {\bibfnamefont{S.}~\bibnamefont{Raghu}}, \bibinfo {author}
  {\bibfnamefont{X.-L.}\ \bibnamefont{Qi}}, \bibinfo {author}
  {\bibfnamefont{C.}~\bibnamefont{Honerkamp}},\ and\ \bibinfo {author}
  {\bibfnamefont{S.-C.}\ \bibnamefont{Zhang}},\ }%
  \bibfield{journal}{%
  \Doi{10.1103/PhysRevLett.100.156401}{\bibinfo {journal} {Phys. Rev. Lett.}}\
  }%
  \textbf{\bibinfo {volume} {100}},\ \bibinfo {pages} {156401} (\bibinfo
  {month} {Apr}\ \bibinfo {year} {2008})%
  \bibAnnoteFile{NoStop}{Raghu:2008}%
\bibitem{Gorbar:2002iw}%
  \BibitemOpen
  \bibfield{author}{%
  \bibinfo {author} {\bibfnamefont{E.~V.}\ \bibnamefont{{Gorbar}}}, \bibinfo
  {author} {\bibfnamefont{V.~P.}\ \bibnamefont{{Gusynin}}}, \bibinfo {author}
  {\bibfnamefont{V.~A.}\ \bibnamefont{{Miransky}}},\ and\ \bibinfo {author}
  {\bibfnamefont{I.~A.}\ \bibnamefont{{Shovkovy}}},\ }%
  \bibfield{journal}{%
  \Doi{10.1103/PhysRevB.66.045108}{\bibinfo {journal} {Phys.Rev.}}\ }%
  \textbf{\bibinfo {volume} {B66}},\ \bibinfo {pages} {045108} (\bibinfo {year}
  {2002}),\
  \Eprint{http://arxiv.org/abs/cond-mat/0202422}{arXiv:cond-mat/0202422
  [cond-mat]}%
  \bibAnnoteFile{NoStop}{Gorbar:2002iw}%
\bibitem{Gamayun:2009em}%
  \BibitemOpen
  \bibfield{author}{%
  \bibinfo {author} {\bibfnamefont{O.~V.}\ \bibnamefont{{Gamayun}}}, \bibinfo
  {author} {\bibfnamefont{E.~V.}\ \bibnamefont{{Gorbar}}},\ and\ \bibinfo
  {author} {\bibfnamefont{V.~P.}\ \bibnamefont{{Gusynin}}},\ }%
  \bibfield{journal}{%
  \Doi{10.1103/PhysRevB.81.075429}{\bibinfo {journal} {Phys.Rev.}}\ }%
  \textbf{\bibinfo {volume} {B81}},\ \bibinfo {pages} {075429} (\bibinfo {year}
  {2010}),\ \Eprint{http://arxiv.org/abs/0911.4878}{arXiv:0911.4878
  [cond-mat.str-el]}%
  \bibAnnoteFile{NoStop}{Gamayun:2009em}%
\bibitem{Drut:2008rg}%
  \BibitemOpen
  \bibfield{author}{%
  \bibinfo {author} {\bibfnamefont{J.~E.}\ \bibnamefont{Drut}}\ and\ \bibinfo
  {author} {\bibfnamefont{T.~A.}\ \bibnamefont{Lahde}},\ }%
  \bibfield{journal}{%
  \Doi{10.1103/PhysRevLett.102.026802}{\bibinfo {journal} {Phys.Rev.Lett.}}\ }%
  \textbf{\bibinfo {volume} {102}},\ \bibinfo {pages} {026802} (\bibinfo {year}
  {2009}),\ \Eprint{http://arxiv.org/abs/0807.0834}{arXiv:0807.0834
  [cond-mat.str-el]}%
  \bibAnnoteFile{NoStop}{Drut:2008rg}%
\bibitem{Drut:2009zi}%
  \BibitemOpen
  \bibfield{author}{%
  \bibinfo {author} {\bibfnamefont{J.~E.}\ \bibnamefont{Drut}}\ and\ \bibinfo
  {author} {\bibfnamefont{T.~A.}\ \bibnamefont{Lahde}},\ }%
  \bibfield{journal}{%
  \Doi{10.1103/PhysRevB.79.241405}{\bibinfo {journal} {Phys.Rev.}}\ }%
  \textbf{\bibinfo {volume} {B79}},\ \bibinfo {pages} {241405} (\bibinfo {year}
  {2009}),\ \Eprint{http://arxiv.org/abs/0905.1320}{arXiv:0905.1320
  [cond-mat.str-el]}%
  \bibAnnoteFile{NoStop}{Drut:2009zi}%
\bibitem{Herbut:2006cs}%
  \BibitemOpen
  \bibfield{author}{%
  \bibinfo {author} {\bibfnamefont{I.~F.}\ \bibnamefont{Herbut}},\ }%
  \bibfield{journal}{%
  \Doi{10.1103/PhysRevLett.97.146401}{\bibinfo {journal} {Phys.Rev.Lett.}}\ }%
  \textbf{\bibinfo {volume} {97}},\ \bibinfo {pages} {146401} (\bibinfo {year}
  {2006}),\
  \Eprint{http://arxiv.org/abs/cond-mat/0606195}{arXiv:cond-mat/0606195
  [cond-mat]}%
  \bibAnnoteFile{NoStop}{Herbut:2006cs}%
\bibitem{Herbut:2009qb}%
  \BibitemOpen
  \bibfield{author}{%
  \bibinfo {author} {\bibfnamefont{I.~F.}\ \bibnamefont{Herbut}}, \bibinfo
  {author} {\bibfnamefont{V.}~\bibnamefont{Juricic}},\ and\ \bibinfo {author}
  {\bibfnamefont{B.}~\bibnamefont{Roy}},\ }%
  \bibfield{journal}{%
  \Doi{10.1103/PhysRevB.79.085116}{\bibinfo {journal} {Phys.Rev.}}\ }%
  \textbf{\bibinfo {volume} {B79}},\ \bibinfo {pages} {085116} (\bibinfo {year}
  {2009}),\ \Eprint{http://arxiv.org/abs/0811.0610}{arXiv:0811.0610
  [cond-mat.str-el]}%
  \bibAnnoteFile{NoStop}{Herbut:2009qb}%
\bibitem{Semenoff:1984dq}%
  \BibitemOpen
  \bibfield{author}{%
  \bibinfo {author} {\bibfnamefont{G.~W.}\ \bibnamefont{Semenoff}},\ }%
  \bibfield{journal}{%
  \Doi{10.1103/PhysRevLett.53.2449}{\bibinfo {journal} {Phys.Rev.Lett.}}\ }%
  \textbf{\bibinfo {volume} {53}},\ \bibinfo {pages} {2449} (\bibinfo {year}
  {1984})%
  \bibAnnoteFile{NoStop}{Semenoff:1984dq}%
\bibitem{Wetterich:1992yh}%
  \BibitemOpen
  \bibfield{author}{%
  \bibinfo {author} {\bibfnamefont{C.}~\bibnamefont{Wetterich}},\ }%
  \bibfield{journal}{%
  \Doi{10.1016/0370-2693(93)90726-X}{\bibinfo {journal} {Phys.Lett.}}\ }%
  \textbf{\bibinfo {volume} {B301}},\ \bibinfo {pages} {90} (\bibinfo {year}
  {1993})%
  \bibAnnoteFile{NoStop}{Wetterich:1992yh}%
\bibitem{Berges:2000ew}%
  \BibitemOpen
  \bibfield{author}{%
  \bibinfo {author} {\bibfnamefont{J.}~\bibnamefont{Berges}}, \bibinfo {author}
  {\bibfnamefont{N.}~\bibnamefont{Tetradis}},\ and\ \bibinfo {author}
  {\bibfnamefont{C.}~\bibnamefont{Wetterich}},\ }%
  \bibfield{journal}{%
  \bibinfo {journal} {Phys.Rept.}\ }%
  \textbf{\bibinfo {volume} {363}},\ \bibinfo {pages} {223} (\bibinfo {year}
  {2002}),\ \Eprint{http://arxiv.org/abs/hep-ph/0005122}{arXiv:hep-ph/0005122
  [hep-ph]}%
  \bibAnnoteFile{NoStop}{Berges:2000ew}%
\bibitem{Haldane:2008a}%
  \BibitemOpen
  \bibfield{author}{%
  \bibinfo {author} {\bibfnamefont{F.~D.~M.}\ \bibnamefont{{Haldane}}}\ and\
  \bibinfo {author} {\bibfnamefont{S.}~\bibnamefont{{Raghu}}},\ }%
  \bibfield{journal}{%
  \Doi{10.1103/PhysRevLett.100.013904}{\bibinfo {journal} {Phys. Rev. Lett.}}\
  }%
  \textbf{\bibinfo {volume} {100}},\ \bibinfo {pages} {013904} (\bibinfo
  {month} {Jan}\ \bibinfo {year} {2008})%
  \bibAnnoteFile{NoStop}{Haldane:2008a}%
\bibitem{Haldane:2008b}%
  \BibitemOpen
  \bibfield{author}{%
  \bibinfo {author} {\bibfnamefont{S.}~\bibnamefont{{Raghu}}}\ and\ \bibinfo
  {author} {\bibfnamefont{F.~D.~M.}\ \bibnamefont{{Haldane}}},\ }%
  \bibfield{journal}{%
  \Doi{10.1103/PhysRevA.78.033834}{\bibinfo {journal} {Phys. Rev. A}}\ }%
  \textbf{\bibinfo {volume} {78}},\ \bibinfo {pages} {033834} (\bibinfo {month}
  {Sep}\ \bibinfo {year} {2008})%
  \bibAnnoteFile{NoStop}{Haldane:2008b}%
\bibitem{Sepkhanov:2007}%
  \BibitemOpen
  \bibfield{author}{%
  \bibinfo {author} {\bibfnamefont{R.~A.}\ \bibnamefont{Sepkhanov}}, \bibinfo
  {author} {\bibfnamefont{Y.~B.}\ \bibnamefont{Bazaliy}},\ and\ \bibinfo
  {author} {\bibfnamefont{C.~W.~J.}\ \bibnamefont{Beenakker}},\ }%
  \bibfield{journal}{%
  \Doi{10.1103/PhysRevA.75.063813}{\bibinfo {journal} {Phys. Rev. A}}\ }%
  \textbf{\bibinfo {volume} {75}},\ \bibinfo {pages} {063813} (\bibinfo {month}
  {Jun}\ \bibinfo {year} {2007})%
  \bibAnnoteFile{NoStop}{Sepkhanov:2007}%
\bibitem{Sepkhanov:2008}%
  \BibitemOpen
  \bibfield{author}{%
  \bibinfo {author} {\bibfnamefont{R.~A.}\ \bibnamefont{Sepkhanov}}, \bibinfo
  {author} {\bibfnamefont{J.}~\bibnamefont{Nilsson}},\ and\ \bibinfo {author}
  {\bibfnamefont{C.~W.~J.}\ \bibnamefont{Beenakker}},\ }%
  \bibfield{journal}{%
  \Doi{10.1103/PhysRevB.78.045122}{\bibinfo {journal} {Phys. Rev. B}}\ }%
  \textbf{\bibinfo {volume} {78}},\ \bibinfo {pages} {045122} (\bibinfo {month}
  {Jul}\ \bibinfo {year} {2008})%
  \bibAnnoteFile{NoStop}{Sepkhanov:2008}%
\bibitem{Bittner:2010}%
  \BibitemOpen
  \bibfield{author}{%
  \bibinfo {author} {\bibfnamefont{S.}~\bibnamefont{{Bittner}}}, \bibinfo
  {author} {\bibfnamefont{B.}~\bibnamefont{{Dietz}}}, \bibinfo {author}
  {\bibfnamefont{M.}~\bibnamefont{{Miski-Oglu}}}, \bibinfo {author}
  {\bibfnamefont{P.}~\bibnamefont{{Oria Iriarte}}}, \bibinfo {author}
  {\bibfnamefont{A.}~\bibnamefont{{Richter}}},\ and\ \bibinfo {author}
  {\bibfnamefont{F.}~\bibnamefont{{Sch{\"a}fer}}},\ }%
  \bibfield{journal}{%
  \Doi{10.1103/PhysRevB.82.014301}{\bibinfo {journal} {\prb}}\ }%
  \textbf{\bibinfo {volume} {82}},\ \bibinfo {eid} {014301} (\bibinfo {month}
  {Jul.}\ \bibinfo {year} {2010}),\
  \Eprint{http://arxiv.org/abs/1005.4506}{arXiv:1005.4506 [cond-mat.mes-hall]}%
  \bibAnnoteFile{NoStop}{Bittner:2010}%
\bibitem{Bittner:2012}%
  \BibitemOpen
  \bibfield{author}{%
  \bibinfo {author} {\bibfnamefont{S.}~\bibnamefont{{Bittner}}}, \bibinfo
  {author} {\bibfnamefont{B.}~\bibnamefont{{Dietz}}}, \bibinfo {author}
  {\bibfnamefont{M.}~\bibnamefont{{Miski-Oglu}}},\ and\ \bibinfo {author}
  {\bibfnamefont{A.}~\bibnamefont{{Richter}}},\ }%
  \bibfield{journal}{%
  \Doi{10.1103/PhysRevB.85.064301}{\bibinfo {journal} {\prb}}\ }%
  \textbf{\bibinfo {volume} {85}},\ \bibinfo {eid} {064301} (\bibinfo {month}
  {Feb.}\ \bibinfo {year} {2012}),\
  \Eprint{http://arxiv.org/abs/1110.3263}{arXiv:1110.3263 [cond-mat.other]}%
  \bibAnnoteFile{NoStop}{Bittner:2012}%
\bibitem{Roy:2011pg}%
  \BibitemOpen
  \bibfield{author}{%
  \bibinfo {author} {\bibfnamefont{B.}~\bibnamefont{Roy}},\ }%
  \bibfield{journal}{%
  \Doi{10.1103/PhysRev.84.113404}{\bibinfo {journal} {Phys.Rev.}}\ }%
  \textbf{\bibinfo {volume} {84}},\ \bibinfo {pages} {113404} (\bibinfo {year}
  {2011}),\ \Eprint{http://arxiv.org/abs/1106.1419}{arXiv:1106.1419
  [cond-mat.str-el]}%
  \bibAnnoteFile{NoStop}{Roy:2011pg}%
\bibitem{Wetterich:2010ni}%
  \BibitemOpen
  \bibfield{author}{%
  \bibinfo {author} {\bibfnamefont{C.}~\bibnamefont{Wetterich}},\ }%
  \bibfield{journal}{%
  \Doi{10.1016/j.nuclphysb.2011.06.013}{\bibinfo {journal} {Nucl.Phys.}}\ }%
  \textbf{\bibinfo {volume} {B852}},\ \bibinfo {pages} {174} (\bibinfo {year}
  {2011}),\ \Eprint{http://arxiv.org/abs/1002.3556}{arXiv:1002.3556 [hep-th]}%
  \bibAnnoteFile{NoStop}{Wetterich:2010ni}%
\bibitem{Osterwalder:1973dx}%
  \BibitemOpen
  \bibfield{author}{%
  \bibinfo {author} {\bibfnamefont{K.}~\bibnamefont{Osterwalder}}\ and\
  \bibinfo {author} {\bibfnamefont{R.}~\bibnamefont{Schrader}},\ }%
  \bibfield{journal}{%
  \Doi{10.1007/BF01645738}{\bibinfo {journal} {Commun.Math.Phys.}}\ }%
  \textbf{\bibinfo {volume} {31}},\ \bibinfo {pages} {83} (\bibinfo {year}
  {1973})%
  \bibAnnoteFile{NoStop}{Osterwalder:1973dx}%
\bibitem{Gomes:1990ed}%
  \BibitemOpen
  \bibfield{author}{%
  \bibinfo {author} {\bibfnamefont{M.}~\bibnamefont{{Gomes}}}, \bibinfo
  {author} {\bibfnamefont{R.~S.}\ \bibnamefont{{Mendes}}}, \bibinfo {author}
  {\bibfnamefont{R.~F.}\ \bibnamefont{{Ribeiro}}},\ and\ \bibinfo {author}
  {\bibfnamefont{A.~J.}\ \bibnamefont{{da Silva}}},\ }%
  \bibfield{journal}{%
  \Doi{10.1103/PhysRevD.43.3516}{\bibinfo {journal} {Phys.Rev.}}\ }%
  \textbf{\bibinfo {volume} {D43}},\ \bibinfo {pages} {3516} (\bibinfo {year}
  {1991})%
  \bibAnnoteFile{NoStop}{Gomes:1990ed}%
\bibitem{Gusynin:2007}%
  \BibitemOpen
  \bibfield{author}{%
  \bibinfo {author} {\bibfnamefont{V.~P.}\ \bibnamefont{{Gusynin}}}, \bibinfo
  {author} {\bibfnamefont{S.~G.}\ \bibnamefont{{Sharapov}}},\ and\ \bibinfo
  {author} {\bibfnamefont{J.~P.}\ \bibnamefont{{Carbotte}}},\ }%
  \bibfield{journal}{%
  \Doi{10.1142/S0217979207038022}{\bibinfo {journal} {International Journal of
  Modern Physics B}}\ }%
  \textbf{\bibinfo {volume} {21}},\ \bibinfo {pages} {4611} (\bibinfo {year}
  {2007}),\ \Eprint{http://arxiv.org/abs/0706.3016}{arXiv:0706.3016}%
  \bibAnnoteFile{NoStop}{Gusynin:2007}%
\bibitem{Gies:2010st}%
  \BibitemOpen
  \bibfield{author}{%
  \bibinfo {author} {\bibfnamefont{H.}~\bibnamefont{Gies}}\ and\ \bibinfo
  {author} {\bibfnamefont{L.}~\bibnamefont{Janssen}},\ }%
  \bibfield{journal}{%
  \Doi{10.1103/PhysRevD.82.085018}{\bibinfo {journal} {Phys.Rev.}}\ }%
  \textbf{\bibinfo {volume} {D82}},\ \bibinfo {pages} {085018} (\bibinfo {year}
  {2010}),\ \Eprint{http://arxiv.org/abs/1006.3747}{arXiv:1006.3747 [hep-th]}%
  \bibAnnoteFile{NoStop}{Gies:2010st}%
\bibitem{Jackiw:1980kv}%
  \BibitemOpen
  \bibfield{author}{%
  \bibinfo {author} {\bibfnamefont{R.}~\bibnamefont{Jackiw}}\ and\ \bibinfo
  {author} {\bibfnamefont{S.}~\bibnamefont{Templeton}},\ }%
  \bibfield{journal}{%
  \Doi{10.1103/PhysRevD.23.2291}{\bibinfo {journal} {Phys.Rev.}}\ }%
  \textbf{\bibinfo {volume} {D23}},\ \bibinfo {pages} {2291} (\bibinfo {year}
  {1981})%
  \bibAnnoteFile{NoStop}{Jackiw:1980kv}%
\bibitem{Kogut:1999iv}%
  \BibitemOpen
  \bibfield{author}{%
  \bibinfo {author} {\bibfnamefont{J.}~\bibnamefont{Kogut}}, \bibinfo {author}
  {\bibfnamefont{M.~A.}\ \bibnamefont{Stephanov}},\ and\ \bibinfo {author}
  {\bibfnamefont{D.}~\bibnamefont{Toublan}},\ }%
  \bibfield{journal}{%
  \bibinfo {journal} {Phys.Lett.}\ }%
  \textbf{\bibinfo {volume} {B464}},\ \bibinfo {pages} {183} (\bibinfo {year}
  {1999}),\ \Eprint{http://arxiv.org/abs/hep-ph/9906346}{arXiv:hep-ph/9906346
  [hep-ph]}%
  \bibAnnoteFile{NoStop}{Kogut:1999iv}%
\bibitem{Kogut:2000ek}%
  \BibitemOpen
  \bibfield{author}{%
  \bibinfo {author} {\bibfnamefont{J.}~\bibnamefont{Kogut}}, \bibinfo {author}
  {\bibfnamefont{M.~A.}\ \bibnamefont{Stephanov}}, \bibinfo {author}
  {\bibfnamefont{D.}~\bibnamefont{Toublan}}, \bibinfo {author}
  {\bibfnamefont{J.}~\bibnamefont{Verbaarschot}},\ and\ \bibinfo {author}
  {\bibfnamefont{A.}~\bibnamefont{Zhitnitsky}},\ }%
  \bibfield{journal}{%
  \Doi{10.1016/S0550-3213(00)00242-X}{\bibinfo {journal} {Nucl.Phys.}}\ }%
  \textbf{\bibinfo {volume} {B582}},\ \bibinfo {pages} {477} (\bibinfo {year}
  {2000}),\ \Eprint{http://arxiv.org/abs/hep-ph/0001171}{arXiv:hep-ph/0001171
  [hep-ph]}%
  \bibAnnoteFile{NoStop}{Kogut:2000ek}%
\bibitem{vonSmekal:2012vx}%
  \BibitemOpen
  \bibfield{author}{%
  \bibinfo {author} {\bibfnamefont{L.}~\bibnamefont{von Smekal}},\ }%
  \bibfield{journal}{%
  \bibinfo {journal} {Nucl. Phys.}\ }%
  \textbf{\bibinfo {volume} {B}},\ \bibinfo {pages} {228 (2012) 179} (\bibinfo
  {year} {Proc. Suppl.}),\
  \Eprint{http://arxiv.org/abs/1205.4205}{arXiv:1205.4205 [hep-ph]}%
  \bibAnnoteFile{NoStop}{vonSmekal:2012vx}%
\bibitem{Heinzner:2004xj}%
  \BibitemOpen
  \bibfield{author}{%
  \bibinfo {author} {\bibfnamefont{P.}~\bibnamefont{Heinzner}}, \bibinfo
  {author} {\bibfnamefont{A.}~\bibnamefont{Huckleberry}},\ and\ \bibinfo
  {author} {\bibfnamefont{M.}~\bibnamefont{Zirnbauer}},\ }%
  \bibfield{journal}{%
  \Doi{10.1007/s00220-005-1330-9}{\bibinfo {journal} {Commun.Math.Phys.}}\ }%
  \textbf{\bibinfo {volume} {257}},\ \bibinfo {pages} {725} (\bibinfo {year}
  {2005}),\ \Eprint{http://arxiv.org/abs/math-ph/0411040}{arXiv:math-ph/0411040
  [math-ph]}%
  \bibAnnoteFile{NoStop}{Heinzner:2004xj}%
\bibitem{Altland:2006}%
  \BibitemOpen
  \bibfield{author}{%
  \bibinfo {author} {\bibfnamefont{A.}~\bibnamefont{{Altland}}},\ }%
  \bibfield{journal}{%
  \Doi{10.1103/PhysRevLett.97.236802}{\bibinfo {journal} {Physical Review
  Letters}}\ }%
  \textbf{\bibinfo {volume} {97}},\ \bibinfo {eid} {236802} (\bibinfo {month}
  {Dec.}\ \bibinfo {year} {2006}),\
  \Eprint{http://arxiv.org/abs/arXiv:cond-mat/0607247}{arXiv:cond-mat/0607247}%
  \bibAnnoteFile{NoStop}{Altland:2006}%
\bibitem{McCann:2006}%
  \BibitemOpen
  \bibfield{author}{%
  \bibinfo {author} {\bibfnamefont{E.}~\bibnamefont{{McCann}}}, \bibinfo
  {author} {\bibfnamefont{K.}~\bibnamefont{{Kechedzhi}}}, \bibinfo {author}
  {\bibfnamefont{V.~I.}\ \bibnamefont{{Falko}}}, \bibinfo {author}
  {\bibfnamefont{H.}~\bibnamefont{{Suzuura}}}, \bibinfo {author}
  {\bibfnamefont{T.}~\bibnamefont{{Ando}}},\ and\ \bibinfo {author}
  {\bibfnamefont{B.~L.}\ \bibnamefont{{Altshuler}}},\ }%
  \bibfield{journal}{%
  \Doi{10.1103/PhysRevLett.97.146805}{\bibinfo {journal} {Physical Review
  Letters}}\ }%
  \textbf{\bibinfo {volume} {97}},\ \bibinfo {eid} {146805} (\bibinfo {month}
  {Oct.}\ \bibinfo {year} {2006}),\
  \Eprint{http://arxiv.org/abs/arXiv:cond-mat/0604015}{arXiv:cond-mat/0604015}%
  \bibAnnoteFile{NoStop}{McCann:2006}%
\bibitem{Ostrovsky:2007}%
  \BibitemOpen
  \bibfield{author}{%
  \bibinfo {author} {\bibfnamefont{P.~M.}\ \bibnamefont{{Ostrovsky}}}, \bibinfo
  {author} {\bibfnamefont{I.~V.}\ \bibnamefont{{Gornyi}}},\ and\ \bibinfo
  {author} {\bibfnamefont{A.~D.}\ \bibnamefont{{Mirlin}}},\ }%
  \bibfield{journal}{%
  \Doi{10.1103/PhysRevLett.98.256801}{\bibinfo {journal} {Physical Review
  Letters}}\ }%
  \textbf{\bibinfo {volume} {98}},\ \bibinfo {eid} {256801} (\bibinfo {month}
  {Jun.}\ \bibinfo {year} {2007}),\
  \Eprint{http://arxiv.org/abs/arXiv:cond-mat/0702115}{arXiv:cond-mat/0702115}%
  \bibAnnoteFile{NoStop}{Ostrovsky:2007}%
\bibitem{DasSarma:2010}%
  \BibitemOpen
  \bibfield{author}{%
  \bibinfo {author} {\bibfnamefont{S.}~\bibnamefont{{Das Sarma}}}, \bibinfo
  {author} {\bibfnamefont{S.}~\bibnamefont{{Adam}}}, \bibinfo {author}
  {\bibfnamefont{E.~H.}\ \bibnamefont{{Hwang}}},\ and\ \bibinfo {author}
  {\bibfnamefont{E.}~\bibnamefont{{Rossi}}},\ }%
  \bibfield{journal}{%
  \Doi{10.1103/RevModPhys.83.407}{\bibinfo {journal} {Reviews of Modern
  Physics}}\ }%
  \textbf{\bibinfo {volume} {83}},\ \bibinfo {pages} {407} (\bibinfo {month}
  {Apr.}\ \bibinfo {year} {2011}),\
  \Eprint{http://arxiv.org/abs/1003.4731}{arXiv:1003.4731 [cond-mat.mes-hall]}%
  \bibAnnoteFile{NoStop}{DasSarma:2010}%
\bibitem{Verbaarschot:1993pm}%
  \BibitemOpen
  \bibfield{author}{%
  \bibinfo {author} {\bibfnamefont{J.}~\bibnamefont{Verbaarschot}}\ and\
  \bibinfo {author} {\bibfnamefont{I.}~\bibnamefont{Zahed}},\ }%
  \bibfield{journal}{%
  \Doi{10.1103/PhysRevLett.70.3852}{\bibinfo {journal} {Phys.Rev.Lett.}}\ }%
  \textbf{\bibinfo {volume} {70}},\ \bibinfo {pages} {3852} (\bibinfo {year}
  {1993}),\ \Eprint{http://arxiv.org/abs/hep-th/9303012}{arXiv:hep-th/9303012
  [hep-th]}%
  \bibAnnoteFile{NoStop}{Verbaarschot:1993pm}%
\bibitem{Verbaarschot:1994qf}%
  \BibitemOpen
  \bibfield{author}{%
  \bibinfo {author} {\bibfnamefont{J.~J.~M.}\ \bibnamefont{Verbaarschot}},\ }%
  \bibfield{journal}{%
  \Doi{10.1103/PhysRevLett.72.2531}{\bibinfo {journal} {Phys.Rev.Lett.}}\ }%
  \textbf{\bibinfo {volume} {72}},\ \bibinfo {pages} {2531} (\bibinfo {year}
  {1994}),\ \Eprint{http://arxiv.org/abs/hep-th/9401059}{arXiv:hep-th/9401059
  [hep-th]}%
  \bibAnnoteFile{NoStop}{Verbaarschot:1994qf}%
\bibitem{Verbaarschot:2000dy}%
  \BibitemOpen
  \bibfield{author}{%
  \bibinfo {author} {\bibfnamefont{J.}~\bibnamefont{Verbaarschot}}\ and\
  \bibinfo {author} {\bibfnamefont{T.}~\bibnamefont{Wettig}},\ }%
  \bibfield{journal}{%
  \Doi{10.1146/annurev.nucl.50.1.343}{\bibinfo {journal}
  {Ann.Rev.Nucl.Part.Sci.}}\ }%
  \textbf{\bibinfo {volume} {50}},\ \bibinfo {pages} {343} (\bibinfo {year}
  {2000}),\ \Eprint{http://arxiv.org/abs/hep-ph/0003017}{arXiv:hep-ph/0003017
  [hep-ph]}%
  \bibAnnoteFile{NoStop}{Verbaarschot:2000dy}%
\bibitem{Zirnbauer:1996zz}%
  \BibitemOpen
  \bibfield{author}{%
  \bibinfo {author} {\bibfnamefont{M.~R.}\ \bibnamefont{Zirnbauer}},\ }%
  \bibfield{journal}{%
  \bibinfo {journal} {J.Math.Phys.}\ }%
  \textbf{\bibinfo {volume} {37}},\ \bibinfo {pages} {4986} (\bibinfo {year}
  {1996})%
  \bibAnnoteFile{NoStop}{Zirnbauer:1996zz}%
\bibitem{Zirnbauer:2010gg}%
  \BibitemOpen
  \bibfield{author}{%
  \bibinfo {author} {\bibfnamefont{M.~R.}\ \bibnamefont{Zirnbauer}}}%
   (\bibinfo {year} {2010}),\
  \Eprint{http://arxiv.org/abs/1001.0722}{arXiv:1001.0722 [math-ph]}%
  \bibAnnoteFile{NoStop}{Zirnbauer:2010gg}%
\bibitem{Gonzalez:1993uz}%
  \BibitemOpen
  \bibfield{author}{%
  \bibinfo {author} {\bibfnamefont{J.}~\bibnamefont{Gonzalez}}, \bibinfo
  {author} {\bibfnamefont{F.}~\bibnamefont{Guinea}},\ and\ \bibinfo {author}
  {\bibfnamefont{M.}~\bibnamefont{Vozmediano}},\ }%
  \bibfield{journal}{%
  \Doi{10.1016/0550-3213(94)90410-3}{\bibinfo {journal} {Nucl.Phys.}}\ }%
  \textbf{\bibinfo {volume} {B424}},\ \bibinfo {pages} {595} (\bibinfo {year}
  {1994}),\ \Eprint{http://arxiv.org/abs/hep-th/9311105}{arXiv:hep-th/9311105
  [hep-th]}%
  \bibAnnoteFile{NoStop}{Gonzalez:1993uz}%
\bibitem{Elias:2011}%
  \BibitemOpen
  \bibfield{author}{%
  \bibinfo {author} {\bibfnamefont{D.~C.}\ \bibnamefont{{Elias}}}, \bibinfo
  {author} {\bibfnamefont{R.~V.}\ \bibnamefont{{Gorbachev}}}, \bibinfo {author}
  {\bibfnamefont{A.~S.}\ \bibnamefont{{Mayorov}}}, \bibinfo {author}
  {\bibfnamefont{S.~V.}\ \bibnamefont{{Morozov}}}, \bibinfo {author}
  {\bibfnamefont{A.~A.}\ \bibnamefont{{Zhukov}}}, \bibinfo {author}
  {\bibfnamefont{P.}~\bibnamefont{{Blake}}}, \bibinfo {author}
  {\bibfnamefont{L.~A.}\ \bibnamefont{{Ponomarenko}}}, \bibinfo {author}
  {\bibfnamefont{I.~V.}\ \bibnamefont{{Grigorieva}}}, \bibinfo {author}
  {\bibfnamefont{K.~S.}\ \bibnamefont{{Novoselov}}}, \bibinfo {author}
  {\bibfnamefont{F.}~\bibnamefont{{Guinea}}},\ and\ \bibinfo {author}
  {\bibfnamefont{A.~K.}\ \bibnamefont{{Geim}}},\ }%
  \bibfield{journal}{%
  \Doi{10.1038/nphys2049}{\bibinfo {journal} {Nature Physics}}\ }%
  \textbf{\bibinfo {volume} {7}},\ \bibinfo {pages} {701} (\bibinfo {month}
  {Sep.}\ \bibinfo {year} {2011}),\
  \Eprint{http://arxiv.org/abs/1104.1396}{arXiv:1104.1396 [cond-mat.mes-hall]}%
  \bibAnnoteFile{NoStop}{Elias:2011}%
\bibitem{Pisarski:1984dj}%
  \BibitemOpen
  \bibfield{author}{%
  \bibinfo {author} {\bibfnamefont{R.~D.}\ \bibnamefont{Pisarski}},\ }%
  \bibfield{journal}{%
  \Doi{10.1103/PhysRevD.29.2423}{\bibinfo {journal} {Phys.Rev.}}\ }%
  \textbf{\bibinfo {volume} {D29}},\ \bibinfo {pages} {2423} (\bibinfo {year}
  {1984})%
  \bibAnnoteFile{NoStop}{Pisarski:1984dj}%
\bibitem{Polychronakos:1987rf}%
  \BibitemOpen
  \bibfield{author}{%
  \bibinfo {author} {\bibfnamefont{A.~P.}\ \bibnamefont{Polychronakos}},\ }%
  \bibfield{journal}{%
  \Doi{10.1103/PhysRevLett.60.1920}{\bibinfo {journal} {Phys.Rev.Lett.}}\ }%
  \textbf{\bibinfo {volume} {60}},\ \bibinfo {pages} {1920} (\bibinfo {year}
  {1988})%
  \bibAnnoteFile{NoStop}{Polychronakos:1987rf}%
\bibitem{Dunne:2003ji}%
  \BibitemOpen
  \bibfield{author}{%
  \bibinfo {author} {\bibfnamefont{G.~V.}\ \bibnamefont{Dunne}}\ and\ \bibinfo
  {author} {\bibfnamefont{S.~M.}\ \bibnamefont{Nishigaki}},\ }%
  \bibfield{journal}{%
  \Doi{10.1016/j.nuclphysb.2003.07.024}{\bibinfo {journal} {Nucl.Phys.}}\ }%
  \textbf{\bibinfo {volume} {B670}},\ \bibinfo {pages} {307} (\bibinfo {year}
  {2003}),\ \Eprint{http://arxiv.org/abs/hep-ph/0306220}{arXiv:hep-ph/0306220
  [hep-ph]}%
  \bibAnnoteFile{NoStop}{Dunne:2003ji}%
\bibitem{Araki:2012gs}%
  \BibitemOpen
  \bibfield{author}{%
  \bibinfo {author} {\bibfnamefont{Y.}~\bibnamefont{Araki}}\ and\ \bibinfo
  {author} {\bibfnamefont{G.}~\bibnamefont{Semenoff}}}%
   (\bibinfo {year} {2012}),\
  \Eprint{http://arxiv.org/abs/1204.4531}{arXiv:1204.4531 [cond-mat.str-el]}%
  \bibAnnoteFile{NoStop}{Araki:2012gs}%
\bibitem{Hou:2006qc}%
  \BibitemOpen
  \bibfield{author}{%
  \bibinfo {author} {\bibfnamefont{C.-Y.}\ \bibnamefont{Hou}}, \bibinfo
  {author} {\bibfnamefont{C.}~\bibnamefont{Chamon}},\ and\ \bibinfo {author}
  {\bibfnamefont{C.}~\bibnamefont{Mudry}},\ }%
  \bibfield{journal}{%
  \Doi{10.1103/PhysRevLett.98.186809}{\bibinfo {journal} {Phys.Rev.Lett.}}\ }%
  \textbf{\bibinfo {volume} {98}},\ \bibinfo {pages} {186809} (\bibinfo {year}
  {2007}),\
  \Eprint{http://arxiv.org/abs/cond-mat/0609740}{arXiv:cond-mat/0609740
  [cond-mat.mes-hall]}%
  \bibAnnoteFile{NoStop}{Hou:2006qc}%
\bibitem{Haldane:1988zza}%
  \BibitemOpen
  \bibfield{author}{%
  \bibinfo {author} {\bibfnamefont{F.~D.~M.}\ \bibnamefont{{Haldane}}},\ }%
  \bibfield{journal}{%
  \Doi{10.1103/PhysRevLett.61.2015}{\bibinfo {journal} {Phys.Rev.Lett.}}\ }%
  \textbf{\bibinfo {volume} {61}},\ \bibinfo {pages} {2015} (\bibinfo {year}
  {1988})%
  \bibAnnoteFile{NoStop}{Haldane:1988zza}%
\bibitem{Bernevig:2006zz}%
  \BibitemOpen
  \bibfield{author}{%
  \bibinfo {author} {\bibfnamefont{B.~A.}\ \bibnamefont{Bernevig}}\ and\
  \bibinfo {author} {\bibfnamefont{S.-C.}\ \bibnamefont{Zhang}},\ }%
  \bibfield{journal}{%
  \Doi{10.1103/PhysRevLett.96.106802}{\bibinfo {journal} {Phys.Rev.Lett.}}\ }%
  \textbf{\bibinfo {volume} {96}},\ \bibinfo {pages} {106802} (\bibinfo {year}
  {2006})%
  \bibAnnoteFile{NoStop}{Bernevig:2006zz}%
\bibitem{Chamon:2007hx}%
  \BibitemOpen
  \bibfield{author}{%
  \bibinfo {author} {\bibfnamefont{C.}~\bibnamefont{Chamon}}, \bibinfo {author}
  {\bibfnamefont{C.-Y.}\ \bibnamefont{Hou}}, \bibinfo {author}
  {\bibfnamefont{R.}~\bibnamefont{Jackiw}}, \bibinfo {author}
  {\bibfnamefont{C.}~\bibnamefont{Mudry}}, \bibinfo {author}
  {\bibfnamefont{S.-Y.}\ \bibnamefont{Pi}}, \emph{et~al.},\ }%
  \bibfield{journal}{%
  \Doi{10.1103/PhysRevB.77.235431}{\bibinfo {journal} {Phys.Rev.}}\ }%
  \textbf{\bibinfo {volume} {B77}},\ \bibinfo {pages} {235431} (\bibinfo {year}
  {2008}),\ \Eprint{http://arxiv.org/abs/0712.2439}{arXiv:0712.2439 [hep-th]}%
  \bibAnnoteFile{NoStop}{Chamon:2007hx}%
\bibitem{Ryu:2009}%
  \BibitemOpen
  \bibfield{author}{%
  \bibinfo {author} {\bibfnamefont{S.}~\bibnamefont{{Ryu}}}, \bibinfo {author}
  {\bibfnamefont{C.}~\bibnamefont{{Mudry}}}, \bibinfo {author}
  {\bibfnamefont{C.-Y.}\ \bibnamefont{{Hou}}},\ and\ \bibinfo {author}
  {\bibfnamefont{C.}~\bibnamefont{{Chamon}}},\ }%
  \bibfield{journal}{%
  \Doi{10.1103/PhysRevB.80.205319}{\bibinfo {journal} {\prb}}\ }%
  \textbf{\bibinfo {volume} {80}},\ \bibinfo {eid} {205319} (\bibinfo {month}
  {Nov.}\ \bibinfo {year} {2009}),\
  \Eprint{http://arxiv.org/abs/0908.3054}{arXiv:0908.3054 [cond-mat.str-el]}%
  \bibAnnoteFile{NoStop}{Ryu:2009}%
\bibitem{Wilson:1971bg}%
  \BibitemOpen
  \bibfield{author}{%
  \bibinfo {author} {\bibfnamefont{K.~G.}\ \bibnamefont{Wilson}},\ }%
  \bibfield{journal}{%
  \Doi{10.1103/PhysRevB.4.3174}{\bibinfo {journal} {Phys.Rev.}}\ }%
  \textbf{\bibinfo {volume} {B4}},\ \bibinfo {pages} {3174} (\bibinfo {year}
  {1971})%
  \bibAnnoteFile{NoStop}{Wilson:1971bg}%
\bibitem{Wilson:1973jj}%
  \BibitemOpen
  \bibfield{author}{%
  \bibinfo {author} {\bibfnamefont{K.}~\bibnamefont{Wilson}}\ and\ \bibinfo
  {author} {\bibfnamefont{J.~B.}\ \bibnamefont{Kogut}},\ }%
  \bibfield{journal}{%
  \Doi{10.1016/0370-1573(74)90023-4}{\bibinfo {journal} {Phys.Rept.}}\ }%
  \textbf{\bibinfo {volume} {12}},\ \bibinfo {pages} {75} (\bibinfo {year}
  {1974})%
  \bibAnnoteFile{NoStop}{Wilson:1973jj}%
\bibitem{Bagnuls:2000ae}%
  \BibitemOpen
  \bibfield{author}{%
  \bibinfo {author} {\bibfnamefont{C.}~\bibnamefont{Bagnuls}}\ and\ \bibinfo
  {author} {\bibfnamefont{C.}~\bibnamefont{Bervillier}},\ }%
  \bibfield{journal}{%
  \bibinfo {journal} {Phys.Rept.}\ }%
  \textbf{\bibinfo {volume} {348}},\ \bibinfo {pages} {91} (\bibinfo {year}
  {2001}),\ \Eprint{http://arxiv.org/abs/hep-th/0002034}{arXiv:hep-th/0002034
  [hep-th]}%
  \bibAnnoteFile{NoStop}{Bagnuls:2000ae}%
\bibitem{Aoki:2000wm}%
  \BibitemOpen
  \bibfield{author}{%
  \bibinfo {author} {\bibfnamefont{K.}~\bibnamefont{Aoki}},\ }%
  \bibfield{journal}{%
  \Doi{10.1016/S0217-9792(00)00092-3}{\bibinfo {journal} {Int.J.Mod.Phys.}}\ }%
  \textbf{\bibinfo {volume} {B14}},\ \bibinfo {pages} {1249} (\bibinfo {year}
  {2000})%
  \bibAnnoteFile{NoStop}{Aoki:2000wm}%
\bibitem{Polonyi:2001se}%
  \BibitemOpen
  \bibfield{author}{%
  \bibinfo {author} {\bibfnamefont{J.}~\bibnamefont{Polonyi}},\ }%
  \bibfield{journal}{%
  \Doi{10.2478/BF02475552}{\bibinfo {journal} {Central Eur.J.Phys.}}\ }%
  \textbf{\bibinfo {volume} {1}},\ \bibinfo {pages} {1} (\bibinfo {year}
  {2003}),\ \Eprint{http://arxiv.org/abs/hep-th/0110026}{arXiv:hep-th/0110026
  [hep-th]}%
  \bibAnnoteFile{NoStop}{Polonyi:2001se}%
\bibitem{Pawlowski:2005xe}%
  \BibitemOpen
  \bibfield{author}{%
  \bibinfo {author} {\bibfnamefont{J.~M.}\ \bibnamefont{Pawlowski}},\ }%
  \bibfield{journal}{%
  \Doi{10.1016/j.aop.2007.01.007}{\bibinfo {journal} {Annals Phys.}}\ }%
  \textbf{\bibinfo {volume} {322}},\ \bibinfo {pages} {2831} (\bibinfo {year}
  {2007}),\ \Eprint{http://arxiv.org/abs/hep-th/0512261}{arXiv:hep-th/0512261
  [hep-th]}%
  \bibAnnoteFile{NoStop}{Pawlowski:2005xe}%
\bibitem{Gies:2006wv}%
  \BibitemOpen
  \bibfield{author}{%
  \bibinfo {author} {\bibfnamefont{H.}~\bibnamefont{Gies}}}%
   (\bibinfo {year} {2006}),\
  \Eprint{http://arxiv.org/abs/hep-ph/0611146}{arXiv:hep-ph/0611146 [hep-ph]}%
  \bibAnnoteFile{NoStop}{Gies:2006wv}%
\bibitem{Delamotte:2007pf}%
  \BibitemOpen
  \bibfield{author}{%
  \bibinfo {author} {\bibfnamefont{B.}~\bibnamefont{Delamotte}}}%
   (\bibinfo {year} {2007}),\
  \Eprint{http://arxiv.org/abs/cond-mat/0702365}{arXiv:cond-mat/0702365
  [COND-MAT]}%
  \bibAnnoteFile{NoStop}{Delamotte:2007pf}%
\bibitem{Rosten:2010vm}%
  \BibitemOpen
  \bibfield{author}{%
  \bibinfo {author} {\bibfnamefont{O.~J.}\ \bibnamefont{Rosten}},\ }%
  \bibfield{journal}{%
  \Doi{10.1016/j.physrep.2011.12.003}{\bibinfo {journal} {Phys.Rept.}}\ }%
  \textbf{\bibinfo {volume} {511}},\ \bibinfo {pages} {177} (\bibinfo {year}
  {2012}),\ \Eprint{http://arxiv.org/abs/1003.1366}{arXiv:1003.1366 [hep-th]}%
  \bibAnnoteFile{NoStop}{Rosten:2010vm}%
\bibitem{Kopietz:2010zz}%
  \BibitemOpen
  \bibfield{author}{%
  \bibinfo {author} {\bibfnamefont{P.}~\bibnamefont{Kopietz}}, \bibinfo
  {author} {\bibfnamefont{L.}~\bibnamefont{Bartosch}},\ and\ \bibinfo {author}
  {\bibfnamefont{F.}~\bibnamefont{Schutz}},\ }%
  \bibfield{journal}{%
  \Doi{10.1007/978-3-642-05094-7}{\bibinfo {journal} {Lect.Notes Phys.}}\ }%
  \textbf{\bibinfo {volume} {798}},\ \bibinfo {pages} {1} (\bibinfo {year}
  {2010})%
  \bibAnnoteFile{NoStop}{Kopietz:2010zz}%
\bibitem{Metzner:2011cw}%
  \BibitemOpen
  \bibfield{author}{%
  \bibinfo {author} {\bibfnamefont{W.}~\bibnamefont{Metzner}}, \bibinfo
  {author} {\bibfnamefont{M.}~\bibnamefont{Salmhofer}}, \bibinfo {author}
  {\bibfnamefont{C.}~\bibnamefont{Honerkamp}}, \bibinfo {author}
  {\bibfnamefont{V.}~\bibnamefont{Meden}},\ and\ \bibinfo {author}
  {\bibfnamefont{K.}~\bibnamefont{Schonhammer}},\ }%
  \bibfield{journal}{%
  \bibinfo {journal} {Rev.Mod.Phys.}\ }%
  \textbf{\bibinfo {volume} {84}},\ \bibinfo {pages} {299} (\bibinfo {year}
  {2012}),\ \Eprint{http://arxiv.org/abs/1105.5289}{arXiv:1105.5289
  [cond-mat.str-el]}%
  \bibAnnoteFile{NoStop}{Metzner:2011cw}%
\bibitem{Canet:2002gs}%
  \BibitemOpen
  \bibfield{author}{%
  \bibinfo {author} {\bibfnamefont{L.}~\bibnamefont{Canet}}, \bibinfo {author}
  {\bibfnamefont{B.}~\bibnamefont{Delamotte}}, \bibinfo {author}
  {\bibfnamefont{D.}~\bibnamefont{Mouhanna}},\ and\ \bibinfo {author}
  {\bibfnamefont{J.}~\bibnamefont{Vidal}},\ }%
  \bibfield{journal}{%
  \Doi{10.1103/PhysRevD.67.065004}{\bibinfo {journal} {Phys.Rev.}}\ }%
  \textbf{\bibinfo {volume} {D67}},\ \bibinfo {pages} {065004} (\bibinfo {year}
  {2003}),\ \Eprint{http://arxiv.org/abs/hep-th/0211055}{arXiv:hep-th/0211055
  [hep-th]}%
  \bibAnnoteFile{NoStop}{Canet:2002gs}%
\bibitem{Morris:1994ie}%
  \BibitemOpen
  \bibfield{author}{%
  \bibinfo {author} {\bibfnamefont{T.~R.}\ \bibnamefont{Morris}},\ }%
  \bibfield{journal}{%
  \Doi{10.1016/0370-2693(94)90767-6}{\bibinfo {journal} {Phys.Lett.}}\ }%
  \textbf{\bibinfo {volume} {B329}},\ \bibinfo {pages} {241} (\bibinfo {year}
  {1994}),\ \Eprint{http://arxiv.org/abs/hep-ph/9403340}{arXiv:hep-ph/9403340
  [hep-ph]}%
  \bibAnnoteFile{NoStop}{Morris:1994ie}%
\bibitem{Morris:1997xj}%
  \BibitemOpen
  \bibfield{author}{%
  \bibinfo {author} {\bibfnamefont{T.~R.}\ \bibnamefont{Morris}}\ and\ \bibinfo
  {author} {\bibfnamefont{M.~D.}\ \bibnamefont{Turner}},\ }%
  \bibfield{journal}{%
  \Doi{10.1016/S0550-3213(97)00640-8}{\bibinfo {journal} {Nucl.Phys.}}\ }%
  \textbf{\bibinfo {volume} {B509}},\ \bibinfo {pages} {637} (\bibinfo {year}
  {1998}),\ \Eprint{http://arxiv.org/abs/hep-th/9704202}{arXiv:hep-th/9704202
  [hep-th]}%
  \bibAnnoteFile{NoStop}{Morris:1997xj}%
\bibitem{Litim:2000ci}%
  \BibitemOpen
  \bibfield{author}{%
  \bibinfo {author} {\bibfnamefont{D.~F.}\ \bibnamefont{Litim}},\ }%
  \bibfield{journal}{%
  \Doi{10.1016/S0370-2693(00)00748-6}{\bibinfo {journal} {Phys.Lett.}}\ }%
  \textbf{\bibinfo {volume} {B486}},\ \bibinfo {pages} {92} (\bibinfo {year}
  {2000}),\ \Eprint{http://arxiv.org/abs/hep-th/0005245}{arXiv:hep-th/0005245
  [hep-th]}%
  \bibAnnoteFile{NoStop}{Litim:2000ci}%
\bibitem{Litim:2001up}%
  \BibitemOpen
  \bibfield{author}{%
  \bibinfo {author} {\bibfnamefont{D.~F.}\ \bibnamefont{Litim}},\ }%
  \bibfield{journal}{%
  \Doi{10.1103/PhysRevD.64.105007}{\bibinfo {journal} {Phys.Rev.}}\ }%
  \textbf{\bibinfo {volume} {D64}},\ \bibinfo {pages} {105007} (\bibinfo {year}
  {2001}),\ \Eprint{http://arxiv.org/abs/hep-th/0103195}{arXiv:hep-th/0103195
  [hep-th]}%
  \bibAnnoteFile{NoStop}{Litim:2001up}%
\bibitem{Latorre:2000qc}%
  \BibitemOpen
  \bibfield{author}{%
  \bibinfo {author} {\bibfnamefont{J.~I.}\ \bibnamefont{Latorre}}\ and\
  \bibinfo {author} {\bibfnamefont{T.~R.}\ \bibnamefont{Morris}},\ }%
  \bibfield{journal}{%
  \bibinfo {journal} {JHEP}\ }%
  \textbf{\bibinfo {volume} {0011}},\ \bibinfo {pages} {004} (\bibinfo {year}
  {2000}),\ \Eprint{http://arxiv.org/abs/hep-th/0008123}{arXiv:hep-th/0008123
  [hep-th]}%
  \bibAnnoteFile{NoStop}{Latorre:2000qc}%
\bibitem{Braun:2010tt}%
  \BibitemOpen
  \bibfield{author}{%
  \bibinfo {author} {\bibfnamefont{J.}~\bibnamefont{Braun}}, \bibinfo {author}
  {\bibfnamefont{H.}~\bibnamefont{Gies}},\ and\ \bibinfo {author}
  {\bibfnamefont{D.~D.}\ \bibnamefont{Scherer}},\ }%
  \bibfield{journal}{%
  \Doi{10.1103/PhysRevD.83.085012}{\bibinfo {journal} {Phys.Rev.}}\ }%
  \textbf{\bibinfo {volume} {D83}},\ \bibinfo {pages} {085012} (\bibinfo {year}
  {2011}),\ \Eprint{http://arxiv.org/abs/1011.1456}{arXiv:1011.1456 [hep-th]}%
  \bibAnnoteFile{NoStop}{Braun:2010tt}%
\bibitem{Wetterich:1991be}%
  \BibitemOpen
  \bibfield{author}{%
  \bibinfo {author} {\bibfnamefont{C.}~\bibnamefont{Wetterich}},\ }%
  \bibfield{journal}{%
  \Doi{10.1007/BF01474340}{\bibinfo {journal} {Z.Phys.}}\ }%
  \textbf{\bibinfo {volume} {C57}},\ \bibinfo {pages} {451} (\bibinfo {year}
  {1993})%
  \bibAnnoteFile{NoStop}{Wetterich:1991be}%
\bibitem{Tetradis:1993ts}%
  \BibitemOpen
  \bibfield{author}{%
  \bibinfo {author} {\bibfnamefont{N.}~\bibnamefont{Tetradis}}\ and\ \bibinfo
  {author} {\bibfnamefont{C.}~\bibnamefont{Wetterich}},\ }%
  \bibfield{journal}{%
  \Doi{10.1016/0550-3213(94)90446-4}{\bibinfo {journal} {Nucl.Phys.}}\ }%
  \textbf{\bibinfo {volume} {B422}},\ \bibinfo {pages} {541} (\bibinfo {year}
  {1994}),\ \Eprint{http://arxiv.org/abs/hep-ph/9308214}{arXiv:hep-ph/9308214
  [hep-ph]}%
  \bibAnnoteFile{NoStop}{Tetradis:1993ts}%
\bibitem{Jungnickel:1995fp}%
  \BibitemOpen
  \bibfield{author}{%
  \bibinfo {author} {\bibfnamefont{D.~U.}\ \bibnamefont{{Jungnickel}}}\ and\
  \bibinfo {author} {\bibfnamefont{C.}~\bibnamefont{{Wetterich}}},\ }%
  \bibfield{journal}{%
  \Doi{10.1103/PhysRevD.53.5142}{\bibinfo {journal} {Phys.Rev.}}\ }%
  \textbf{\bibinfo {volume} {D53}},\ \bibinfo {pages} {5142} (\bibinfo {year}
  {1996}),\ \Eprint{http://arxiv.org/abs/hep-ph/9505267}{arXiv:hep-ph/9505267
  [hep-ph]}%
  \bibAnnoteFile{NoStop}{Jungnickel:1995fp}%
\bibitem{Berges:1996ja}%
  \BibitemOpen
  \bibfield{author}{%
  \bibinfo {author} {\bibfnamefont{J.}~\bibnamefont{Berges}}\ and\ \bibinfo
  {author} {\bibfnamefont{C.}~\bibnamefont{Wetterich}},\ }%
  \bibfield{journal}{%
  \Doi{10.1016/S0550-3213(96)00670-0}{\bibinfo {journal} {Nucl.Phys.}}\ }%
  \textbf{\bibinfo {volume} {B487}},\ \bibinfo {pages} {675} (\bibinfo {year}
  {1997}),\ \Eprint{http://arxiv.org/abs/hep-th/9609019}{arXiv:hep-th/9609019
  [hep-th]}%
  \bibAnnoteFile{NoStop}{Berges:1996ja}%
\bibitem{Rosa:2000ju}%
  \BibitemOpen
  \bibfield{author}{%
  \bibinfo {author} {\bibfnamefont{L.}~\bibnamefont{Rosa}}, \bibinfo {author}
  {\bibfnamefont{P.}~\bibnamefont{Vitale}},\ and\ \bibinfo {author}
  {\bibfnamefont{C.}~\bibnamefont{Wetterich}},\ }%
  \bibfield{journal}{%
  \Doi{10.1103/PhysRevLett.86.958}{\bibinfo {journal} {Phys.Rev.Lett.}}\ }%
  \textbf{\bibinfo {volume} {86}},\ \bibinfo {pages} {958} (\bibinfo {year}
  {2001}),\ \Eprint{http://arxiv.org/abs/hep-th/0007093}{arXiv:hep-th/0007093
  [hep-th]}%
  \bibAnnoteFile{NoStop}{Rosa:2000ju}%
\bibitem{Hofling:2002hj}%
  \BibitemOpen
  \bibfield{author}{%
  \bibinfo {author} {\bibfnamefont{F.}~\bibnamefont{Hofling}}, \bibinfo
  {author} {\bibfnamefont{C.}~\bibnamefont{Nowak}},\ and\ \bibinfo {author}
  {\bibfnamefont{C.}~\bibnamefont{Wetterich}},\ }%
  \bibfield{journal}{%
  \Doi{10.1103/PhysRevB.66.205111}{\bibinfo {journal} {Phys.Rev.}}\ }%
  \textbf{\bibinfo {volume} {B66}},\ \bibinfo {pages} {205111} (\bibinfo {year}
  {2002}),\
  \Eprint{http://arxiv.org/abs/cond-mat/0203588}{arXiv:cond-mat/0203588
  [cond-mat]}%
  \bibAnnoteFile{NoStop}{Hofling:2002hj}%
\bibitem{Strack:2009ia}%
  \BibitemOpen
  \bibfield{author}{%
  \bibinfo {author} {\bibfnamefont{P.}~\bibnamefont{Strack}}, \bibinfo {author}
  {\bibfnamefont{S.}~\bibnamefont{Takei}},\ and\ \bibinfo {author}
  {\bibfnamefont{W.}~\bibnamefont{Metzner}},\ }%
  \bibfield{journal}{%
  \Doi{10.1103/PhysRevB.81.125103}{\bibinfo {journal} {Phys.Rev.}}\ }%
  \textbf{\bibinfo {volume} {B81}},\ \bibinfo {pages} {125103} (\bibinfo {year}
  {2010}),\ \Eprint{http://arxiv.org/abs/0905.3894}{arXiv:0905.3894
  [cond-mat.str-el]}%
  \bibAnnoteFile{NoStop}{Strack:2009ia}%
\bibitem{Obert:2011}%
  \BibitemOpen
  \bibfield{author}{%
  \bibinfo {author} {\bibfnamefont{B.}~\bibnamefont{{Obert}}}, \bibinfo
  {author} {\bibfnamefont{S.}~\bibnamefont{{Takei}}},\ and\ \bibinfo {author}
  {\bibfnamefont{W.}~\bibnamefont{{Metzner}}},\ }%
  \bibfield{journal}{%
  \Doi{10.1002/andp.201100039}{\bibinfo {journal} {Annalen der Physik}}\ }%
  \textbf{\bibinfo {volume} {523}},\ \bibinfo {pages} {621} (\bibinfo {month}
  {Aug.}\ \bibinfo {year} {2011}),\
  \Eprint{http://arxiv.org/abs/1104.2988}{arXiv:1104.2988 [cond-mat.str-el]}%
  \bibAnnoteFile{NoStop}{Obert:2011}%
\bibitem{Herbut:1996ut}%
  \BibitemOpen
  \bibfield{author}{%
  \bibinfo {author} {\bibfnamefont{I.~F.}\ \bibnamefont{Herbut}}\ and\ \bibinfo
  {author} {\bibfnamefont{Z.}~\bibnamefont{Tesanovic}},\ }%
  \bibfield{journal}{%
  \Doi{10.1103/PhysRevLett.76.4588}{\bibinfo {journal} {Phys.Rev.Lett.}}\ }%
  \textbf{\bibinfo {volume} {76}},\ \bibinfo {pages} {4588} (\bibinfo {year}
  {1996}),\
  \Eprint{http://arxiv.org/abs/cond-mat/9605185}{arXiv:cond-mat/9605185
  [cond-mat]}%
  \bibAnnoteFile{NoStop}{Herbut:1996ut}%
\bibitem{Hove:2000zz}%
  \BibitemOpen
  \bibfield{author}{%
  \bibinfo {author} {\bibfnamefont{J.}~\bibnamefont{Hove}}\ and\ \bibinfo
  {author} {\bibfnamefont{A.}~\bibnamefont{Sudbo}},\ }%
  \bibfield{journal}{%
  \Doi{10.1103/PhysRevLett.84.3426}{\bibinfo {journal} {Phys.Rev.Lett.}}\ }%
  \textbf{\bibinfo {volume} {84}},\ \bibinfo {pages} {3426} (\bibinfo {year}
  {2000})%
  \bibAnnoteFile{NoStop}{Hove:2000zz}%
\end{thebibliography}%

\end{document}